\documentclass[aps,prb,twocolumn,amsmath,amssymb,nofootinbib,superscriptaddress,floatfix,eqsecnum]{revtex4}

\usepackage{amsmath}
\usepackage{amssymb}
\usepackage{amsthm}
\usepackage[pdftex]{color}
\usepackage{graphicx}% Include figure files
\usepackage{dcolumn} % Align table columns on decimal point
\usepackage{bm} % bold math
\usepackage{longtable}
\usepackage{ulem}   % to strike things out
\newcommand{\bs}[1]{{\boldsymbol{#1}}}

\begin{document}

\title{
Noncommutative geometry for three-dimensional topological insulators
      }

\author{Titus Neupert} 
\affiliation{
Condensed Matter Theory Group, 
Paul Scherrer Institute, CH-5232 Villigen PSI,
Switzerland
            } 

\author{Luiz Santos} 
\affiliation{
Department of Physics, 
Harvard University, 
17 Oxford St., 
Cambridge, Massachusetts 02138, USA
            } 

\author{Shinsei Ryu} 
\affiliation{
Department of Physics, University of Illinois, 
1110 West Green St, Urbana Illinois 61801, USA
            } 

\author{Claudio Chamon} 
\affiliation{
Physics Department, 
Boston University, 
Boston, Massachusetts 02215, USA
            } 

\author{Christopher Mudry} 
\affiliation{
Condensed Matter Theory Group, 
Paul Scherrer Institute, CH-5232 Villigen PSI,
Switzerland
            } 

\date{\today}

\begin{abstract}
We generalize the noncommutative relations obeyed by the
guiding centers in the two-dimensional quantum Hall effect
to those obeyed by the projected position operators in 
three-dimensional~(3D) topological band insulators.
The noncommutativity in 3D space is tied to the
integral over the 3D Brillouin zone of a Chern-Simons invariant in
momentum-space. We provide an example of a model on the cubic lattice 
for which the chiral symmetry guarantees a macroscopic number 
of zero-energy modes that form a perfectly flat band. 
This lattice model realizes a chiral 3D noncommutative geometry.
Finally, we find conditions on the density-density
structure factors that lead to a gapped 3D fractional chiral
topological insulator within Feynman's single-mode approximation.
\end{abstract}

\maketitle

%\tableofcontents
%\newpage

\medskip
\section{Introduction}

The integer quantum Hall effect (IQHE)%
\cite{Klitzing80}
is the first known example of a
fermionic phase of matter
characterized by a topological index that is directly
connected to a physical observable.
The index in this case is the sum of
the (first) Chern numbers obtained for each of the fully filled Landau
bands, and the associated physical observable
is the Hall conductance.%
~\cite{Thouless82,Avron83,Simon83} 
The fractional quantum Hall effect (FQHE) %
~\cite{Tsui82}
results from the effects of electron-electron interactions when the
Landau levels are partially filled with electrons, for certain
rational filling fractions.%
~\cite{Laughlin83} 
More examples of
topological states of matter that are comprised of noninteracting
fermions have been discovered recently,%
~\cite{Kane05a,Kane05b,Bernevig06a,Bernevig06b,Konig07,Fu07,Moore07,Qi08,Hsieh08,Hsieh09}
and have been classified
according to discrete symmetries they respect or not, 
and the dimensionality of space in which the particles propagate.%
~\cite{Schnyder08,Kitaev09,Ryu10} 
Such classification is sometimes
referred to as the ``periodic table'' of topological insulators.%
~\cite{Kitaev09}  
Among these states are $\mathbb{Z}^{\ }_{2}$ 
topological ones associated with the
presence of time-reversal symmetry (TRS) in 
two-dimensional (2D) as well as three-dimensional (3D)
systems. It is natural to then question what the ``fractional''
version of these phases should be, and how they could be described. In
particular, it is interesting to ask what are the possible fractional
topological phases of interacting fermionic systems in three spatial
dimensions.

One approach to capture universal physics arising from topological
interacting electron systems in (2+1)%
~\cite{Jain89,Wen91,Wen99,Levin09,Barkeshli10,Vaezi11,Lu11,McGreevy11}  
and (3+1)%
~\cite{{Maciejko10,Swingle11,Maciejko11}}
dimensions of space and time is via the parton construction: 
a fractional phase of electrons is obtained by
constructing integer filled bands of ``partons'', 
which are then ``glued'' together by very strong
gauge-mediated interactions so as to assemble together the physical electron. 
This approach is a generalization of
theories that capture the universal physics of the FQHE, and yields,
for instance, wavefunctions describing states which have fractional
magneto-electric effects %
~\cite{Maciejko10,Swingle11,Maciejko11}
in the case of the $\mathbb Z^{\ }_{2}$
topological insulators. The parton construction is one way to
obtain an effective topological quantum field theory (TQFT) to
describe fractional topological insulators.

However, TQFTs do not capture the dynamics of the systems beyond their
topological properties. As emphasized by
Haldane,%
~\cite{Haldane12} TQFTs are incomplete theories
of the FQHE, for while they characterize the quantum numbers of the
elementary excitations (topological defects), such as their charges and
statistics, they do not contain any information about their
energies. The information about the fundamental
length scale in the FQHE, the magnetic length, is lost
in its  TQFT treatment. Recently Haldane has proposed in Refs.%
~\onlinecite{Haldane12} and \onlinecite{Haldane11} 
a geometric description of the FQHE based on the algebra 
obeyed by
the density operators projected to the lowest Landau level that was
originally introduced by Girvin, MacDonald and Platzman (GMP)
in Ref.~\onlinecite{Girvin85}.

When projected to the lowest Landau level, the density operators do
not commute. However, the algebra closes in that the commutation of
two density operators is proportional to a third one.  Using this
algebra, GMP were able to employ an approach that parallels that of
Feynman and Bijl in their study of excitations in $^4$He.%
~\cite{Feynman72} Their approach allows to place a variational
estimate on the excitation gap, if the static structure factor is
known. The algebraic approach to the FQHE pioneered by GMP has also
been useful to understand the hydrodynamic description of the edge
states in the IQHE and FQHE.%
\cite{Iso92,Cappelli93,Martinez93}
More recently, Parameswaran, Roy, and Sondhi in
Ref.~\onlinecite{Parameswaran11} 
have initiated a study of the algebra obeyed by the density operators
in two-dimensional Chern band insulators
(see also Refs.~\onlinecite{Goerbig12} and \onlinecite{Bernevig11}). 
Our work in 3D is motivated by this successful approach in 2D.

The main objective of this work is to identify the 
noncommutative geometry
that can emerge from
3D topological insulators, 
its relation to topological invariants, and its relevance to possible
interaction-driven topological fractional phases in fermionic 3D systems. 
Armed with this noncommutative geometry, 
one can forge ahead in trying to
construct a dynamical theory of 3D fractional topological insulators
that could perhaps parallel the solid understanding of the FQHE in
2D. In particular, the approach might suggest 
which types of interactions can give rise to incompressible gapped phases.

There are important symmetry considerations that need to be carefully
taken into account when searching for interacting topological insulators
in 3D. The FQHE descends from the IQHE when Landau or Chern bands are
partially filled. In turn, the 2D IQHE is a stable class of states
characterized by a $\mathbb Z$ index 
(symmetry class A in the terminology of Ref.~\onlinecite{Schnyder08}), 
which has no symmetry left out to be broken. If the logic is that we are
also to start from a noninteracting topological insulator in 3D when
constructing the interacting fractional counterpart, we need to look
at systems which are topologically nontrivial in 3D space. One
possibility is to start with $\mathbb{Z}^{\ }_{2}$ topological insulators.
This has been the choice in most works so far. Here, instead, we shall
start from systems that have chiral symmetry, but that lack TRS (symmetry 
class AIII in the terminology of Ref.~\onlinecite{Schnyder08}). 
The rational for this choice is twofold. First, from experience
working on strongly interacting 2D $\mathbb Z^{\ }_{2}$ topological
insulators, we have observed that TRS is easily broken in favor of
magnetized states due to the Stoner instability, which is enhanced in
bands with nonzero topological invariant.%
~\cite{Neupert11b,Neupert12,Xiao11}
Second, because the 3D chiral systems are characterized by a 
$\mathbb{Z}$-valued topological invariant, 
it might keep a closer parallel to the FQHE. Indeed,
we shall show that the noncommutative geometry
for this 3D model does
depend on this $\mathbb Z$-valued topological invariant.

The approach of GMP is ideally suited to the situation 
where density operators are projected
into
a dispersionless band (for example the lowest Landau level in the case
of the FQHE). Here, we shall give a concrete lattice model with chiral
symmetry that contains an exactly flat topological band, on which we
construct the projected density operators. The resulting 
operator product expansions
will depend on the nonzero integral over the 3D Brillouin zone
of a Chern-Simons action in momentum-space. 
For this lattice model, the average Berry
curvature over the entire Brillouin zone is zero. Hence, the type of
3D fractional topological insulator that we discuss is qualitatively
different from the FQHE, where the average Berry curvature over the
Brillouin zone is nonzero. The nature of the fractional states we
discuss are intrinsically 3D, and not layered 2D 
(i.e., weak topological insulators).

This paper is organized as follows.
We show in Sec.~\ref{sec: Noncommutative geometry} how the
position and density operators for noninteracting
fermions, if projected onto the occupied bands 
of their insulating ground state,
can generate a noncommutative geometry. 
Although this is done explicitly in 2D and 3D space,
the method applies to any dimension of space.
A minimal microscopic 3D noninteracting lattice model that realizes
the conditions necessary to establish the noncommutative geometry
of Sec.~\ref{sec: Noncommutative geometry}
is presented in Sec.%
~\ref{sec: Noninteracting three-band tight-binding model}.
The role of interactions is then discussed in
Sec.~\ref{sec: Interactions within the single-mode approximation}.

\medskip
\section{
Noncommutative geometry
        }
\label{sec: Noncommutative geometry}

We begin by recalling some elementary facts about the 
quantum motion of a spinless electron
confined to move in the plane 
spanned by the orthonormal unit vectors 
$\bs{e}^{\ }_{1}$ and $\bs{e}^{\ }_{2}$
perpendicular to an applied uniform magnetic field 
$\bs{B}=B\,\bs{e}^{\ }_{3}$,
whereby
$\bs{e}^{\ }_{3}=\bs{e}^{\ }_{1}\wedge\bs{e}^{\ }_{2}$.

Its quantum dynamics is governed by the single-particle (Landau)
Hamiltonian
\begin{equation}
\widehat{\mathcal{H}}:=
\frac{1}{2m^{\ }_{\mathrm{e}}}
\left[
\widehat{\bs{P}}
+
\frac{e}{c}
\bs{A}(\widehat{\bs{R}})
\right]^{2},
\qquad
\bs{B}=\bs{\nabla}\wedge\bs{A}(\bs{r}),
\label{eq: def Landau Hamiltonian}
\end{equation}
where the momentum 
$\widehat{\bs{P}}^{\mathsf{T}}\equiv(\widehat{P}^{\ }_{1},\widehat{P}^{\ }_{2})$ 
and position 
$\widehat{\bs{R}}^{\mathsf{T}}\equiv(\widehat{R}^{\ }_{1},\widehat{R}^{\ }_{2})$
operators obey the canonical commutation relation
\begin{equation}
\left[\widehat{R}^{\ }_{\mu},\widehat{P}^{\ }_{\nu}\right]=
\mathrm{i}\hbar\,\delta^{\ }_{\mu,\nu}
\end{equation}
with $\mu,\nu=1,2$. 

Hence, neither do the components of the 
covariant derivative in position space
\begin{subequations}
\begin{equation}
\widehat{\bs{\Pi}}:=
\frac{\mathrm{i}m^{\ }_{\mathrm{e}}}{\hbar}
\left[\widehat{\mathcal{H}},\widehat{\bs{R}}\right]=
\widehat{\bs{P}}
+
\frac{e}{c}
\bs{A}(\widehat{\bs{R}})
\label{velocity}
\end{equation}
nor do the components of the conserved guiding center
\begin{equation}
\widehat{\bs{X}}:=
\widehat{\bs{R}}
-
\frac{\ell^2_B}{\hbar}
\bs{e}^{\ }_{3}\wedge\widehat{\bs{\Pi}}
\label{guiding center}
\end{equation}
\end{subequations}
commute, for
\begin{subequations}
\label{eq: noncommuting algebra in QHE}
\begin{equation}
\left[
\widehat{\Pi}^{\ }_{1},
\widehat{\Pi}^{\ }_{2}
\right]=
-\mathrm{i}\frac{\hbar^{2}}{\ell^2_B}
\end{equation}  
and
\begin{equation}
\left[
\widehat{X}^{\ }_{1},
\widehat{X}^{\ }_{2}
\right]=
+\mathrm{i}\,\ell^2_B
\label{eq: noncommuting algebra in QHE b}
\end{equation}  
\end{subequations}
with $\ell^{\ }_B=\sqrt{\hbar c/(eB)}$ the magnetic length.

An orthonormal basis of energy eigenstates of the Landau Hamiltonian%
~\eqref{eq: def Landau Hamiltonian} is made of the kets
\begin{subequations}
\begin{equation}
|n,m\rangle:=
\frac{1}{\sqrt{n!m!}} 
(\widehat{a}^{\dag})^n (\widehat{b}^{\dag})^m |0\rangle,
\end{equation}
with
\begin{equation}
\widehat{a}^{\dag}:=
\frac{\ell^{\ }_B}{\sqrt{2}\hbar}
\left(\widehat{\Pi}^{\ }_1+\mathrm{i}\widehat{\Pi}^{\ }_2\right),
\quad
\widehat{b}^{\dag}:=
\frac{1}{\sqrt{2}\,\ell^{\ }_B}
\left(\widehat{X}^{\ }_1+\mathrm{i}\widehat{X}^{\ }_2\right),
\end{equation}
\end{subequations}
and where $n=0,1,2,\cdots$ labels the Landau levels with energy
$\varepsilon^{\ }_n=\hbar \omega_c\;(n+1/2)$ and
$m=0,1,2,\cdots,[(\Phi/\Phi^{\ }_{0})-1]$ labels the orbital angular momentum.
Here, $\Phi=AB$ is the magnetic flux threading the area $A$ of the
system, $\Phi^{\ }_{0}=hc/e$ is the flux quantum, and
$\omega_c=eB/m^{\ }_{\mathrm{e}} c$ is the cyclotron frequency.

Defining the projector on the $n$-th Landau level 
\begin{equation}
\widehat{\mathcal{P}}^{\ }_n:=
\sum_{m}
|n,m\rangle\langle n,m|,
\end{equation}
one finds that the guiding center defined in Eq.~\eqref{guiding center} 
is the position operator projected on any single Landau level
\begin{equation}
\begin{split}
\widehat{\bs{X}}=&\,
\frac{\ell^{\ }_B}{\sqrt{2}}
\begin{pmatrix}
\widehat{b}+\widehat{b}^\dag
\\ 
\mathrm{i}\widehat{b}-\mathrm{i}\widehat{b}^\dag
\end{pmatrix}
\\
=&\,
\frac{\ell^{\ }_B}{\sqrt{2}}
\widehat{\mathcal{P}}^{\ }_n
\left[
\begin{pmatrix}
\widehat{b}+\widehat{b}^\dag\\ 
\mathrm{i}\widehat{b}-\mathrm{i}\widehat{b}^\dag
\end{pmatrix}
-
\begin{pmatrix}
\mathrm{i}\widehat{a}^\dag-\mathrm{i}\widehat{a}
\\
\widehat{a}^\dag+\widehat{a}
\end{pmatrix}
\right]\widehat{\mathcal{P}}^{\ }_n
\\
=&\,
\widehat{\mathcal{P}}^{\ }_n\,\widehat{\bs{R}}\,\widehat{\mathcal{P}}^{\ }_n,
\end{split}
\end{equation}
since 
$\widehat{\mathcal{P}}^{\ }_n\,\widehat{a}\,\widehat{\mathcal{P}}^{\ }_n=
0$ 
and 
$\widehat{\mathcal{P}}^{\ }_n\,\widehat{a}^{\dag}\,\widehat{\mathcal{P}}^{\ }_n
=0$,
while
$\widehat{\mathcal{P}}^{\ }_n\,\widehat{b}\,\widehat{\mathcal{P}}^{\ }_n
=\widehat{b}$ 
and 
$\widehat{\mathcal{P}}^{\ }_n\,\widehat{b}^{\dag}\,\widehat{\mathcal{P}}^{\ }_n
=\widehat{b}^{\dag}$.
Thus, the position operators projected to any given Landau level 
satisfy the noncommutative   
geometry~\eqref{eq: noncommuting algebra in QHE b}.

This noncommutative geometry
is at the heart of both the IQHE and the FQHE.
For example, it is intimately related to the quantized Hall conductivity 
$\sigma^{\mathrm{H}}$.
The Kubo formula for the contribution of the $n$-th Landau level 
($n=0,1,2,\cdots$)
to the Hall conductivity is 
\begin{equation}
\sigma^{\mathrm{H}}_n:=
\frac{e^2\hbar}{\mathrm{i}m^{2}_{\mathrm{e}}}
\frac{1}{A}
\sum_{n'\neq n} \sum_{m}
\frac{
\langle n,m|\widehat{\Pi}^{\ }_1\widehat{\mathcal{P}}^{\ }_{n'}\widehat{\Pi}^{\ }_2|n,m\rangle
-
(1 \leftrightarrow 2)
}{(\varepsilon^{\ }_{n}-\varepsilon^{\ }_{n'})^2},
\end{equation}
where $A$ is the area of the Hall droplet.
This can be rewritten using 
Eq.~\eqref{velocity}
as
\begin{equation}
\begin{split}
\sigma^{\mathrm{H}}_n=&\,
\frac{\mathrm{i}e^2}{A\hbar}
\sum_{n'\neq n} \sum_{m}
\left[
\langle n,m|\widehat{R}^{\ }_1\widehat{\mathcal{P}}^{\ }_{n'}\widehat{R}^{\ }_2|n,m\rangle
-
(1 \leftrightarrow 2)
\right]
\\
=&\,-
\frac{\mathrm{i}e^2}{A\hbar}
\sum_{m}
\left[
\langle n,m|\widehat{R}^{\ }_1\widehat{\mathcal{P}}^{\ }_{n}\widehat{R}^{\ }_2|n,m\rangle
-
(1 \leftrightarrow 2)
\right]
\\
=&\,
-
\frac{\mathrm{i}e^2}{A\hbar}
\sum_{m}
\left\langle n,m\left|
\left[\widehat{X}^{\ }_1,\widehat{X}^{\ }_2\right]
\right|n,m\right\rangle
\\
=&\,
\frac{e^2}{h},
\end{split}
\label{eq: Hall conductivity from noncommutative}
\end{equation}
where we used that $A=2\pi\sum_m\ell^2_B$.
The role of the noncommutative position-operator algebra 
is apparent in the penultimate line.

To quadratic order in $\ell^{\ }_B$, 
the algebra of the projected position operators~\eqref{guiding center}
is maintained if a coordinate transformation 
$r^{\ }_{\mu}\to f^{\ }_{\mu}(\bs{r})$, $\mu=1,2$, 
that varies on length scales larger than $\ell^{\ }_B$, is area preserving. 
Indeed, we can then expand
\begin{subequations}
\begin{equation}
\label{eq:2D_f_commutator}
\left[
f^{\ }_{1}(\widehat{\bs{X}}),
f^{\ }_{2}(\widehat{\bs{X}})
\right]=
+\mathrm{i}\,\ell^2_B
\,\{f^{\ }_{1},f^{\ }_{2}\}_{\mathrm{P}}\,(\widehat{\bs{X}})
+
\mathcal{O}\left(\ell^4_B\right),
\end{equation}
where the classical Poisson bracket is defined as
\begin{equation}
\left\{f^{\ }_{1},f^{\ }_{2}\right\}^{\ }_{\mathrm{P}}(\bs{r}):=
\epsilon^{\mu\nu}
\left(
\frac{\partial f^{\ }_1}{\partial r^{\ }_\mu}
\frac{\partial f^{\ }_2}{\partial r^{\ }_\nu}
\right)(\bs{r}).
\label{eq: commutator f1 f2 b intro}
\end{equation}
\end{subequations}
The condition for this coordinate transformation to 
locally preserve area is that its Jacobian equals unity, 
or equivalently that 
$\left\{f^{\ }_{1},f^{\ }_{2}\right\}^{\ }_{\mathrm{P}}(\bs{r})=1$. 
In this case, it follows that
$\left[f^{\ }_{1}(\widehat{\bs{X}}),f^{\ }_{2}(\widehat{\bs{X}})\right]=
+\mathrm{i}\,\ell^2_B+\mathcal{O}(\ell^4_B)$. 

From the projected coordinate algebra,
one can obtain a (projected) density algebra, 
by defining the projected density
\begin{subequations}
\begin{equation}
\widehat{\rho}(\bs{r}):=
\widehat{\mathcal{P}}^{\ }_n\,
\widehat{\varrho}(\bs{r})\,
\widehat{\mathcal{P}}^{\ }_n,
\end{equation}
where the unprojected density operator is 
\begin{equation}
\widehat{\varrho}(\bs{r}):=
\delta\left(\bs{r}-\widehat{\bs{R}}\right).
\end{equation}
\end{subequations}
One can also construct the guiding center operators%
~\eqref{guiding center} from the projected density operators through
\begin{equation}
\widehat X^{\ }_{\mu}=
\int\mathrm{d}^2\bs{r}\,
r^{\ }_{\mu}\,
\widehat{\mathcal{P}}^{\ }_{n}\,
\widehat{\varrho}(\bs{r})\,\widehat{\mathcal{P}}^{\ }_{n},
\qquad
\mu=1,2.
\label{eq: rewriting guiding center}
\end{equation}

In momentum space, the projected normal ordered density operators 
\begin{equation}
\begin{split}
\widehat{\rho}(\bs{q}):=&\,
e^{\ell^2_B\bs{q}^2/4}\,
\widehat{\mathcal{P}}^{\ }_0\!
:\!e^{\mathrm{i}\bs{q}\cdot\widehat{\bs{R}}}\!:\!
\widehat{\mathcal{P}}^{\ }_0
\\
=&\,
e^{\mathrm{i}\bs{q}\cdot\widehat{\bs{X}}}
\end{split}
\end{equation}
in the lowest Landau level $n=0$
satisfy the commutation relations\ %
~\cite{Girvin85}
\begin{subequations}
\label{eq: closed commutation rho q projected_intro}
\begin{equation}
\left[
\widehat{\rho}(\bs{q}^{\ }_1),
\widehat{\rho}(\bs{q}^{\ }_2)
\right]=
-2\,\mathrm{i}\,
\sin
\left(
\frac{\ell^2_B}{2}
\left(\bs{q}^{\ }_1\wedge \bs{q}^{\ }_2\right) 
\cdot
\bs{e}^{\ }_{3}
\right)
\widehat{\rho}(\bs{q}^{\ }_1+\bs{q}^{\ }_2)
\label{eq: closed commutation rho q projected_intro a}
\end{equation}
or, in the limit of small momenta
$\bs{q}^{\ }_{1}$ and $\bs{q}^{\ }_{2}$,
\begin{equation}
\left[
\widehat{\rho}(\bs{q}^{\ }_1),
\widehat{\rho}(\bs{q}^{\ }_2)
\right]\approx
-
\mathrm{i}\,\ell^2_B\,
\left(\bs{q}^{\ }_1\wedge \bs{q}^{\ }_2\right) 
\cdot
\bs{e}^{\ }_{3}\,
\widehat{\rho}(\bs{q}^{\ }_1+\bs{q}^{\ }_2),
\label{eq: closed commutation rho q projected_intro b}
\end{equation}
or, equivalently, to lowest order in the $\bm{q}$'s,
\begin{equation}
\left[
\partial^{\ }_{q^{\mu}_{1}} 
\widehat{\rho}\,
(\bs{q}^{\ }_{1}),
\partial^{\ }_{q^{\nu}_{2}} 
\widehat{\rho}\,
(\bs{q}^{\ }_2)
\right]
\approx-\mathrm{i}\,
\ell^{2}_{B}\,
\epsilon^{\ }_{\mu\nu}\,
\widehat{\rho}(\bs{q}^{\ }_1+\bs{q}^{\ }_2).
\label{eq: closed commutation rho q projected_intro c}
\end{equation}
\end{subequations}
This algebra, the GMP algebra,%
~\cite{GMP vs W algebra,Moyal49,Fairlie89,Bakas89,Hoppe90} 
plays two crucial roles. 
First, within the SMA approximation,~\cite{Girvin85}
it dictates under what conditions interactions open a spectral
gap between the many-body interacting ground state and its excitations
upon lowering the chemical potential within the first Landau level.
Second, it also dictates the universal properties of
the low-energy and long-distance dynamics
at the edge in an open geometry.%
~\cite{Iso92,Cappelli93,Martinez93}

The goal of the work presented in the remainder of this section 
is to generalize the 
noncommutative geometry encoded by Eqs%
~\eqref{eq: noncommuting algebra in QHE b}
and%
~(\ref{eq: closed commutation rho q projected_intro c})
to noninteracting many-body fermionic Hamiltonians in 3D space. 
Before carrying out this program, let us motivate what it
is to come by first presenting what would constitute a natural
extension of the algebra in the QHE to 3D problems.

First, instead of the commutator, consider the case where the 3-bracket
of the 3D projected position operators equals a $\mathbb{C}$-number
\begin{subequations}
\label{eq:3-bracket-intro}
\begin{equation}
\left[\widehat{X}_1,\widehat{X}_2,\widehat{X}_3\right]=
\mathrm{i}\,\ell^{3},
\label{eq:3-bracket-intro a}
\end{equation}
where, following Nambu,~\cite{Nambu73} we have defined the 3-bracket
\begin{equation}
\begin{split}
[\widehat{A}^{\ }_1,\widehat{A}^{\ }_2,\widehat{A}^{\ }_3]:=&\,
\epsilon^{ijk} \widehat{A}^{\ }_i \widehat{A}^{\ }_j \widehat{A}^{\ }_k
\\
=&\,
[\widehat{A}^{\ }_1,\widehat{A}^{\ }_2]\,\widehat{A}^{\ }_3
+
[\widehat{A}^{\ }_2,\widehat{A}^{\ }_3]\,\widehat{A}^{\ }_1
+
[\widehat{A}^{\ }_3,\widehat{A}^{\ }_1]\,\widehat{A}^{\ }_2.
\end{split}
\label{eq:3-bracket-intro b}
\end{equation}
\end{subequations}
The characteristic length scale $\ell$
of the 3D noninteracting many-body Hamiltonian, 
not to be confused with the magnetic length $\ell^{\ }_B$ 
of the 2D Landau Hamiltonian,
is the signature of a spectral gap separating the ground state from
the excited states.
Similarly to the 2D case, for which area preserving coordinate
transformations leave the commutation relations unchanged, we would
like volume preserving transformations not to change the
3-bracket. Under generic 
transformations $r^{\ }_{\mu}\to f^{\ }_{\mu}(\bs{r})$, $\mu=1,2,3$,
that vary on length scales larger than $\ell$,
\begin{subequations}
\label{eq: commutator f1 f2 f3 intro}
\begin{equation}
\left[
f^{\ }_{1}(\widehat{\bs{X}}),
f^{\ }_{2}(\widehat{\bs{X}}),
f^{\ }_{2}(\widehat{\bs{X}})
\right]=
\mathrm{i}\,\ell^{3}
\,\{f^{\ }_1,f^{\ }_2,f^{\ }_3\}^{\ }_{\mathrm{N}}\,(\widehat{\bs{X}})
+
\mathcal{O}(\ell^5),
\label{eq: commutator f1 f2 f3 a intro}
\end{equation}
where the classical Nambu bracket is defined as\ %
~\cite{Nambu73}
\begin{equation}
\left\{f^{\ }_1,f^{\ }_2,f^{\ }_3\right\}^{\ }_{\mathrm{N}}(\bs{r}):=
\epsilon^{\mu\nu\lambda}
\left(
\frac{\partial f^{\ }_{1}}{\partial r^{\mu}}
\frac{\partial f^{\ }_{2}}{\partial r^{\nu}}
\frac{\partial f^{\ }_{3}}{\partial r^{\lambda}}
\right)(\bs{r}).
\label{eq: commutator f1 f2 f3 b intro}
\end{equation}
\end{subequations}
The condition for this coordinate transformation to 
locally preserve volume is that its Jacobian equals unity, 
or equivalently that 
$
\left\{f^{\ }_{1},f^{\ }_{2},f^{\ }_{3}\right\}^{\ }_{\mathrm{N}}(\bs{r})=1
$. 
In this case, it follows that
$\left[f^{\ }_{1}(\widehat{\bs{X}}),f^{\ }_{2}(\widehat{\bs{X}}),
f^{\ }_{3}(\widehat{\bs{X}})\right]=
\mathrm{i}\,\ell^{3}+\mathcal{O}(\ell^{5})$.

Second, we claim (and show in this paper) that the 3D counterpart to
the operator product expansion%
~(\ref{eq: closed commutation rho q projected_intro})
of the projected densities is, 
to lowest order in the $\bm{q}$'s,
\begin{equation}
\left[
\partial^{\ }_{q^{\mu}_{1}} 
\widehat{\rho}\,
(\bs{q}^{\ }_{1}),
\partial^{\ }_{q^{\nu}_{2}} 
\widehat{\rho}\,
(\bs{q}^{\ }_{2}),
\partial^{\ }_{q^{\lambda}_{3}} 
\widehat{\rho}\,
(\bs{q}^{\ }_{3})
\right] 
\approx
\epsilon_{\mu\nu\lambda}\,\ell^3\,
\widehat{\rho}(\bs{q}^{\ }_1+\bs{q}^{\ }_2+\bs{q}^{\ }_3).
\label{eq: 3D_commutation rho q projected_intro_linear}
\end{equation}

The algebra defined by Eqs.~(\ref{eq:3-bracket-intro})
and (\ref{eq: 3D_commutation rho q projected_intro_linear}),
if it can be realized by a 3D fermionic noninteracting many-body 
Hamiltonian, might then deliver two results. 
First, within the SMA approximation,
it might dictate under what conditions interactions open a spectral
gap between the many-body interacting ground state and its excitations
upon lowering the chemical potential below the single-particle gap.
Second, it might also dictate the universal properties of the
low-energy and long-distance dynamics
at the boundary in an open geometry.

The key idea to realize 
the algebra defined by Eqs.~(\ref{eq:3-bracket-intro})
and (\ref{eq: 3D_commutation rho q projected_intro_linear})
is to replace the effect of the magnetic field
in the Landau Hamiltonian by that
of the projection of suitable
operators on a suitable subspace of the fermionic Fock space.
The construction of this suitable subspace
presumes the existence of fermionic Bloch bands as occurs
in condensed matter physics and assumes that a subset of these
bands are fully occupied, while the complementary set
are empty and separated from the filled subset by an energy gap.

Now, carrying out this program for some Bloch bands
will not yield immaculately the 3-brackets \eqref{eq:3-bracket-intro} and
\eqref{eq: 3D_commutation rho q projected_intro_linear}. 
It will yield these relations approximately 
in the long-wavelength limit. 
The situation here is similar to the case of
the quantized Hall effect in flat Chern bands of 2D models.%
~\cite{Neupert11a,Sheng11,Wang11a,Regnault11}
As discussed by Parameswaran, Roy, and Sondhi in
Ref.~\onlinecite{Parameswaran11} 
(see also Refs.~\onlinecite{Goerbig12} and \onlinecite{Bernevig11}), 
the algebra~\eqref{eq: closed commutation rho q projected_intro b}
follows if the fluctuations in the Berry curvature over the Brillouin
zone are neglected, or equivalently if the local curvature is approximated
by its average over the entire Brillouin zone. 
Without this approximation, however, 
the noncommutative relations obeyed by the
projected position operator 
will not be as simple
as in Eq.~\eqref{eq: noncommuting algebra in QHE b} 
and may instead be 
represented as
\begin{equation}
\left[\widehat{X}_1,\widehat{X}_2\right]=\mathrm{i}\ell^{2}+\cdots,
\label{eq:2-bracket-intro nonimmaculate}
\end{equation}
where $\cdots$ stands for operators that appear as a result 
of the inhomogeneities in the Berry curvature.
The central question is how to distinguish universal 
from nonuniversal contributions to the 
right-hand side of Eq.~\eqref{eq:2-bracket-intro nonimmaculate}.
To answer this question, we propose to
consider the ground state expectation value 
$\left\langle\left[\widehat{X}^{\ }_1,\widehat{X}^{\ }_2\right]\right\rangle$,
that encodes the quantized Hall conductivity, as seen in 
Eq.~\eqref{eq: Hall conductivity from noncommutative}.
We show in Sec.~\ref{subsec: Algebra of the position operators} 
that
\begin{equation}
\frac{1}{N^{\ }_{\mathrm{p}}}
\left\langle\left[\widehat{X}_1,\widehat{X}_2\right]\right\rangle=
\frac{2\pi\,\mathrm{i}}{\bar{\rho}}\mathrm{Ch}^{(1)}.
\label{eq:2-bracket-intro nonimmaculate expecta}
\end{equation}
Here, $\mathrm{Ch}^{(1)}$ is the first Chern number of the topological band 
that sustains the IQHE in the lattice, 
and will be defined in Eq.~\eqref{eq: Chern numbers 1}, 
while $N^{\ }_{\mathrm{p}}$ is the total particle number and $\bar{\rho}$
is the average particle density.
This suggests that the universal physical properties are captured
by the $\mathbb{C}$-number contribution to the right-hand side of 
Eq.~\eqref{eq:2-bracket-intro nonimmaculate}.

As with the commutator~\eqref{eq:2-bracket-intro nonimmaculate},
the 3-bracket~\eqref{eq:3-bracket-intro}
will also acquire extra terms in 3D space
\begin{equation}
\left[\widehat{X}^{\ }_1,\widehat{X}^{\ }_2,\widehat{X}^{\ }_3\right]=
\mathrm{i}\,\ell^{3}
+
\cdots.
\label{eq:3-bracket-intro nonimmaculate}
\end{equation}
We are thus lead to consider its normal ordered expectation value instead,
which, as we show 
in Sec.~\ref{subsec: Algebra of the position operators}, 
is given by
\begin{equation}
\frac{1}{N^{\ }_{\mathrm{p}}}
\left\langle 
:\left[\widehat{X}^{\ }_1,\widehat{X}^{\ }_2,\widehat{X}^{\ }_3\right]:
\right\rangle=\,
\frac{12\pi^2\,\mathrm{i}}{\bar{\rho}}\,
\mathrm{CS}^{(3)}.
\label{eq:3-bracket-intro nonimmaculate expectation}
\end{equation}
Here, the symbol $\mathrm{CS}^{(3)}$ 
stands for the 3D Chern-Simons invariant 
defined in Eq.~\eqref{eq: CS 3}.
If the discrete chiral symmetry or
time-reversal symmetry holds,
$\mathrm{CS}^{(3)}$ is a
quantized topological invariant that takes
half-integer values.
It
is related to the dimensionless 
coupling 
\begin{subequations}
\begin{equation}
\theta=2\pi\,(\mathrm{CS}^{(3)}\,\mathrm{mod}\,1)
\end{equation}
that enters the effective action 
\begin{equation}
\mathcal{L}^{\mathrm{eff}}_{\theta}:=
\frac{\theta\,e^{2}}{4\pi^{2}}\,
\bm{E}\cdot\bm{B}
\label{eq: def theta term}
\end{equation}
\end{subequations}
obtained from integrating
out noninteracting fermions of a 3D topological insulator 
in the background of external electric $\bm{E}$
and magnetic $\bm{B}$ fields within linear response theory. 
This electro-magnetic coupling was derived
by Xi, Hughes, and Zhang in Ref.~\onlinecite{Qi08}
by dimensional reduction from a topological insulator 
in 4D displaying an integer quantum Hall effect 
to a 3D $\mathbb{Z}^{\ }_{2}$ topological insulator 
(see also Ref.~\onlinecite{Wang10}
for a generalization that accounts for moderate interactions).
For a  3D $\mathbb{Z}^{\ }_{2}$ topological insulator,
time-reversal symmetry holds. In turn, time-reversal symmetry 
restricts $\theta$ to the two values $\theta=0$
and $\theta=\pi$ that distinguish
``ordinary'' from topological 3D insulators,
respectively.~\cite{Qi08} 
Several derivations of the magneto-electric response, 
of which the $\theta$ term~(\ref{eq: def theta term})
is an example, have been proposed without time-reversal symmetry.%
\cite{Malashevich10,Essin10,Coh11,Chen11a,Chen11b}

Equation (\ref{eq:3-bracket-intro nonimmaculate expectation})
relates a nonvanishing $\mathrm{CS}^{(3)}$ to the noncommutative
algebra obeyed by the components of the
projected position operator through the noninteracting groundstate
expectation value of their $3$-bracket. Because the position
operator and its projection are unbounded operators
and because Wannier states may not be exponentially localized 
if the Bloch states have a topological character,%
~\cite{Thouless84,Soluyanov11} 
a regularization procedure is needed to compute
Eq.~(\ref{eq:3-bracket-intro nonimmaculate expectation}).
We shall choose a regularization that preserves gauge invariance
under pure gauge transformation of the Bloch states, but that breaks
a discrete translation symmetry. In doing so, we shall make a connection
with Ref.~\onlinecite{Coh11}, where a representation of the $\theta$ term 
is given in terms of expectation
values of the position operators in the Wannier basis.

We start by deriving the conditions under which
Eq.~\eqref{eq:3-bracket-intro nonimmaculate expectation} 
holds in
Sec.~\ref{subsec: Algebra of the position operators}
for any Hamiltonian that 
is endowed with translation invariance, a spectral gap, and
describes the motion of noninteracting fermions in flat Euclidean space
$\mathbb{R}^{3}$. 
We draw a connection between the 3-bracket and the (classical) Nambu
bracket in Sec.%
~\ref{subsec: Regularized 3-bracket and the Nambu bracket}.
We then specialize in
Sec.~\ref{subsec: massive Dirac fermions}
to the case of massive noninteracting
Dirac Hamiltonians for which 
some analytical results can be obtained
in the long-wavelength limit. Finally, 
Sec.~\ref{subsec: Algebra of the density operator}
is devoted to the operator product expansion
of single-particle density operators in 3D lattice models
and the conditions under which Eq.%
~(\ref{eq: 3D_commutation rho q projected_intro_linear}) 
holds.

\begin{figure}
\includegraphics[angle=0,scale=0.5]{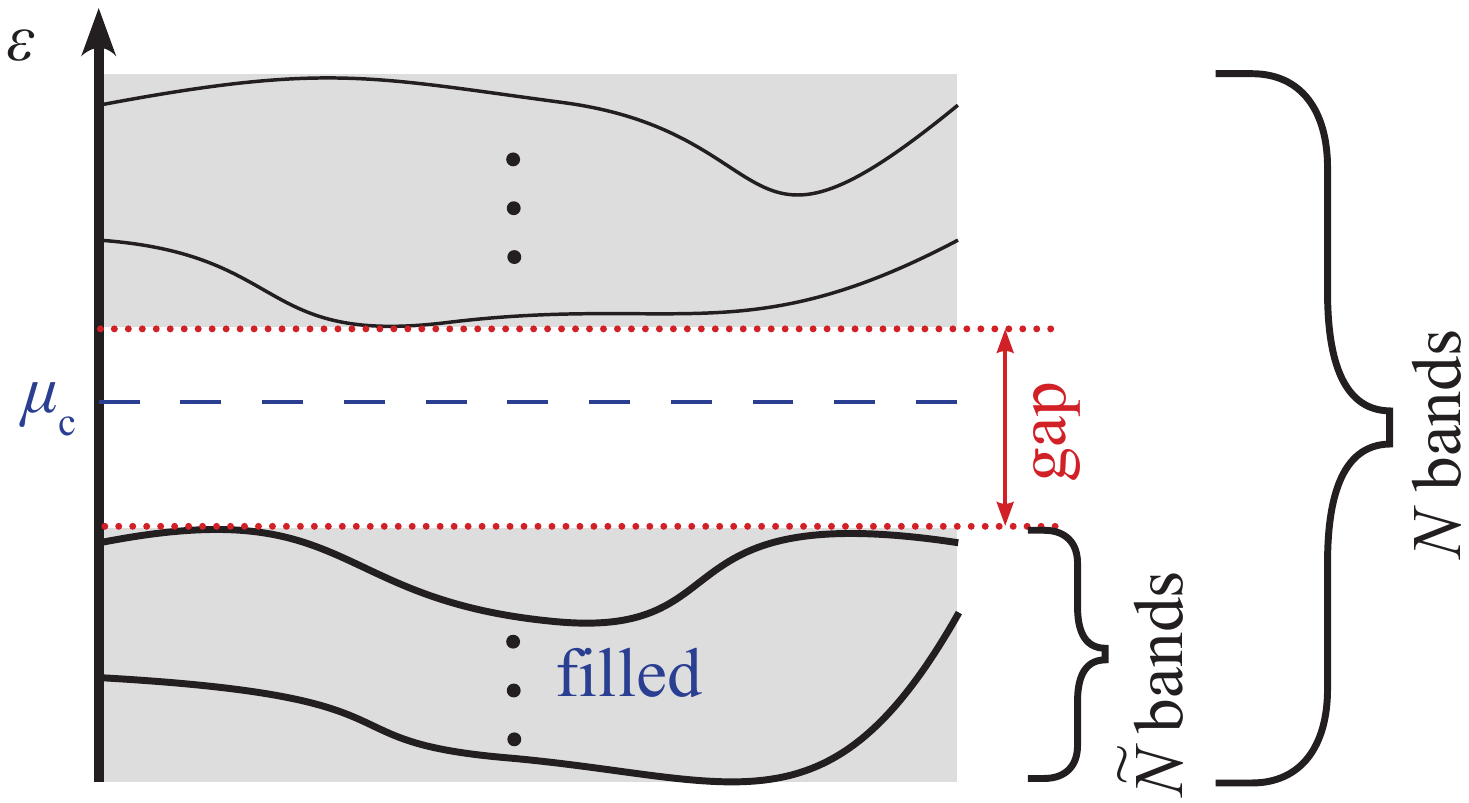}
\caption{\label{fig: spectral single-particle gap}
(Color online)
Assumed spectral gap in the single-particle energy spectrum. Here,
$\mu^{\ }_{\mathrm{c}}$ denotes the chemical potential and the 
insulating noninteracting many-body ground state $|\Phi\rangle$ is
obtained by filling all the states in the $\widetilde{N}$ 
bands below the spectral gap.
        }
\end{figure}

\subsection{
Noncommutative geometry for the projected
position operators
           }
\label{subsec: Algebra of the position operators}

We shall consider noninteracting fermions whose
dynamics are governed by the translation invariant Bloch Hamiltonian 
\begin{subequations}
\label{eq: def widehat H in chi basis}
\begin{equation}
\widehat{H}=
\int\limits_{\Lambda^{\star}_{\mathrm{BZ}}}\mathrm{d}^{d}\bs{k}\,
\sum_{a=1}^{N}\,
\widehat{\chi}^{\dag}_{a}(\bs{k})\,
\varepsilon^{\ }_{a}(\bs{k})\,
\widehat{\chi}^{\ }_{a}(\bs{k}).
\label{eq: def widehat H in chi basis a}
\end{equation}
We are reserving the latin index
$a=1,\cdots,N$ for the band label. The momentum
$\bs{k}=(k^{\mu})$ belongs to the Brillouin zone
\begin{equation}
\Lambda^{\star}_{\mathrm{BZ}}:=
\left\{
(k^{\mu})\in\mathbb{R}^{d}\left|
-\frac{\pi}{\mathfrak{a}}\leq
k^{\mu}<
\frac{\pi}{\mathfrak{a}},
\quad
\mu=1,\cdots,d
\right.\right\}
\label{eq: def widehat H in chi basis b}
\end{equation}
with $\pi/\mathfrak{a}$ playing the role of the upper momentum cutoff.
Each band $a=1,\cdots,N$ is characterized by the single-particle
energy dispersion $\varepsilon^{\ }_{a}(\bs{k})$, a real-valued
function over the Brillouin zone.
The creation and annihilation operators obey the fermionic algebra
\begin{equation}
\begin{split}
&
\left\{
\chi^{\ }_{a}(\bs{k}),
\chi^{\ }_{a'}(\bs{k}')
\vphantom{\chi^{\dag}_{a'}(\bs{k}')}
\right\}=
\left\{
\chi^{\dag}_{a}(\bs{k}),
\chi^{\dag}_{a'}(\bs{k}')
\right\}=0,
\\
&
\left\{
\chi^{\ }_{a}(\bs{k}),
\chi^{\dag}_{a'}(\bs{k}')
\right\}=
\delta^{\ }_{a,a'}\,
\delta(\bs{k}-\bs{k}'),
\end{split}
\label{eq: def widehat H in chi basis c}
\end{equation}
for all pairs of bands and for all pairs of momenta in the Brillouin zone.
Finally, 
for any band $a=1,\cdots,N$, 
for any momentum $\bs{k}$ from the Brillouin zone, and
for any Cartesian unit vector $\bs{e}^{\mu}$ from $\mathbb{R}^{d}$,
we impose twisted boundary conditions across the Brillouin zone through
\begin{equation}
\chi^{\ }_{a}\big(\bs{k}+(2\pi/\mathfrak{a})\,\bs{e}^{\mu}\big)=
e^{+\mathrm{i}(2\pi\,\theta^{\mu})}\,
\chi^{\ }_{a}(\bs{k}).
\label{eq: def widehat H in chi basis d}
\end{equation}
\end{subequations}
These twisted boundary conditions are parametrized by the
real numbers $0\leq\theta^{\mu}<1$ with $\mu=1,\cdots,d$.

We shall also assume that (i) there are $\widetilde{N}$ 
lower bands out of the $N$ bands that are separated 
by an energy gap from the $N-\widetilde{N}$ remaining upper bands 
and (ii) the chemical potential lies in this spectral gap
(see Fig.~\ref{fig: spectral single-particle gap}).
We denote with 
\begin{equation}
\widehat{P}^{\ }_{\widetilde{N}}\equiv
\int\limits_{\Lambda^{\star}_{\mathrm{BZ}}}\mathrm{d}^{d}\bs{k}\,
\sum_{\tilde{a}=1}^{\widetilde{N}}\,
\widehat{\chi}^{\dag}_{\tilde{a}}(\bs{k})\,
\widehat{\chi}^{\ }_{\tilde{a}}(\bs{k})
\label{eq: def projection operator in chi basis}
\end{equation} 
the projection operator on the single-particle states 
spanned by these gapped lower bands.
We are reserving the latin index with a tilde sign
$\tilde{a}=1,\cdots,\widetilde{N}$ 
for the lower band labels.

In analogy to the guiding center coordinates%
~\eqref{eq: rewriting guiding center}
from the IQHE, we would like to define projected position operators.
However, projected position operators are associated to gauge
fields as we now explain.

On the one hand,
we may define the Wannier creation operator through the Fourier transform
\begin{subequations}
\label{eq: def Wannier Fock space}
\begin{equation}
\widehat{W}^{\dag}_{a;\bs{R}}:=
\int\limits_{\Lambda^{\star}_{\mathrm{BZ}}}
\frac{\mathrm{d}^{d}\bs{k}}{(2\pi/\mathfrak{a})^{d/2}}\,
e^{-\mathrm{i}\bs{k}\cdot\bs{R}}\,
\widehat{\chi}^{\dag}_{a}(\bs{k}),
\label{eq: def Wannier Fock space a}
\end{equation}
or, equivalently, the inverse Fourier transform
\begin{equation}
\widehat{\chi}^{\dag}_{a}(\bs{k})=:
\frac{1}{(2\pi/\mathfrak{a})^{d/2}}\,
\sum_{\bs{R}\in\Lambda^{\ }_{R}}
e^{+\mathrm{i}\bs{k}\cdot\bs{R}}\,
\widehat{W}^{\dag}_{a;\bs{R}},
\label{eq: def Wannier Fock space b}
\end{equation}
for any band index $a=1,\cdots,N$ and for any lattice point
$\bs{R}=(R^{\mu})\in\Lambda^{\ }_{R}$ whereby
\begin{equation}
\Lambda^{\ }_{R}:=
\left\{
(R^{\mu})\in\mathbb{R}^{d}
\left|
\frac{R^{\mu}}{\mathfrak{a}}=
\theta^{\mu}
\hbox{ mod } 1,
\
\mu=1,\cdots,d
\right.\right\}.
\label{eq: def Wannier Fock space c}
\end{equation}
\end{subequations}
The length scale $\mathfrak{a}$ can thus be interpreted as
a lattice spacing. Consequently, 
creation and annihilation Wannier operators
obey the fermionic algebra
\begin{equation}
\begin{split}
&
\left\{
\widehat{W}^{\ }_{a;\bs{R}},
\widehat{W}^{\ }_{a';\bs{R}'}
\vphantom{\chi^{\dag}_{a',\bs{R}'}}
\right\}=
\left\{
\widehat{W}^{\dag}_{a;\bs{R}},
\widehat{W}^{\dag}_{a';\bs{R}'}
\right\}=0,
\\
&
\left\{
\widehat{W}^{\ }_{a;\bs{R}},
\widehat{W}^{\dag}_{a';\bs{R}'}
\right\}=
\delta^{\ }_{a,a'}\,
\delta^{\ }_{\bs{R},\bs{R}'},
\end{split}
\end{equation}
for all pairs of bands and for all pairs of lattice sites.
Moreover, the projection operator%
~(\ref{eq: def projection operator in chi basis})
remains diagonal in the Wannier representation%
~(\ref{eq: def Wannier Fock space}),
\begin{equation}
\widehat{P}^{\ }_{\widetilde{N}}=
\sum_{\bs{R}\in\Lambda^{\ }_{R}}\,
\sum_{\tilde{a}=1}^{\widetilde{N}}\,
\widehat{W}^{\dag}_{\tilde{a};\bs{R}}\,
\widehat{W}^{\ }_{\tilde{a};\bs{R}}.
\label{eq: def projection operator in W basis}
\end{equation} 
Hence, the Wannier position operator defined by
\begin{subequations}
\label{eq: def R operator}
\begin{equation}
\widehat{\bs{R}}:=
\sum_{\bs{R}\in\Lambda^{\ }_{R}}\,
\sum_{a=1}^{N}\,
\widehat{W}^{\dag}_{a;\bs{R}}\,
\bs{R}\,
\widehat{W}^{\ }_{a;\bs{R}}
\label{eq: def R operator a}
\end{equation}
is projected onto the lower bands by
restricting the band index to the lower ones,
\begin{equation}
\begin{split}
\widehat{\bs{X}}^{\ }_{R}:=&\,
\widehat{P}^{\ }_{\widetilde{N}}\,
\widehat{\bs{R}}\,
\widehat{P}^{\ }_{\widetilde{N}}
\\
=&\,
\sum_{\bs{R}\in\Lambda^{\ }_{R}}\,
\sum_{\tilde{a}=1}^{\widetilde{N}}\,
\widehat{W}^{\dag}_{\tilde{a};\bs{R}}\,
\bs{R}\,
\widehat{W}^{\ }_{\tilde{a};\bs{R}}.
\end{split}
\label{eq: def R operator b}
\end{equation}
\end{subequations}
Hamiltonian%
~(\ref{eq: def widehat H in chi basis a})
in the Wannier basis 
is represented by
\begin{subequations}
\label{eq: def widehat H in W basis}
\begin{equation}
\widehat{H}=
\sum_{\bs{R},\bs{R}'\in\Lambda^{\ }_{R}}\,
\sum_{a=1}^{N}\,
\widehat{W}^{\dag}_{a;\bs{R}}\,
\mathcal{H}^{\ }_{a;\bs{R}-\bs{R}'}\,
\widehat{W}^{\ }_{a;\bs{R}'}.
\label{eq: def widehat H in W basis a}
\end{equation}
The single-particle matrix elements,
\begin{equation}
\mathcal{H}^{\ }_{a;\bs{R}-\bs{R}'}:=
\int\limits_{\Lambda^{\star}_{\mathrm{BZ}}}
\frac{\mathrm{d}^{d}\bs{k}}{(2\pi/\mathfrak{a})^{d}}\,
e^{+\mathrm{i}\bs{k}\cdot(\bs{R}-\bs{R}')}\,
\varepsilon^{\ }_{a}(\bs{k})
\label{eq: def widehat H in W basis b}
\end{equation}
\end{subequations}
may decay slower than exponentially with the separation
$
|\bs{R}-\bs{R}'|
$
on the lattice $\Lambda^{\ }_{R}$ for some of the bands,
i.e., locality in position space is not manifest
in the Wannier basis.%
~\cite{Thouless84,Soluyanov11} 

On the other hand,
we can enforce locality of the Hamiltonian%
~(\ref{eq: def widehat H in chi basis a})
as follows. 
We shall assume that, for any momentum $\bs{k}$ from the Brillouin zone,
there exists a unitary transformation from the band creation operators 
to the so-called orbital creation operators, i.e.,
\begin{subequations}
\label{eq: unitary trsf from chi to psi}
\begin{equation}
\widehat{\psi}^{\dag}_{\alpha}(\bs{k}):=
\sum_{a=1}^{N}\,
u^{(a)*}_{\alpha}(\bs{k})\,
\widehat{\chi}^{\dag}_{a}(\bs{k})
\label{eq: unitary trsf from chi to psi a}
\end{equation}
where we have reserved the greek index $\alpha=1,\cdots,N$ 
for the orbital label. For any $\bs{k}$ from the Brillouin zone, 
the $N\times N$ matrix elements between the band 
$a=1,\cdots,N$
and orbital
$\alpha=1,\cdots,N$
labels obey (i) the periodic boundary conditions
\begin{equation}
u^{(a)}_{\alpha}(\bs{k})=
u^{(a)}_{\alpha}\big(\bs{k}+(2\pi/\mathfrak{a})\,\bs{e}^\mu\big),
\label{eq: unitary trsf from chi to psi b}
\end{equation}
for any $\mu=1,\cdots,d$,
in order for $\widehat{\psi}^{\ }_{\alpha}(\bs{k})$
to share with $\widehat{\chi}^{\ }_{a}(\bs{k})$
the same twisted boundary condition%
~(\ref{eq: def widehat H in chi basis d})
and (ii) the orthonormality conditions
\begin{equation}
\sum_{\alpha=1}^{N}
u^{(a)*}_{\alpha}(\bs{k})\,
u^{(a') }_{\alpha}(\bs{k})=
\delta^{a,a'},
\qquad 
a,a'= 1,\cdots,N,
\label{eq: unitary trsf from chi to psi c}
\end{equation}
\end{subequations}
in order for the pair
$\widehat{\psi}^{\dag}_{\alpha}(\bs{k})$
and
$\widehat{\psi}^{\ }_{\alpha'}(\bs{k}')$
to share the same fermionic algebra%
~(\ref{eq: def widehat H in chi basis c})
as the pair
$\widehat{\chi}^{\dag}_{\alpha}(\bs{k})$
and
$\widehat{\chi}^{\ }_{\alpha'}(\bs{k}')$
does. Finally, we assume that the representation
\begin{subequations}
\label{eq: def widehat H in psi basis}
\begin{equation}
\widehat{H}=
\sum_{\bs{r},\bs{r}'\in\Lambda^{\ }_{r}}\,
\sum_{\alpha,\alpha'=1}^{N}\,
\widehat{\psi}^{\dag}_{\alpha;\bs{r}}\,
\mathcal{H}^{\ }_{a,a';\bs{r}-\bs{r}'}\,
\widehat{\psi}^{\ }_{\alpha';\bs{r}'}
\label{eq: def widehat H in psi basis a}
\end{equation}
in terms of the Fourier transform
\begin{equation}
\widehat{\psi}^{\dag}_{\alpha;\bs{r}}:=
\int\limits_{\Lambda^{\star}_{\mathrm{BZ}}}
\frac{\mathrm{d}^{d}\bs{k}}{(2\pi/\mathfrak{a})^{d/2}}\,
e^{-\mathrm{i}\bs{k}\cdot\bs{r}}\,
\widehat{\psi}^{\dag}_{\alpha}(\bs{k}),
\label{eq: def widehat H in psi basis b}
\end{equation}
or, equivalently, the inverse Fourier transform
\begin{equation}
\widehat{\psi}^{\dag}_{\alpha}(\bs{k})=:
\frac{1}{(2\pi/\mathfrak{a})^{d/2}}\,
\sum_{\bs{r}\in\Lambda^{\ }_{r}}
e^{+\mathrm{i}\bs{k}\cdot\bs{r}}\,
\widehat{\psi}^{\dag}_{\alpha;\bs{r}}
\label{eq: def widehat H in psi basis c}
\end{equation}
for any orbital index $\alpha=1,\cdots,N$ and for any lattice point
$\bs{r}=(r^{\mu})\in\Lambda^{\ }_{r}$,
has the single-particle matrix elements
\begin{equation}
\begin{split}
\mathcal{H}^{\ }_{\alpha,\alpha';\bs{r}-\bs{r}'}:=&\,
\int\limits_{\Lambda^{\star}_{\mathrm{BZ}}}
\frac{\mathrm{d}^{d}\bs{k}}{(2\pi/\mathfrak{a})^{d}}\,
e^{+\mathrm{i}\bs{k}\cdot(\bs{r}-\bs{r}')}\,
\\
&\times
\sum_{a=1}^{N}
u^{(a)}_{\alpha}(\bs{k})\,
\varepsilon^{\ }_{a}(\bs{k})\,
u^{(a)*}_{\alpha'}(\bs{k}),
\end{split}
\label{eq: def widehat H in psi basis d}
\end{equation}
that decay exponentially with increasing distance
$
|\bs{r}-\bs{r}'|
$
for any pair of orbitals.
Thus, locality on the lattice 
\begin{equation}
\Lambda^{\ }_{r}:=
\left\{
(r^{\mu})\in\mathbb{R}^{d}
\left|
\frac{r^{\mu}}{\mathfrak{a}}=
\theta^{\mu}
\hbox{ mod } 1,
\
\mu=1,\cdots,d
\right.\right\}
\label{eq: def widehat H in psi basis e}
\end{equation}
\end{subequations}
is manifest in the orbital basis.
The lattices $\Lambda^{\ }_{r}$ and $\Lambda^{\ }_{R}$
share the same unit cell,
however the two lattices can be shifted relative to each other
in their embedding space $\mathbb{R}^{d}$ by any vector 
$\sum_{\mu=1}^{d}e^{\ }_{\mu}\,\bs{e}^{\mu}$ with $-1\leq e^{\mu}<1$
from their unit cell.

The projection operator%
~(\ref{eq: def projection operator in chi basis})
is not diagonal with respect to the orbital index
while the projection operator%
~(\ref{eq: def projection operator in W basis})
is neither diagonal with respect to the orbital index 
nor with respect to the lattice sites from $\Lambda^{\ }_{r}$.
Hence, the orbital position operator defined by
\begin{subequations}
\label{eq: def r operator}
\begin{equation}
\widehat{\bs{r}}:=
\sum_{\bs{r}\in\Lambda^{\ }_{r}}\,
\sum_{\alpha=1}^{N}\,
\widehat{\psi}^{\dag}_{\alpha;\bs{r}}\,
\bs{r}\,
\widehat{\psi}^{\ }_{\alpha;\bs{r}}
\label{eq: def r operator a}
\end{equation}
turns after projection into 
(see Appendix%
~\ref{appsec: Regularization of non-commuting projected position operators})
\begin{equation}
\begin{split}
\widehat{\bs{X}}^{\ }_{r}:=&\,
\widehat{P}^{\ }_{\widetilde{N}}\,
\widehat{\bs{r}}\,
\widehat{P}^{\ }_{\widetilde{N}}
\\
=&\,
\sum_{\bs{R},\bs{R}'\in\Lambda^{\ }_{R}}\,
\sum_{\tilde{a},\tilde{a}'=1}^{\widetilde{N}}\,
\widehat{W}^{\dag}_{\tilde{a};\bs{R}}\,
\bs{\mathcal{X}}^{\ }_{\tilde{a},\tilde{a}';\bs{R},\bs{R}'}\,
\widehat{W}^{\ }_{\tilde{a}';\bs{R}'}
\end{split}
\label{eq: def r operator b}
\end{equation}
where we have introduced the single-particle kernel
\begin{equation}
\begin{split}
\bs{\mathcal{X}}^{\ }_{\tilde{a},\tilde{a}';\bs{R},\bs{R}'}:=&\,
\int\limits_{\Lambda^{\star}_{\mathrm{BZ}}}
\frac{\mathrm{d}^{d}\bs{k}}{(2\pi/\mathfrak{a})^{d}}\,
e^{+\mathrm{i}\bs{k}\cdot(\bs{R}-\bs{R}')}\\
&\,\times
\left[
\delta^{\ }_{\tilde{a},\tilde{a}'}\,
\bs{R}'\,
+
\mathrm{i}\,
\bs{A}^{\ }_{\tilde{a}\tilde{a}'}(\bs{k})
\right].
\end{split}
\label{eq: def r operator c}
\end{equation}
This kernel depends on the $U(\widetilde{N})$ gauge field
$\bs{A}(\bs{k})$, 
an antihermitean $\widetilde{N}\times\widetilde{N}$
matrix whose components
\begin{equation}
\bs{A}^{\ }_{\tilde{a}\tilde{a}'}(\bs{k}):=
\sum_{\alpha=1}^{N}
u^{(\tilde{a})*}_{\alpha}(\bs{k})
\left(
\frac{\partial u^{(\tilde{a}')}_{\alpha}}{\partial\bs{k}}
\right)(\bs{k}),
\label{eq: def r operator d}
\end{equation}
\end{subequations}
are labeled by the lower band indices 
$\tilde{a},\tilde{a}'=1,\cdots,\widetilde{N}$
and obey periodic boundary conditions across the Brillouin zone.

The gauge field~(\ref{eq: def r operator d})
does not need to be a pure gauge as it originates from projecting
the pure gauge field
\begin{equation}
\bs{A}^{\ }_{aa'}(\bs{k}):=
\sum_{\alpha=1}^{N}
u^{(a)*}_{\alpha}(\bs{k})
\left(
\frac{\partial u^{(a')}_{\alpha}}{\partial\bs{k}}
\right)(\bs{k})
\label{eq: def unprojected gauge field}
\end{equation}
by restricting the band indices $a,a'=1,\cdots,N$
to the lower band indices 
$\tilde{a},\tilde{a}'=1,\cdots,\widetilde{N}$.
Furthermore, the decomposition%
~(\ref{eq: unitary trsf from chi to psi a})
is not unique. Indeed,
for any pair of orbital and band labels $\alpha,a=1,\cdots,N$,
the simultaneous transformations
\begin{subequations} 
\label{eq: sim gauge trsf u's and chi's}
\begin{equation}
u^{(a)}_{\alpha}(\bs{k})=:
\sum_{\mathsf{a}=1}^{N}\,
u^{(\mathsf{a})}_{\alpha}(\bs{k})\,
{G}^{*}_{a\mathsf{a}}(\bs{k}),
\label{eq: sim gauge trsf u's and chi's a}
\end{equation}
on the one hand, and
\begin{equation}
{\widehat{\chi}}^{\ }_{a}(\bs{k})=:
\sum_{\mathsf{a}=1}^{N}\,
{G}^{\ }_{a\mathsf{a}}(\bs{k})\,
{\widehat{\chi}}^{\ }_{\mathsf{a}}(\bs{k}),
\label{eq: sim gauge trsf u's and chi's b}
\end{equation}
\end{subequations}
on the other hand,
leaves $\widehat{\psi}^{\ }_{\alpha}(\bs{k})$ unchanged.
The $N\times N$ matrix 
${G}(\bs{k})$
with the matrix elements
${G}^{\ }_{a\mathsf{a}}(\bs{k})$
is unitary and obeys periodic boundary conditions
across the Brillouin zone.
The sans-serif font for the index 
$\mathsf{a}=1,\cdots,N$
conveys that the vector $u^{(\mathsf{a})}(\bs{k})$
with the $N$ components
$u^{(\mathsf{a}) }_{\alpha}(\bs{k})$
labeled by the orbitals $\alpha=1,\cdots,N$
need not be anymore an eigenstate of the single-particle 
Bloch Hamiltonian.

Observe that for any triplet $a,\alpha,\mathsf{a}=1,\cdots,N$,
for any momentum $\bs{k}$ from the Brillouin zone,
and any Cartesian unit vector $\bs{e}^{\mu}$ from $\mathbb{R}^{d}$,
had we imposed the twisted boundary conditions
\begin{equation}
G^{\ }_{a\mathsf{a}}\big(\bs{k}+(2\pi/\mathfrak{a})\bs{e}^{\mu}\big)=
e^{-\mathrm{i}(2\pi\,\phi^{\mu})}\,\,
G^{\ }_{a\mathsf{a}}(\bs{k})
\end{equation}
parametrized by the
real numbers $0\leq\phi^{\mu}<1$ with $\mu=1,\cdots,d$,
it would then follow that
\begin{equation}
u^{\mathsf{a}}_{\alpha}\big(\bs{k}+(2\pi/\mathfrak{a})\,\bs{e}^{\mu}\big)=
e^{-\mathrm{i}(2\pi\,\phi^{\mu})}\,
u^{\mathsf{a}}_{\alpha}(\bs{k})
\end{equation}
obeys twisted boundary conditions instead of periodic ones, while
\begin{equation}
\chi^{\ }_{\mathsf{a}}\big(\bs{k}+(2\pi/\mathfrak{a})\,\bs{e}^{\mu}\big)=
e^{+\mathrm{i}[2\pi\,(\theta^{\mu}+\phi^{\mu})]}\,
\chi^{\ }_{\mathsf{a}}(\bs{k})
\end{equation}
obeys new twisted boundary conditions. As a corollary, the
gauge field $\bs{A}^{\ }_{\mathsf{a}\mathsf{a}'}$ 
obtained from Eq.~(\ref{eq: def unprojected gauge field})
by substituting the band indices for the sans-serif ones
would not be a pure gauge anymore as a result of this large gauge 
transformation. An example of a large gauge transformation is 
\begin{subequations}
\label{eq: def ex large gauge trsf}
\begin{equation}
G^{\ }_{\bs{R}^{\ }_{0}}=
e^{+\mathrm{i}\,\widehat{\bs{P}}\cdot\bs{R}^{\ }_{0}}
\end{equation}
for some $\bs{R}^{\ }_{0}\in\Lambda^{\ }_{R}$
where $\widehat{\bs{P}}$ is the operator defined by the algebra
\begin{equation}
[\widehat{\bs{R}},\widehat{\bs{P}}]=
\mathrm{i}\,\widehat{Q}
\end{equation}
\end{subequations}
with $\widehat{Q}$ the fermion number operator.
It acts on the single-particle states
$|\chi^{a}(\bs{k})\rangle:=\widehat{\chi}^{\dag}_{a}(\bs{k})\,|0\rangle$,
where $|0\rangle$ is the state annihilated by any band annihilation operator,
by multiplication with the phase 
$e^{+\mathrm{i}\,\bs{k}\cdot\bs{R}^{\ }_{0}}$.
Thus, the action of the large gauge transformation%
~(\ref{eq: def ex large gauge trsf})
on $|\chi^{a}(\bs{k})\rangle$ is to
change the boundary condition obeyed by
$|\chi^{a}(\bs{k})\rangle$
from twisted to periodic.
In turn, the large gauge transformation%
~(\ref{eq: def ex large gauge trsf})
acts on the single-particle states
$|W^{a}_{\bs{R}}\rangle:=\widehat{W}^{\dag}_{a;\bs{R}}\,|0\rangle$
by shifting $\bs{R}$ to  $\bs{R}-\bs{R}^{\ }_{0}$, i.e., 
as a global translation of the lattice $\Lambda^{\ }_{R}$.

Let the insulating noninteracting many-body ground state 
$|\Phi\rangle$ be
obtained by filling all the single-particle states from the $\widetilde{N}$ 
bands below the spectral gap depicted in
Fig.~\ref{fig: spectral single-particle gap}. 
The ground state $|\Phi\rangle$
is an $SU(\widetilde{N})$ singlet under the $U(\widetilde{N})$
gauge transformation
defined by restricting the band index in Eq.%
~(\ref{eq: sim gauge trsf u's and chi's b})
to the subset of occupied band indices.
Consequently, the ground state expectation value
of any polynomial $P$ of the components of the projected
position operator $\widehat{\bs{X}}^{\ }_{r}$
is, if it exists, invariant under the simultaneous 
$U(\widetilde{N})$
gauge transformation defined by restricting the band index in Eq.%
~(\ref{eq: sim gauge trsf u's and chi's})
to the subset of occupied band indices, i.e.,
\begin{subequations}
\label{eq: gauge invariance local density k space bis}
\begin{equation}
{\widehat{\chi}}^{\dag}(\bs{k})\to
{\widehat{\chi}}^{\dag}(\bs{k})\,
{G}^{\dag}(\bs{k}),
\qquad
{\widehat{\chi}}(\bs{k})\to
{G}(\bs{k})\,
{\widehat{\chi}}(\bs{k}),
\label{eq: gauge invariance local density k space bis a}
\end{equation}
on the one hand, and
\begin{equation}
{A}^{\ }_{\mu}(\bs{k})\to
{G}(\bs{k})\,
{A}^{\ }_{\mu}(\bs{k})\,
{G}^{\dag}(\bs{k})
-
\left(
\partial^{\ }_{\mu}
{G}
\right)\!
(\bs{k})\,
{G}^{\dag}(\bs{k}),
\label{eq: gauge invariance local density k space bis b}
\end{equation}
\end{subequations}
on the other hand. Here,
$G(\bs{k})$ is any unitary 
$\widetilde{N}\times\widetilde{N}$ matrix
for any $\bs{k}\in\Lambda^{\star}_{\mathrm{BZ}}$,
including one that changes the boundary conditions across
the Brillouin zone,
and matrix multiplication is implied 
in Eq.~\eqref{eq: gauge invariance local density k space bis}
with the operator-valued column-vectors 
${\widehat{\chi}}(\bs{k})$ 
and row-vectors
${\widehat{\chi}}^{\dag}(\bs{k})$
that have the components
${\widehat{\chi}}^{\   }_{\tilde{a}}(\bs{k})$ 
and 
${\widehat{\chi}}^{\dagger}_{\tilde{a}}(\bs{k})$,
$\tilde{a}=1,\cdots,\widetilde{N}$,
respectively. Existence of
$\langle\Phi|P(\widehat{\bs{X}}^{\ }_{r})|\Phi\rangle$
amounts to constructing
a gauge-invariant regularization of 
$\langle\Phi|P(\widehat{\bs{X}}^{\ }_{r})|\Phi\rangle$.
As we now prove, although not all polynomials $P$
are compatible with a gauge-invariant regularization of 
$\langle\Phi|P(\widehat{\bs{X}}^{\ }_{r})|\Phi\rangle$,
we do find polynomials $P$ that admit such
a gauge-invariant regularization.

To see this, we are going to momentarily dispense with complications arising
from many-body terms and work solely in the single-particle
Hilbert space. We define the pair of single-particle states
\begin{subequations}
\label{eq: def single-particle Hilbert space}
\begin{equation}
\begin{split}
&
|W^{a}_{\bs{R}}\rangle:=
\widehat{W}^{\dag}_{a;\bs{R}}\,
|0\rangle,
\qquad
|\chi^{a}(\bs{k})\rangle:=
\widehat{\chi}^{\dag}_{a}(\bs{k})\,
|0\rangle,
\end{split}
\label{eq: def single particle states W and chi}
\end{equation}
with 
$a=1,\cdots,N$, 
$\bs{R}\in\Lambda^{\ }_{R}$, 
$\bs{k}\in\Lambda^{\star}_{\mathrm{BZ}}$,
and the pair of single-particle states
\begin{equation}
\begin{split}
&
|\psi^{\alpha}_{\bs{r}}\rangle:=
\widehat{\psi}^{\dag}_{\alpha;\bs{r}}\,
|0\rangle,
\qquad
|\psi^{\alpha}(\bs{k})\rangle:=
\widehat{\psi}^{\dag}_{\alpha}(\bs{k})\,
|0\rangle,
\end{split}
\label{eq: def single particle states W and psi}
\end{equation}
\end{subequations}
with 
$\alpha=1,\cdots,N$ and
$\bs{r}\in\Lambda^{\ }_{r}$. 
The single-particle counterparts to the
projected position operators%
~(\ref{eq: def R operator b})
and%
~(\ref{eq: def r operator b})
are defined to be
\begin{subequations}
\begin{equation}
\widehat{\bs{X}}^{\ }_{R}:=
\sum_{\bs{R}\in\Lambda^{\ }_{R}}
\sum_{\tilde{a}=1}^{\widetilde{N}}
|W^{\tilde{a}}_{\bs{R}}\rangle\,
\bs{R}\,
\langle W^{\tilde{a}}_{\bs{R}}|,
\label{eq: def hat X R}
\end{equation}
and, with the help of the single-particle kernel defined
in Eq.~(\ref{eq: def r operator c}),
\begin{equation}
\widehat{\bs{X}}^{\ }_{r}:=
\sum_{\bs{R},\bs{R}'\in\Lambda^{\ }_{R}}
\sum_{\tilde{a},\tilde{a}'=1}^{\widetilde{N}}
|W^{\tilde{a}}_{\bs{R}}\rangle\,
\bs{\mathcal{X}}^{\ }_{\tilde{a},\tilde{a}';\bs{R},\bs{R}'}\,
\langle W^{\tilde{a}'}_{\bs{R}'}|,
\label{eq: def hat X r}
\end{equation}
\end{subequations}
respectively. Evidently, the trace over the (unprojected)
single-particle Hilbert space of
either $\widehat{\bs{X}}^{\ }_{R}$ 
or $\widehat{\bs{X}}^{\ }_{r}$
is ill-defined because of the ill-conditioned sum over the lattice
$\Lambda^{\ }_{R}$. 

The situation is much better with
the commutator between 
$\widehat{X}^{\mu}_{r}$
and 
$\widehat{X}^{\nu}_{r}$
for any $\mu,\nu=1,\cdots,d$
owing to the identity 
\begin{subequations}
\label{eq: 2 bracket Xr}
\begin{equation}
\begin{split}
\left[
\widehat{X}^{\mu}_{r},
\widehat{X}^{\nu}_{r}
\right]
=&\,
-
\int\limits_{\Lambda^{\star}_{\mathrm{BZ}}}
\frac{\mathrm{d}^{d}\bs{k}}{(2\pi/\mathfrak{a})^{d}}\,
|
\chi^{\tilde{a}}(\bs{k})
\rangle
F^{\tilde{a}\tilde{b}}_{\mu\nu}(\bs{k})
\langle
\chi^{\tilde{b}}(\bs{k})
|
\end{split}
\label{eq: 2 bracket Xr a}
\end{equation}
where the summation convention over the repeated band labels
$\tilde{a},\tilde{b}=1,\cdots,\widetilde{N}$ is implied and
\begin{equation}
F^{\ }_{\mu\nu}(\bs{k}):=
\left(
\partial^{\ }_\mu\,
A^{\ }_{\nu}
\right)(\bs{k})
-
\left(
\partial^{\ }_\nu\,
A^{\ }_{\mu}
\right)(\bs{k})
+
\left[
A^{\ }_{\mu},
A^{\ }_{\nu}
\right](\bs{k}).
\label{eq: 2 bracket Xr b}
\end{equation}
\end{subequations}
We refer the reader to
Appendix%
~\ref{appsec: Regularization of non-commuting projected position operators}
for the proof of Eq.~(\ref{eq: 2 bracket Xr}).
Evidently, the components of $\widehat{\bs{X}}^{\ }_{r}$ 
are noncommutative if the non-Abelian gauge field 
$\bs{A}^{\ }_{\mu}(\bs{k})$
has a nonvanishing field strength
$\bs{F}^{\ }_{\mu\nu}(\bs{k})$.%
~\cite{King-Smith93}
We can now safely take the trace of the commutator%
~(\ref{eq: 2 bracket Xr})
over the single-particle Hilbert space,
\begin{widetext}
\begin{equation}
\begin{split}
\frac{1}{N^{\ }_{\mathrm{p}}}
\mathrm{Tr}\,
\left[
\widehat{X}^{\mu}_{r},
\widehat{X}^{\nu}_{r}
\right]
=
-
\frac{1}{\bar\rho}\,
\int\limits_{\Lambda^{\star}_{\mathrm{BZ}}}\!
\frac{\mathrm{d}^{d}\bs{k}}{\left(2\pi/\mathfrak{a}\right)^{d}}\,
\mathrm{tr}\,
F^{\ }_{\mu\nu}(\bs{k})=
-
\frac{1}{\bar\rho}
\int\limits_{\Lambda^{\star}_{\mathrm{BZ}}}\!
\frac{\mathrm{d}^{d}\bs{k}}{\left(2\pi/\mathfrak{a}\right)^{d}}\,
\mathrm{tr}\,
\left(
\partial^{\ }_{\mu} A^{\ }_{\nu}
-
\partial^{\ }_{\nu} A^{\ }_{\mu}
\right)(\bs{k}),
\end{split}
\label{eq: desired intensive single-particle expectation value for com}
\end{equation}
\end{widetext}
provided we multiply the functional trace $\mathrm{Tr}$ 
by the inverse of the total number of particles $N^{\ }_{\mathrm{p}}$ 
in the insulating ground state $|\Phi\rangle$
to obtain an intensive quantity. 
Then, the ratio of the number of particles $N^{\ }_{\mathrm{p}}$
to the single-particle Bloch wavefunction normalization constant
is nothing but the average particle density
$
\bar{\rho}
$.
The symbol
$\mathrm{tr}$ denotes the trace over the lower $\widetilde{N}$
bands.
Equation%
~(\ref{eq: desired intensive single-particle expectation value for com})
is well-defined and 
invariant under both pure and large gauge transformations
of the form%
~(\ref{eq: gauge invariance local density k space bis b}).

For any integer $n=2,3,\cdots,$
we define the $n$-bracket of the $n$ symbols
$B^{\ }_{1}$,
$B^{\ }_{2}$,
$\cdots$,
$B^{\ }_{n}$
equipped with the product $\times$
to be their fully antisymmetrized product
\begin{equation}
[
B^{\ }_{1},
B^{\ }_{2},
\cdots,
B^{\ }_{n}
]\equiv
\epsilon^{i^{\ }_{1}i^{\ }_{2}\cdots i^{\ }_{n}}
B^{\ }_{i^{\ }_{1}}\times
B^{\ }_{i^{\ }_{2}}\times
\cdots\times
B^{\ }_{i^{\ }_{n}}
\end{equation}
where the summation convention over repeated indices is implied and
the symbol $\epsilon^{i^{\ }_{1}i^{\ }_{2}\cdots i^{\ }_{n}}$ implies
antisymmetrization. For convenience, we also introduce the terminology
of the 1-bracket of the symbol $B$ to be the symbol $B$ itself.

\vskip 10 true pt
\begin{widetext}
Observe that any odd-bracket can be rewritten as 
\begin{subequations}
\begin{equation}
\begin{split}
[
B^{\ }_{1},
B^{\ }_{2},
\cdots,
B^{\ }_{2m+1}
]=
\left(
\frac{1}{2}
\right)^{m}
\epsilon^{i^{\ }_{1}i^{\ }_{2}\cdots i^{\ }_{2m+1}}
\left[
B^{\ }_{i^{\ }_{1}},
B^{\ }_{i^{\ }_{2}}
\right]
\times
\cdots
\times
\left[
B^{\ }_{i^{\ }_{2m-1}},
B^{\ }_{i^{\ }_{2m}}
\right]
\times
B^{\ }_{i^{\ }_{2m+1}},
\end{split}
\end{equation}
while any even-bracket can be rewritten as
\begin{equation}
\begin{split}
[
B^{\ }_{1},
B^{\ }_{2},
\cdots,
B^{\ }_{2m+2}
]=
%&\,
\left(
\frac{1}{2}
\right)^{m+1}
\epsilon^{i^{\ }_{1}i^{\ }_{2}\cdots i^{\ }_{2m+2}}
\left[
B^{\ }_{i^{\ }_{1}},
B^{\ }_{i^{\ }_{2}}
\right]
\times
\cdots
\times
\left[
B^{\ }_{i^{\ }_{2m+1}},
B^{\ }_{i^{\ }_{2m+2}}
\right]
\end{split}
\end{equation}
\end{subequations}
for $m=0,1,2,\cdots$.
For any integer $m=0,1,2,\cdots$ such that $2m+2\leq d$,
it then follows that
\begin{equation}
\frac{1}{N^{\ }_{\mathrm{p}}}
\mathrm{Tr}\,
\left[
\widehat{X}^{\mu^{\ }_{1}}_{r},
\cdots,
\widehat{X}^{\mu^{\ }_{2m+1}}_{r},
\widehat{X}^{\mu^{\ }_{2m+2}}_{r}
\right]
=
-
\left(-\frac{1}{2}\right)^{m+1}\,
\frac{1}{\bar\rho}\,
\int\limits_{\Lambda^{\star}_{\mathrm{BZ}}}\!
\frac{\mathrm{d}^{d}\bs{k}}{\left(2\pi/\mathfrak{a}\right)^{d}}\,
\epsilon^{i^{\ }_{1}\cdots i^{\ }_{2m+1}i^{\ }_{2m+2}}\,
\mathrm{tr}\,
\left(
F^{\ }_{i^{\ }_{1}i^{\ }_{2}}\,
\cdots
F^{\ }_{i^{\ }_{2m+1}i^{\ }_{2m+2}}
\right)(\bs{k})
\label{eq: desired intensive single-particle expectation value for 2m+2}
\end{equation}
with $i^{\ }_{1},\cdots,i^{\ }_{2m+2}=\mu^{\ }_{1},\cdots,\mu^{\ }_{2m+2}$.
Equation%
~(\ref{eq: desired intensive single-particle expectation value for 2m+2})
is well-defined and 
invariant under both pure and large gauge transformations
of the form%
~(\ref{eq: gauge invariance local density k space bis b}).
The right-hand side of Eq.%
~(\ref{eq: desired intensive single-particle expectation value for 2m+2})
is proportional to the $(m+1)$-th Chern number.
\vskip 10 true pt
\end{widetext}

For any integer $n$ such that $2\leq n\leq d$,
the single-particle trace
over any $n$-bracket of the components 
$\widehat{X}^{\mu^{\ }_{1}}_{R}$,
$\cdots$,
$\widehat{X}^{\mu^{\ }_{n}}_{R}$
vanishes 
owing to the fact that (i)
$\widehat{\bs{X}}^{\ }_{R}$
is diagonal in the Wannier basis
and (ii) performing the antisymmetrization
$\epsilon^{\ }_{i^{\ }_{1}i^{\ }_{2}\cdots i^{\ }_{n}}\,
R^{i^{\ }_{1}}\,R^{i^{\ }_{2}}\cdots R^{i^{\ }_{n}}=0$
before taking the sum over the lattice $\Lambda^{\ }_{R}$.

In contrast to these brackets,
neither is the single-particle trace over the 1-bracket of the component 
$\widehat{X}^{\mu}_{r}$ 
nor that of the 1-bracket of the component 
$\widehat{X}^{\mu}_{R}$
with $\mu=1,\cdots,d$  
well-defined. More generally, for any integer 
$m=0,1,2,\cdots$ such that $2m+1\leq d$,
the single-particle trace
over any $(2m+1)$-bracket of the components 
$\widehat{X}^{\mu^{\ }_{1}}_{r}$,
$\cdots$,
$\widehat{X}^{\mu^{\ }_{2m+1}}_{r}$
is ill-defined because there always remain
ill-conditioned sums over the lattice 
$\Lambda^{\ }_{R}$.
We are going to construct explicitly
a suitable regularization of
the single-particle trace
over any $(2m+1)$-bracket of the components 
$\widehat{X}^{\mu^{\ }_{1}}_{r}$,
$\cdots$,
$\widehat{X}^{\mu^{\ }_{2m+1}}_{r}$
for $m=0$ and $m=1$ that can be nonvanishing and is invariant under
any pure gauge transformation of the form%
~(\ref{eq: gauge invariance local density k space bis b}).

To this end, we need the important identity
\begin{equation}
\widehat{\bs{X}}^{\ }_{r}
-
\widehat{\bs{X}}^{\ }_{R}=
\int\limits_{\Lambda^{\star}_{\mathrm{BZ}}}
\frac{\mathrm{d}^{d}\bs{k}}{\left(2\pi/\mathfrak{a}\right)^{d}}\,
|
\chi^{\tilde{a}}(\bs{k})
\rangle\,
\mathrm{i}\,
\bs{A}^{\tilde{a}\tilde{b}}(\bs{k})
\langle
\chi^{\tilde{b}}(\bs{k})
|
\label{eq: fundamental identity for odd bracket}
\end{equation}
which is proved in
Appendix%
~\ref{appsec: Regularization of non-commuting projected position operators}.
We can now safely take the trace of Eq.%
~(\ref{eq: fundamental identity for odd bracket})
over the single-particle Hilbert space
as we did in Eq.%
~(\ref{eq: desired intensive single-particle expectation value for com}). 
We find 
\begin{equation}
\frac{1}{N^{\ }_{\mathrm{p}}}
\mathrm{Tr}\,
\left(
\widehat{\bs{X}}^{\ }_{r}
-
\widehat{\bs{X}}^{\ }_{R}
\right)
=
\frac{\mathrm{i}}{\bar\rho}\,
\int\limits_{\Lambda^{\star}_{\mathrm{BZ}}}\!
\frac{\mathrm{d}^{d}\bs{k}}{\left(2\pi/\mathfrak{a}\right)^{d}}\,
\mathrm{tr}\,
\bs{A}(\bs{k}).
\label{eq: desired intensive single-particle expectation value for 1-bracket}
\end{equation}
Equation%
~(\ref{eq: desired intensive single-particle expectation value for 1-bracket})
is invariant under pure (but not large) gauge transformations
of the form%
~(\ref{eq: gauge invariance local density k space bis b}).
The loss of the invariance under the large gauge transformations
of the form%
~(\ref{eq: gauge invariance local density k space bis b})
is to be attributed to the fact that the regularization%
~(\ref{eq: desired intensive single-particle expectation value for 1-bracket})
breaks translation invariance in that there are gauge nonequivalent ways
of defining eigenstates of the projected position operator at short distances.
In other words, it is not possible to construct a wave packet that
can resolve distances smaller than the lattice spacing $\mathfrak{a}$.
This fuzziness survives the limit $\mathfrak{a}\to0$ 
as the breakdown of gauge symmetry under large
gauge transformations of the form%
~(\ref{eq: gauge invariance local density k space bis b}).

The regularization%
~(\ref{eq: desired intensive single-particle expectation value for 1-bracket})
is not unique. For example, we could have chosen a regularization
of the single-particle trace over any $(2m+1)$-bracket of the components 
$\widehat{X}^{\mu^{\ }_{1}}_{r}$,
$\cdots$,
$\widehat{X}^{\mu^{\ }_{2m+1}}_{r}$
that preserves this translation invariance through the heat kernel
method. The heat kernel regularization yields zero for all odd-brackets,
a manifestly gauge invariant result! However, we reject this
regularization because enforcing invariance under 
large gauge transformations of the form%
~(\ref{eq: gauge invariance local density k space bis b})
is not required by general symmetry arguments.

Yet another example of a regularization 
of the single-particle trace over
any $n$-bracket of the components 
$\widehat{X}^{\mu^{\ }_{1}}_{r}$,
$\cdots$,
$\widehat{X}^{\mu^{\ }_{n}}_{r}$
is to do the replacement
$\widehat{X}^{\mu^{\ }_{1}}_{r}\to
 \widehat{X}^{\mu^{\ }_{1}}_{r}-\widehat{X}^{\mu^{\ }_{1}}_{R}$,
$\cdots$,
$\widehat{X}^{\mu^{\ }_{n}}_{r}\to
 \widehat{X}^{\mu^{\ }_{n}}_{r}-\widehat{X}^{\mu^{\ }_{n}}_{R}$.
With this substitution, the single-particle trace
is well-defined, for it does not contain anymore ill-conditioned
sums over the lattice $\Lambda^{\ }_{R}$.
However, whenever the single-particle trace over this $n$-bracket
is nonvanishing, it breaks the invariance under pure
$SU(\widetilde{N})$ 
gauge transformations of the form%
~(\ref{eq: gauge invariance local density k space bis b})
for any $n\geq2$. For this reason, we reject this regularization.

\medskip
\begin{widetext}
We are now in position to state the main result of
Sec.~\ref{subsec: Algebra of the position operators}.
When $d\geq3$ and for any choice of the triplet
$\mu^{\ }_{1},\mu^{\ }_{2},\mu^{\ }_{3}=1,\cdots,d$,
we define the regularized 3-bracket of the components
$\widehat{X}^{\mu^{\ }_{1}}_{r}$,
$\widehat{X}^{\mu^{\ }_{2}}_{r}$,
and
$\widehat{X}^{\mu^{\ }_{3}}_{r}$
of the projected position operator%
~(\ref{eq: def r operator b}) to be%
~\cite{footnote reg 3-bracket,Estienne12}
\begin{equation}
\begin{split}
2
\left[
\widehat{X}^{\mu^{\ }_{1}}_{r},
\widehat{X}^{\mu^{\ }_{2}}_{r},
\widehat{X}^{\mu^{\ }_{3}}_{r}
\right]^{\ }_{\mathrm{reg}}:=&\,
\left[
\widehat{X}^{\mu^{\ }_{1}}_{r},
\widehat{X}^{\mu^{\ }_{2}}_{r},
\left(
\widehat{X}^{\mu^{\ }_{3}}_{r}
-
\widehat{X}^{\mu^{\ }_{3}}_{R}
\right)
\right]
+
\left[
\widehat{X}^{\mu^{\ }_{1}}_{r},
\left(
\widehat{X}^{\mu^{\ }_{2}}_{r}
-
\widehat{X}^{\mu^{\ }_{2}}_{R}
\right),
\widehat{X}^{\mu^{\ }_{3}}_{r}
\right]
+
\left[
\left(
\widehat{X}^{\mu^{\ }_{1}}_{r}
-
\widehat{X}^{\mu^{\ }_{1}}_{R}
\right),
\widehat{X}^{\mu^{\ }_{2}}_{r},
\widehat{X}^{\mu^{\ }_{3}}_{r}
\right]
\\
&\,
-
\left[
\left(
\widehat{X}^{\mu^{\ }_{1}}_{r}
-
\widehat{X}^{\mu^{\ }_{1}}_{R}
\right),
\left(
\widehat{X}^{\mu^{\ }_{2}}_{r}
-
\widehat{X}^{\mu^{\ }_{2}}_{R}
\right),
\left(
\widehat{X}^{\mu^{\ }_{3}}_{r}
-
\widehat{X}^{\mu^{\ }_{3}}_{R}
\right)
\right].
\end{split}
\label{eq: def reg 3 bracket}
\end{equation}
We have introduced the multiplicative factor $2$ 
on the left-hand side in order to preserve the 
number of $3$-brackets under regularization, 
namely one prior to regularization. Indeed,
since we add 3-brackets that include one 
substitution $\bs{X}^{\ }_{r}\to \bs{X}^{\ }_{r}- \bs{X}^{\ }_{R}$
and remove one 3-bracket that include 3 substitutions
$\bs{X}^{\ }_{r}\to \bs{X}^{\ }_{r}- \bs{X}^{\ }_{R}$
on the right-hand side, we are left with $3-1=2$ 3-brackets
on the right-hand side.
It is shown in
Appendix%
~\ref{appsec: Regularization of non-commuting projected position operators}
that we can safely take the single-particle
trace over the regularized 3-bracket%
~(\ref{eq: def reg 3 bracket}) 
after accounting
for the same normalization as for the 1- and 2-brackets, 
\begin{equation}
\frac{1}{N^{\ }_{\mathrm{p}}}
\mathrm{Tr}\,
\left[
\widehat{X}^{\mu^{\ }_{1}}_{r},
\widehat{X}^{\mu^{\ }_{2}}_{r},
\widehat{X}^{\mu^{\ }_{3}}_{r}
\right]^{\ }_{\mathrm{reg}}=
-
\frac{3}{4}\times
\frac{\mathrm{i}}{\bar\rho}\,
\int\limits_{\Lambda^{\star}_{\mathrm{BZ}}}\!
\frac{\mathrm{d}^{d}\bs{k}}{\left(2\pi/\mathfrak{a}\right)^{d}}\,
\epsilon^{i^{\ }_{1}i^{\ }_{2}i^{\ }_{3}}\,
\mathrm{tr}\,
\left(
F^{\ }_{i^{\ }_{1}i^{\ }_{2}}\,
A^{\ }_{i^{\ }_{3}}
-
\frac{2}{3}\,
A^{\ }_{i^{\ }_{1}}\,
A^{\ }_{i^{\ }_{2}}\,
A^{\ }_{i^{\ }_{3}}
\right)(\bs{k})
\label{eq: desired intensive single-particle expectation value for 3-bracket}
\end{equation}
where $i^{\ }_{1},i^{\ }_{2},i^{\ }_{3}=\mu^{\ }_{1},\mu^{\ }_{2},\mu^{\ }_{3}$.
As was the case with
Eq.%
~(\ref{eq: desired intensive single-particle expectation value for 1-bracket})
and for the same reason,
Eq.%
~(\ref{eq: desired intensive single-particle expectation value for 3-bracket})
is invariant under pure (but not large) gauge transformations
of the form%
~(\ref{eq: gauge invariance local density k space bis b}).
\end{widetext}
\medskip

The generalization to the case of any integer $m$ such that $2m+1\leq d$
consists in defining the regularized $(2m+1)$-bracket
$
\left[
\widehat{X}^{\mu^{\ }_{1}}_{r},
\widehat{X}^{\mu^{\ }_{2}}_{r},
\cdots,
\widehat{X}^{\mu^{\ }_{2m}}_{r},
\widehat{X}^{\mu^{\ }_{2m+1}}_{r}
\right]^{\ }_{\mathrm{reg}}
$
by replacing the $(2m+1)$-bracket
$
\left[
\widehat{X}^{\mu^{\ }_{1}}_{r},
\widehat{X}^{\mu^{\ }_{2}}_{r},
\cdots,
\widehat{X}^{\mu^{\ }_{2m}}_{r},
\widehat{X}^{\mu^{\ }_{2m+1}}_{r}
\right]
$
with the sum of all $(2m+1)$-brackets
obtained by doing all the possible substitution
$\widehat{X}^{\mu^{\ }_{i}}_{r}\to 
\widehat{X}^{\mu^{\ }_{i}}_{r}-\widehat{X}^{\mu^{\ }_{i}}_{R}$
$(2l+1)$ times with $l=0,1,\cdots,m$ and adding all
resulting $(2m+1)$-brackets weighted with the sign
$(-)^{l}$. We then define a normal ordering
by which all $\widehat{X}^{\mu^{\ }_{i}}_{r}$
are placed to the left of all 
$\widehat{X}^{\mu^{\ }_{i}}_{r}-\widehat{X}^{\mu^{\ }_{i}}_{R}$
as if they were commuting numbers.
Finally, we divide the resulting linear
combination of $(2m+1)$-brackets
by the integer equal to the absolute value of
the alternating sum of the binomials coefficients
$\left(\begin{array}{c}2m+1\\1\end{array}\right)
-
\left(\begin{array}{c}2m+1\\3\end{array}\right)
\pm
\cdots$.
The single-particle trace over the regularized $(2m+1)$-bracket
after accounting for the same normalization as for the even-brackets and 
the 1- and 3-brackets is intensive and proportional
to the Chern-Simons invariant obtained from integrating
over the $d$-dimensional Brillouin zone with $d\geq 2m+1$
the Chern-Simons $(2m+1)$ form.

It is time to draw a precise connection between
the single-particle traces over the (regularized)
brackets of the components of the position operator
$\widehat{\bs{X}}^{\ }_{r}$
and topological invariants.

We define the $d$ Chern-Simons invariants
built from Chern-Simons 1 forms
in $d$-dimensional momentum space 
for any choice of $\mu=1,\cdots,d$ as
\begin{subequations}
\begin{equation}
\mathrm{CS}^{(1)}_{\mu}:=
\mathrm{i}\,
\int
\frac{\mathrm{d}^{d}\bs{k}}{(2\pi)^d}\,
{\mathrm{tr}}\,
{A}^{\ }_{\mu}.
\label{eq: CS 1}
\end{equation}
We also define the $d(d-1)(d-2)/6$ Chern-Simons invariants
built from Chern-Simons 3 forms
in $d$-dimensional momentum space 
for any choice of $\mu,\mu,\lambda=1,\cdots,d$ as
\begin{equation}
\mathrm{CS}^{(3)}_{\mu\nu\rho}:=
\frac{\epsilon^{IJK}}{4}
\int
\frac{\mathrm{d}^{d}\bs{k}}{(2\pi)^{d-1}}\,
{\mathrm{tr}}
\left(
{A}^{\ }_{I}
{F}^{\ }_{JK}
-\frac{2}{3}
{A}^{\ }_{I}
{A}^{\ }_{J}
{A}^{\ }_{K}
\right),
\label{eq: CS 3}
\end{equation}
\end{subequations}
where $I,J,K=\mu,\nu,\rho$.
The integral on the right-hand side of
Eq.~\eqref{eq: CS 1} and Eq.~\eqref{eq: CS 3}
is quantized to half-integer values
if the single-particle Hamiltonian
obeys the chiral symmetry
and the domain of integration is that of a $d$-dimensional torus
$T^{d}$ with the volume $(2\pi)^{d}$, i.e., $\mathfrak{a}=1$.%
~\cite{Qi08}
The 1D and 3D Chern-Simons invariants in
$d$-dimensional momentum space carry the engineering
dimensions of length raised to the powers $(1-d)$ and $(3-d)$,
respectively. They are thus dimensionless if and only if
$d=1$ and $d=3$, respectively.

The Chern-Simons invariants~\eqref{eq: CS 1} 
and~\eqref{eq: CS 3}
are only well-defined modulo integer values under the 
$U(\widetilde{N})$ 
gauge transformations%
~(\ref{eq: gauge invariance local density k space bis b})
since the latter can change the former by their winding numbers, 
namely the numbers
\begin{subequations}
\begin{equation}
\mathrm{i}
\int\frac{\mathrm{d}^{d}\bs{k}}{(2\pi)^d}\
{\mathrm{tr}}^{\ }\,
{G}^{\dag}\,
\partial^{\ }_{\mu}
{G},
\qquad
\mu=1,\cdots,d,
\end{equation}
and
\begin{equation}
\frac{\mathrm{i}
\epsilon^{IJK}}{6}
\int\frac{\mathrm{d}^{d}\bs{k}}{(2\pi)^{d-1}}\
{\mathrm{tr}}^{\ }
\left[
\left(
{G}^{\dag}\,
\partial^{\ }_{I}
{G}
\right)
\left(
{G}^{\dag}\,
\partial^{\ }_{J}
{G}
\right)
\left(
{G}^{\dag}\,
\partial^{\ }_{K}
{G}
\right)
\right]
,
\end{equation}
with $I,J,K=\mu,\nu,\rho$
and $\mu,\nu,\rho=1,\cdots,d$, respectively.
\end{subequations}

In contrast, the $d(d-1)/2$ first Chern numbers
defined in $d$-dimensional momentum space as
\begin{equation}
\begin{split}
\mathrm{Ch}^{(1)}_{\mu\nu}:=&\,
\mathrm{i}
\int
\frac{\mathrm{d}^{d}\bs{k}}{(2\pi)^{d-1}}\,
{\mathrm{tr}}^{\ }\,
{F}^{\ }_{\mu\nu},
\end{split}
\qquad 
\mu,\nu=1,\cdots, d,
\label{eq: Chern numbers 1}
\end{equation}
can only take integer values
if the domain of integration is that of a $d$-dimensional torus
$T^{d}$ with the volume $(2\pi)^{d}$ in momentum space,%
~\cite{footnote on Stokes thm} 
irrespective of whether or not
the single-particle Hamiltonian
obeys the chiral symmetry. However, when chiral symmetry holds, 
the 1D Chern-Simons invariants~(\ref{eq: CS 1}) are quantized.%
~\cite{Ryu10}
Therefore, derivatives of these quantities vanish, which 
in turn implies that all first Chern numbers~(\ref{eq: Chern numbers 1}) 
vanish. Furthermore, the first Chern numbers~\eqref{eq: Chern numbers 1}
are invariant under the  
$U(\widetilde{N})$ 
gauge transformations%
~(\ref{eq: gauge invariance local density k space bis b}).
The first Chern numbers
defined in $d$-dimensional momentum space
carry the engineering dimensions of $(2-d)$.

In closing, we reexpress our main result
using second quantization and
for the case of $d=3$ dimensions.
We shall use the standard notation
$:(\cdots):$ for normal ordering under which it is 
understood that creation operators are to be moved to the left
of annihilation operators within the symbol $(\cdots)$ 
as if they were Grassman numbers. After identifying
the single-particle states defined in 
Eq.~(\ref{eq: def single particle states W and chi})
with the single-particle holes resulting from annihilating
a single-particle state from the insulating
ground state $|\Phi\rangle$, we find that
\begin{equation}
\begin{split}
&
\frac{1}{N^{\ }_{\mathrm{p}}}
\left\langle{\Phi}\left|
:
\left[
{\widehat{X}}^{1}_{r},
{\widehat{X}}^{2}_{r},
{\widehat{X}}^{3}_{r}
\right]^{\ }_{\mathrm{reg}}
:
\right|{\Phi}\right\rangle
=\\
&\quad
-
\frac{(2\pi)^2\mathrm{i}}{\bar{\rho}}\,
\left[
\frac{(2\pi)^3}{2}\,
\frac{N^{\ }_{\mathrm{p}}}{\bar{\rho}}\,
\epsilon^{\mu\nu\rho}\,
\mathrm{CS}^{(1)}_\mu\,
\mathrm{Ch}^{(1)}_{\nu\rho}
+
3
\mathrm{CS}^{(3)}
\right].
\label{eq: Main result}
\end{split}
\end{equation}
Again, it should be noted that the right-hand side of
Eq.~\eqref{eq: Main result} is entirely determined by
its quantized topological numbers
if the single-particle Hamiltonian
obeys the chiral symmetry, 
and if the equality is understood 
modulo contributions from large
$U(\widetilde{N})$ 
gauge transformations%
~(\ref{eq: gauge invariance local density k space bis b})
in which case
\begin{equation}
\frac{1}{N^{\ }_{\mathrm{p}}}
\left\langle{\Phi}\left|
:
\left[
{\widehat{X}}^{1}_{r},
{\widehat{X}}^{2}_{r},
{\widehat{X}}^{3}_{r}
\right]^{\ }_{\mathrm{reg}}
:
\right|{\Phi}\right\rangle
=
-
\frac{3(2\pi)^2\mathrm{i}}{\bar{\rho}}\,
\mathrm{CS}^{(3)}
\label{eq: Main result with chiral symmetry}
\end{equation}
owing to $\mathrm{Ch}^{(1)}_{\nu\rho}=0$
for any $\nu,\rho=1,\cdots,d$.

\subsection{
Regularized 3-bracket and the Nambu bracket
           }
\label{subsec: Regularized 3-bracket and the Nambu bracket}

On the one hand,
we may define for any momentum $\bs{q}$ from the Brillouin zone
the projected operator
\begin{equation}
\widehat{T}^{\ }_{R}(\bs{q}):=
e^{
-\mathrm{i}\bs{q}\cdot\widehat{\bs{X}}^{\ }_{R}
  }
\end{equation}
with $\widehat{\bs{X}}^{\ }_{R}$
defined in Eq.~(\ref{eq: def hat X R})
that acts on the single-particle Hilbert space defined in
Eq.~(\ref{eq: def single particle states W and chi}).
This operator is a projected
translation operator in the Brillouin zone,
\begin{equation}
\widehat{T}^{\ }_{R}(\bs{q})\,
|\chi^{a}(\bs{k})\rangle=
\sum_{\tilde{a}=1}^{\widetilde{N}}
\delta^{a,\tilde{a}}\,
|\chi^{\tilde{a}}(\bs{k}+\bs{q})\rangle
\end{equation}
for any $a=1,\cdots,N$.
Its algebra under composition closes,
\begin{equation}
\widehat{T}^{\ }_{R}(\bs{q}^{\ }_{1})\,
\widehat{T}^{\ }_{R}(\bs{q}^{\ }_{2})=
\widehat{T}^{\ }_{R}(\bs{q}^{\ }_{1}+\bs{q}^{\ }_{2})
\end{equation}
for any pair of momenta from the Brillouin zone.

On the other hand, we may also want to define
for any momentum $\bs{q}$ from the Brillouin zone
the projected operator
\begin{equation}
\widehat{T}^{\ }_{r}(\bs{q}):=
e^{
-\mathrm{i}\bs{q}\cdot\widehat{\bs{X}}^{\ }_{r}
  }
\end{equation}
with $\widehat{\bs{X}}^{\ }_{r}$
defined in Eq.~(\ref{eq: def hat X r})
that acts on the single-particle Hilbert space defined in
Eq.~(\ref{eq: def single particle states W and psi}).

Be aware that $\widehat{T}^{\ }_{r}(\bs{q})$
differs from the projection 
\begin{subequations}
\label{eq: def Fourier trsf density operator}
\begin{equation}
\begin{split}
\widehat{\rho}(\bs{q}):=&\,
\int\limits_{\Lambda^{\star}_{\mathrm{BZ}}}\,
\frac{\mathrm{d}^{d}\bs{k}}{\left(2\pi/\mathfrak{a}\right)^{d}}\,
\sum_{\alpha=1}^{N}
\sum_{\tilde{a},\tilde{a}'=1}^{\widetilde{N}}
u^{(\tilde{a} )*}_{\alpha}(\bs{k})\,
u^{(\tilde{a}') }_{\alpha}(\bs{k}+\bs{q})
\\
&\,
\times
|\chi^{\tilde{a}}(\bs{k})\rangle
\langle\chi^{\tilde{a}'}(\bs{k}+\bs{q})|
\\
\equiv&\,
\frac{1}{(2\pi/\mathfrak{a})^{d}}\,
\widehat{P}^{\ }_{\widetilde{N}}\,
e^{-\mathrm{i}\bs{q}\cdot\widehat{\bs{r}}}
\widehat{P}^{\ }_{\widetilde{N}}
\end{split}
\end{equation}
on the $\widetilde{N}$ lower bands of
the momentum-resolved
density operator $\widehat{\varrho}(\bs{q})$
defined through the Fourier expansion
\begin{equation}
\widehat{\varrho}_{\bs{r}}=:
\int\limits_{\Lambda^{\star}_{\mathrm{BZ}}}\,
\mathrm{d}^{d}\bs{q}\,
e^{+\mathrm{i}\bs{q}\cdot\bs{r}}\,
\widehat{\varrho}(\bs{q})
\equiv
\int\limits_{\Lambda^{\star}_{\mathrm{BZ}}}\,
\frac{\mathrm{d}^{d}\bs{q}}{(2\pi/\mathfrak{a})^{d}}\,
e^{+\mathrm{i}\bs{q}\cdot(\bs{r}-\widehat{\bs{r}})}
\end{equation}
of the unprojected density operator
\begin{equation}
\widehat{\varrho}_{\bs{r}}:=
\sum_{\alpha=1}^{N}
|\psi^{\alpha}_{\bs{r}}\rangle
\langle\psi^{\alpha}_{\bs{r}}|
\end{equation}
\end{subequations}
defined for any site $\bs{r}$ of lattice $\Lambda^{\ }_{r}$.

The task of computing the regularized
$n$-bracket of the operators
$
\widehat{T}^{\ }_{r}(\bs{q}^{\ }_{1}),
\cdots,
\widehat{T}^{\ }_{r}(\bs{q}^{\ }_{n})
$
is formidable for arbitrary momenta
$\bs{q}^{\ }_{1},\cdots,\bs{q}^{\ }_{n}$
from the Brillouin zone. However, 
an expansion in the momenta
up to order $n$ is feasible in the limit of small momenta.
We undertake such an expansion for the $3$-bracket with the help
of the (classical) Nambu bracket.

To simplify notation, we work with $d=3$.
Let $f^{\ }_{i}(\bs{x})$ with $i=1,2,3$ denote three functions
with the Taylor expansions 
\begin{equation}
f^{\ }_{i}(\bs{x})=
f^{\ }_{i}(\bs{0})
+
\sum_{\mu=1}^{3}
\left(
\partial^{\ }_{\mu}\,
f^{\ }_{i}
\right)(\bs{0})\,
x^{\mu}
+
\cdots
\end{equation}
at the origin of $\bs{x}\in\mathbb{R}^{3}$.
For any pair of functions 
$f^{\ }_{1}$ and $f^{\ }_{2}$,
or for any triplet of functions
$f^{\ }_{1}$, $f^{\ }_{2}$, and $f^{\ }_{3}$
their classical Poisson and Nambu brackets were defined
in Eqs.%
~(\ref{eq: commutator f1 f2 b intro})
and%
~(\ref{eq: commutator f1 f2 f3 b intro}), 
respectively. For any pairs $\mu,\nu=1,2,3$
and
$f^{\ }_{i},f^{\ }_{j}$,
with $i,j=1,2,3$, we shall also need the variant
\begin{equation}
\left\{
f^{\ }_{i},
f^{\ }_{j}
\right\}^{\mu\nu}_{\mathrm{P}}(\bs{0}):=
\sum_{I,J=\mu,\nu}
\epsilon^{\ }_{IJ}\,
\left(
\partial^{I}
f^{\ }_{i}
\right)
\left(
\partial^{J}
f^{\ }_{j}
\right)(\bs{0})
\end{equation}
of the Poisson bracket, respectively. From the operator
identity~(\ref{appeq: operator identity for reg 3 bracket expansion})
follows that the single-particle trace over the regularized 3-bracket of 
$f^{\ }_{1}(\widehat{\bs{X}})$,
$f^{\ }_{2}(\widehat{\bs{X}})$,
and
$f^{\ }_{3}(\widehat{\bs{X}})$
admits the Taylor expansion
\begin{widetext}
\begin{equation}
\begin{split}
\mathrm{Tr}\,
\left[
f^{\ }_{1}(\widehat{\bs{X}}^{\ }_{r}),
f^{\ }_{2}(\widehat{\bs{X}}^{\ }_{r}),
f^{\ }_{3}(\widehat{\bs{X}}^{\ }_{r})
\right]^{\ }_{\mathrm{reg}}=&\,
\epsilon^{ijk}\,
f^{\ }_{i}
\left\{
f^{\ }_{j},
f^{\ }_{k}
\right\}^{\mu\nu}_{\mathrm{P}}(\bs{0})\,
\mathrm{Tr}\,
\left[
\widehat{X}^{\mu}_{r},
\widehat{X}^{\nu}_{r}
\right]
+
\left\{
f^{\ }_{1},
f^{\ }_{2},
f^{\ }_{3}
\right\}^{\ }_{\mathrm{N}}(\bs{0})\,
\mathrm{Tr}\,
\left[
\widehat{X}^{1}_{r},
\widehat{X}^{2}_{r},
\widehat{X}^{3}_{r}
\right]^{\ }_{\mathrm{reg}}
+\cdots.
\end{split}
\label{eq: master formula 3 bracket f1 X f2 X f3 X}
\end{equation}
(A summation convention is implied over the repeated indices
$\mu,\nu=1,2,3$ and $i,j,k=1,2,3$ on the right-hand side.)
This expansion preserves the invariance under
the pure gauge transformations of the form%
~(\ref{eq: gauge invariance local density k space bis b}).
Moreover, in the chiral class,
\begin{equation}
\begin{split}
\mathrm{Tr}\,
\left[
f^{\ }_{1}(\widehat{\bs{X}}^{\ }_{r}),
f^{\ }_{2}(\widehat{\bs{X}}^{\ }_{r}),
f^{\ }_{3}(\widehat{\bs{X}}^{\ }_{r})
\right]^{\ }_{\mathrm{reg}}=&\,
\left\{
f^{\ }_{1},
f^{\ }_{2},
f^{\ }_{3}
\right\}^{\ }_{\mathrm{N}}(\bs{0})\,
\mathrm{Tr}\,
\left[
\widehat{X}^{1}_{r},
\widehat{X}^{2}_{r},
\widehat{X}^{3}_{r}
\right]^{\ }_{\mathrm{reg}}
+\cdots
\end{split}
\label{eq: chiral class 3-bracket of f}
\end{equation}
and
\begin{equation}
\begin{split}
\mathrm{Tr}\,
\left[
\widehat{T}^{\ }_{r}(\bs{q}^{\ }_{1}),
\widehat{T}^{\ }_{r}(\bs{q}^{\ }_{2}),
\widehat{T}^{\ }_{r}(\bs{q}^{\ }_{3})
\right]^{\ }_{\mathrm{reg}}=&\,
+
\mathrm{i}\,
\left(
\bs{q}^{\ }_{1}
\wedge
\bs{q}^{\ }_{2}
\right)
\cdot
\bs{q}^{\ }_{3}\,
\mathrm{Tr}\,
\left[
\widehat{X}^{1}_{r},
\widehat{X}^{2}_{r},
\widehat{X}^{3}_{r}
\right]^{\ }_{\mathrm{reg}}
+\cdots.
\end{split}
\end{equation}
\end{widetext}

Equation~\eqref{eq: chiral class 3-bracket of f} 
admits the following interpretation.
The functions $f^{\ }_{1}$, $f^{\ }_{2}$, and $f^{\ }_{3}$ 
may represent a coordinate transformation in 3D space.
If this transformation preserves volume,
its Jacobian, i.e., the Nambu bracket, equals 1. 
If chiral symmetry holds, 
the trace over the regularized 3-bracket 
of the projected
position operator $\widehat{\bs{X}}^{\ }_{r}$
is to lowest order in the Taylor expansion invariant 
under volume preserving diffeomorphisms, 
while quantum corrections appear at higher order.

Had we restricted ourselfs to $d=2$, 
the transformation property of the 2-bracket (commutator)
under the smooth coordinate transformation
\begin{equation}
(x^{1},x^{2})\to
\Big(
f^{\ }_{1}(\bs{x}),f^{\ }_{2}(\bs{x})
\Big),
\qquad
\bs{x}\equiv(x^{1},x^{2})\in\mathbb{R}^{2},
\end{equation}
is
\begin{equation}
\left[
f^{\ }_{1}(\widehat{\bs{X}}^{\ }_{r}),
f^{\ }_{2}(\widehat{\bs{X}}^{\ }_{r})
\right]=
\left\{
f^{\ }_{1},
f^{\ }_{2}
\right\}^{\ }_{\mathrm{P}}(\bs{0})\,
\left[
\widehat{X}^{1}_{r},
\widehat{X}^{2}_{r}
\right]
+
\cdots.
\label{eq: 2-bracket vs Poisson bracket}
\end{equation}
Except for the quantum corrections contained in $\cdots$, 
the 2-bracket of the projected
position operator $\widehat{\bs{X}}^{\ }_{r}$
is thus invariant under area-preserving diffeomorphisms.
The difference between the 2-bracket and the regularized 3-bracket is,
according to Eq.%
~(\ref{appeq: operator identity for reg 3 bracket expansion}),
that for the latter it is necessary to invoke chiral
symmetry and taking the single-particle trace in order to guarantee
invariance under volume-preserving diffeomorphisms.

The algebra obeyed by the set of diffeomorphisms
of the Euclidean plane that leave the Poisson bracket
invariant realizes the so-called classical
$w^{\ }_{\infty}$ algebra.
Thus, Eq.~(\ref{eq: 2-bracket vs Poisson bracket})
draws the connection to a quantum version of
the $w^{\ }_{\infty}$ algebra. For the quantum Hall effect 
the relevant quantum version
is the $W^{\ }_{\infty}$ algebra 
(see Refs.~\onlinecite{Moyal49},
\onlinecite{Fairlie89},
\onlinecite{Bakas89}, and \onlinecite{Hoppe90})
obeyed by the 
projected density operators in a Landau level.%
~\cite{Girvin85,Iso92,Cappelli93,Martinez93}
A manifestation of the connection between the
$w^{\ }_{\infty}$ and $W^{\ }_{\infty}$ algebras
is found in 
the nondissipative Hall viscosity, which can be viewed as the 
response function of the quantum fluid to an infinitesimal
area-preserving deformation.%
~\cite{Avron95}
In turn, an incompressible 2D classical fluid
may be described in terms of a one-form gauge field,
as appears in the Chern-Simons theory relevant to 
the quantum Hall effect (QHE).%
~\cite{Bachcall91,Jackiw04,Polychronakos07}

In 3D and for Bloch Hamiltonians belonging to the chiral
symmetry class,
the invariance under volume-preserving diffeomorphisms
of 3D Euclidean space displayed
in Eq.~(\ref{eq: chiral class 3-bracket of f})
to lowest order in the Taylor expansion
draws a similiar connection to a quantum algebra that generalizes the
classical algebra obeyed by volume-preserving diffeomorphisms.
In the description of ideal 3D classical fluids 
a two-form gauge field naturally arises as a consequence of 
volume preserving diffeomorphisms.%
~\cite{footnote on ideal 3D classical fluids}
Such a two-form gauge field 
also appears in the 3D BF theory that is believed to be relevant to 
3D topological insulators.%
~\cite{Cho11}

\subsection{
Massive Dirac fermions 
}
\label{subsec: massive Dirac fermions}

In Sec.~\ref{subsec: Algebra of the position operators},
we have related the ground state expectation values  
of the commutator and 
of the regularized 3-bracket of the projected position operators 
$\widehat{\bs{X}}^{\ }_{r}$
to quantized topological numbers, 
namely the Chern numbers and Chern-Simons invariants. 
By contrast,
we have recalled in Eq.~\eqref{eq: noncommuting algebra in QHE b}
that a Landau level has the special property that the commutator 
of projected position operators \emph{itself} is nothing but 
an imaginary number
\begin{equation}
\begin{split}
[\widehat{X}^{\mu},\widehat{X}^{\nu}]
=&\,
-F^{\ }_{\mu\nu}\\
=&\,
\mathrm{i}\,\epsilon^{\ }_{\mu\nu}\,\ell^2_B
\label{eq: closing algebra position operator}
\end{split}
\end{equation}
where $\mu,\nu=1,2$.
In other words, the Berry curvature is constant in a Landau level.

Here, we are going to show that the same is true for massive Dirac electrons 
in 2D, if the limit of small momenta $\bs{k}\to 0$ is considered. 
We then extend the discussion to massive Dirac electrons in 3D, where we 
consider the 3-bracket of projected position operators in the same limit of 
small momenta.

In 2D Euclidean flat space, a single flavor of Dirac fermions
with mass $m$ and
in the fundamental representation of the Lorentz group
is governed by the single-particle Hamiltonian in momentum space
\begin{equation}
\mathcal{H}^{\ }_{2\mathrm{D}}(\bs{k}):=
k^{\ }_1\sigma^{\ }_1
+
k^{\ }_2\sigma^{\ }_2
+
m\sigma^{\ }_3.
\end{equation}
As usual, we use $\sigma^{\ }_{0}$ for the $2\times 2$ unit matrix, while 
$\sigma^{\ }_{1}$,
$\sigma^{\ }_{2}$,
and
$\sigma^{\ }_{3}$
are the three Pauli matrices.

This Hamiltonian supports two bands with 
the Bloch states $|\chi^{\pm}(\bs{k})\rangle$,
the nondegenerate energy eigenvalues
\begin{equation}
\varepsilon^{(\pm)}(\bs{k})=
\pm\sqrt{\bs{k}^{2}+m^{2}},
\end{equation}
and the Berry curvatures
\begin{equation}
F^{(\pm)}_{\mu\nu}(\bs{k})=
\mathrm{i}\,\epsilon^{\ }_{\mu\nu}\,
\frac{m}{2\left[\varepsilon^{(\pm)}(\bs{k})\right]^{3}},
\label{eq: F mu nu for 2D massive Dirac}
\end{equation}
for $\mu,\nu=1,2$.
Upon projection to the lower band $\varepsilon^{(-)}(\bs{k})$,
we can combine
Eq.~(\ref{eq: 2 bracket Xr a})
with
Eq.~(\ref{eq: F mu nu for 2D massive Dirac})
to deduce that
\begin{subequations}
\label{eq: algebra Dirac 2D}
\begin{equation}
\begin{split}
\langle\chi^{-}(\bs{k})|
[\widehat{X}^{\mu}_{r},\widehat{X}^{\nu}_{r}]
|\chi^{-}(\bs{k})\rangle
=&\,
-F^{(-)}_{\mu\nu}(\bs{k})\\
=&\,
\mathrm{i}\,\epsilon^{\ }_{\mu\nu}\,\ell^{2}_{\mathrm{D}}\,
\mathrm{sgn}\,m 
+\mathcal{O}(\bs{k}^2)
\label{eq: algebra Dirac 2D a}
\end{split}
\end{equation}
for $\mu,\nu=1,2$.
The Dirac counterpart to the magnetic length in the QHE is here
\begin{equation}
\ell^{\ }_{\mathrm{D}}:=\frac{1}{\sqrt{2}\,m}.
\label{eq: algebra Dirac 2D b}
\end{equation}
\end{subequations}
As announced,
the algebra~(\ref{eq: algebra Dirac 2D})
reproduces the algebra~\eqref{eq: closing algebra position operator}
in the limit $\bs{k}\to 0$.
The first Chern number of the lower band is given by
\begin{equation}
\begin{split}
\mathrm{Ch}^{(1)}:=&\,
\frac{\mathrm{i}}{2\pi }
\int\limits_{\mathbb{R}^2} \mathrm{d}^2 \bs{k}
\,\mathrm{tr}\,F^{(-)}_{12}\\
=&\,
\frac{\mathrm{sgn}\,m}{2}.
\end{split}
\label{eq: Chern numbers flat}
\end{equation}

In 3D Euclidean flat space, a single flavor of Dirac fermions 
with mass $m$ and in the fundamental representation of the Lorentz group
is governed by the single-particle Hamiltonian in momentum space
\begin{subequations}
\begin{equation}
\mathcal{H}^{\ }_{3\mathrm{D}}(\bs{k}):=
\sum_{\mu=1}^3
k^{\ }_\mu\alpha^{\ }_\mu
-\mathrm{i}m\beta\gamma^{\ }_5,
\label{eq:H3D}
\end{equation}
where we have defined the 
Hermitian $4\times 4$ matrices 
\begin{equation}
\alpha^{\ }_\mu=
\begin{pmatrix}
0&\sigma^{\ }_\mu\\
\sigma^{\ }_\mu&0
\end{pmatrix},
\quad
\beta=
\begin{pmatrix}
\sigma^{\ }_0&0\\
0&-\sigma^{\ }_0
\end{pmatrix},
\quad
\gamma^{\ }_5=
\begin{pmatrix}
0&\sigma^{\ }_0\\
\sigma^{\ }_0&0
\end{pmatrix}.
\end{equation}
\end{subequations}
Observe that this Hamiltonian has the chiral symmetry
\begin{equation}
\gamma^{\ }_5\,\mathcal{H}_{3\mathrm{D}}(\bs{k})\,\gamma^{\ }_5=
-\mathcal{H}_{3\mathrm{D}}(\bs{k})
\end{equation}
for all $\bs{k}\in\mathbb{R}^3$.
The spectrum of Hamiltonian~\eqref{eq:H3D} consists 
of two doubly degenerate bands 
with the Bloch states $|\chi^{\pm,a}(\bs{k})\rangle$,
the energy eigenvalues
\begin{equation}
\varepsilon^{(\pm)}(\bs{k})=\pm\sqrt{\bs{k}^2+m^2},
\end{equation}
and the non-Abelian Berry field strengths 
\begin{equation}
F^{(\pm)}_{\mu\nu}(\bs{k})=
\pm\mathrm{i}\,\ell^{2}_{\mathrm{D}}\,
\epsilon^{\ }_{\mu\nu\lambda}\,
\gamma^{\lambda}
+
\mathcal{O}(|\bs{k}|)
\label{eq: nonAbelian Berry field strengths in 3D}
\end{equation}
for $\mu,\nu=1,2,3$,
where 
$\bs{\gamma}^{\mathsf{T}}=(-\sigma^{\ }_1,\sigma^{\ }_2,\sigma^{\ }_3)$.
Upon projection to the lower band $\varepsilon^{(-)}(\bs{k})$,
we can combine
Eq.~(\ref{eq: 2 bracket Xr a})
with
Eq.~(\ref{eq: nonAbelian Berry field strengths in 3D})
to deduce that
\begin{equation}
\begin{split}
\mathrm{tr}\,
\Big(
\langle\chi^{-,a}(\bs{k})|
[\widehat{X}^{\mu}_{r},\widehat{X}^{\nu}_{r}]
|\chi^{-,b}(\bs{k})\rangle
\Big)
=
0
+
\mathcal{O}(|\bs{k}|)
\end{split}
\end{equation}
for $\mu,\nu=1,2,3$,
as expected for a system with chiral symmetry.
In contrast, 
Eq.%
~(\ref{eq: desired intensive single-particle expectation value for 3-bracket})
delivers
\begin{equation}
\begin{split}
\mathrm{tr}\,
\Big(
\langle\chi^{-,a}(\bs{k})|
[\widehat{X}^{1}_{r},\widehat{X}^{2}_{r},\widehat{X}^{3}_{r}]^{\ }_{\mathrm{reg}}
|\chi^{-,b}(\bs{k})\rangle
\Big)
=&\,
\mathrm{i}\,3\,\sqrt{2}\,\ell^{3}_{\mathrm{D}}
\\
&\,
+
\mathcal{O}(|\bs{k}|).
\label{eq: algebra Dirac 3D}
\end{split}
\end{equation}
The definition~(\ref{eq: algebra Dirac 2D b})
of $\ell^{\ }_{\mathrm{D}}$
has carried over.

\subsection{
Operator product expansions for the projected single-particle
density operators
           }
\label{subsec: Algebra of the density operator}

Until now, we have considered the 
noncommutative relations obeyed by the 
projected position operator assuming translation invariance
in Euclidean flat spaces. This noncommutative geometry 
encodes topological properties of the
noninteracting many-body ground state
in view of the expectation values
(\ref{eq: desired intensive single-particle expectation value for com}),
(\ref{eq: desired intensive single-particle expectation value for 1-bracket}),
and
(\ref{eq: desired intensive single-particle expectation value for 3-bracket}).
Moreover, according to Eq.~(\ref{eq: def r operator}),
it is also predicated on some underlying noncommutative relations 
obeyed by the second-quantized fermion density operator projected onto
the occupied bands of the insulating ground state.

On the other hand, GMP were able to derive for the 2D QHE
the closed algebra obeyed by the single-particle electronic density projected
onto the lowest Landau level. Can we do the same for the 
single-particle fermionic density projected onto one band, say,
of a 3D topological band insulator?

To answer this question, we resort to a tight-binding model
defined on a lattice $\Lambda$ with a Brillouin zone BZ, and
on which we impose periodic boundary conditions. We assume,
without loss of generality, that the lattice is three dimensional.
In this spirit,
we turn our attention to the single-particle electronic density defined 
on a given site $\bs{r}$ of a lattice $\Lambda$ as
\begin{subequations}
\begin{equation}
\widehat{\varrho}^{\ }_{\bs{r}}:=
\sum_{\alpha=1}^{N}
|\bs{r},\alpha\rangle\langle\bs{r},\alpha|,
\label{eq: def rho r}
\end{equation}
where $\alpha=1,\cdots,N$ 
labels degrees of freedom on every lattice site, e.g., spin or orbitals.
These operators obey the closed algebra
\begin{equation}
\widehat{\varrho}^{\ }_{\bs{r}^{\ }_{1}}\,
\widehat{\varrho}^{\ }_{\bs{r}^{\ }_{2}}=
\delta^{\ }_{\bs{r}^{\ }_{1},\bs{r}^{\ }_{2}}\,
\widehat{\varrho}^{\ }_{\bs{r}^{\ }_{1}}
\end{equation}
owing to the orthonormality of the single-particle states
\begin{equation}
\langle\bs{r}^{\ }_{1},\alpha^{\ }_{1}|
       \bs{r}^{\ }_{2},\alpha^{\ }_{2}\rangle=
\delta^{\ }_{\bs{r}^{\ }_{1},\bs{r}^{\ }_{2}}\,
\delta^{\ }_{\alpha^{\ }_{1},\alpha^{\ }_{2}}
\end{equation}
for any pair of sites $\bs{r}^{\ }_{1}$ and $\bs{r}^{\ }_{2}$ 
from the lattice $\Lambda$
and for any pair of orbitals 
$\alpha^{\ }_{1},\alpha^{\ }_{2}=1,\cdots N$.
As a consequence, these operators commute pairwise.
\end{subequations}

The Fourier transform of $\widehat{\varrho}^{\ }_{\bs{r}}$ 
in terms of the orthonormal Bloch states $|\bs{k},\alpha\rangle$ 
labeled by the momentum $\bs{k}$ 
from the BZ and orbital index $\alpha=1,\cdots,N$ 
reads
\begin{subequations}
\begin{equation}
\widehat{\varrho}^{\ }_{\bs{q}}=
\sum_{\bs{k}\in\mathrm{BZ}}
\sum_{\alpha=1}^{N}
|\bs{k},\alpha\rangle\langle\bs{k}+\bs{q},\alpha|
\label{eq: def rho q}
\end{equation}
for any $\bs{q}\in\mathrm{BZ}$. These operators
obey the closed algebra
\begin{equation}
\widehat{\varrho}^{\ }_{\bs{q}^{\ }_1}\,
\widehat{\varrho}^{\ }_{\bs{q}^{\ }_2}=
\widehat{\varrho}^{\ }_{\bs{q}^{\ }_1+\bs{q}^{\ }_2},
\qquad 
\label{eq: algebra rho q}
\end{equation}
owing to the orthonormality of the single-particle states
\begin{equation}
\langle\bs{q}^{\ }_{1},\alpha^{\ }_{1}|
       \bs{q}^{\ }_{2},\alpha^{\ }_{2}\rangle=
\delta^{\ }_{\bs{q}^{\ }_{1},\bs{q}^{\ }_{2}}\,
\delta^{\ }_{\alpha^{\ }_{1},\alpha^{\ }_{2}}
\end{equation}
for any pair of momentum $\bs{q}^{\ }_{1}$ and $\bs{q}^{\ }_{2}$ 
from the BZ and for any pair of orbitals 
$\alpha^{\ }_{1},\alpha^{\ }_{2}=1,\cdots N$.
As a consequence, these operators commute pairwise.
\end{subequations}

Consider now a basis transformation in the $\alpha$ degrees of freedom 
for every $\bs{k}\in\mathrm{BZ}$ that is parametrized
by the  $N\times N$
complex-valued numbers
$u^{(b)}_{\bs{k},\alpha}$
with $\alpha,b=1,\cdots,N$,
i.e.,
\begin{equation}
\left|u^{(b)}_\bs{k}\right\rangle:=
\sum_{\alpha=1}^{N}
u^{(b)}_{\bs{k},\alpha}
|\bs{k},\alpha\rangle,
\qquad b=1,\cdots, N.
\end{equation}
The ket $|u^{\ }_\bs{k},b\rangle$ labeled by
$\bs{k}\in\mathrm{BZ}$ for any given $b=1,\cdots,N$
should be thought of as Bloch state of the $b$-th band
of a single-particle Hamiltonian. This Hamiltonian shares
the translational symmetry of 
$\Lambda$ and periodic boundary conditions are imposed.
For any $\bs{q}\in\mathrm{BZ}$,
we define the density operator
projected on a single (nondegenerate) band 
$\tilde{b}$ by
\begin{equation}
{\widehat{\rho}}^{\ }_{\bs{q}}:=
\sum_{\bs{k}\in\mathrm{BZ}}
\sum_{\alpha=1}^{N}
u^{(\tilde{b})*}_{\bs{k},\alpha}
u^{(\tilde{b})}_{\bs{k}+\bs{q},\alpha}
\left|u^{(\tilde{b})}_{\bs{k}}\right\rangle
\left\langle u^{(\tilde{b})}_{\bs{k}+\bs{q}}\right|.
\label{eq: def rho q projected}
\end{equation}
The projected operators ${\widehat{\rho}}^{\ }_{\bs{q}}$ 
with $\bs{q}\in\mathrm{BZ}$
are invariant under the 
simultaneous local U(1) gauge transformations
defined by
\begin{subequations}
\label{eq: Abelian gauge trafo lattice}
\begin{equation}
u^{(\tilde{b})}_{\bs{k},\alpha}\to
e^{\mathrm{i}\varphi^{\ }_{\bs{k}}}
u^{(\tilde{b})}_{\bs{k},\alpha}
\label{eq: Abelian gauge trafo lattice a}
\end{equation}
on the one hand, and
\begin{equation}
\left|u^{(\tilde{b})}_{\bs{k}}\right\rangle
\rightarrow
e^{\mathrm{i}\varphi^{\ }_{\bs{k}}}
\left|u^{(\tilde{b})}_{\bs{k}}\right\rangle
\label{eq: Abelian gauge trafo lattice b}
\end{equation}
on the other hand,
\end{subequations}
for all $\alpha=1,\cdots, N$, $\bs{k}\in\mathrm{BZ}$, 
and any real-valued function $\varphi^{\ }_{\bs{k}}$.
They do not obey anymore the algebra~\eqref{eq: algebra rho q}.
In the long-wavelength limit $\bs{q}^{\ }_1,\bs{q}^{\ }_2\ll 1$ 
(the lattice spacing of $\Lambda$ is set to unity),
their commutation relation is
~\cite{Parameswaran11,Goerbig12,Bernevig11}
\begin{subequations}
\begin{equation}
\left[
{\widehat{\rho}}^{\ }_{\bs{q}^{\ }_1},
{\widehat{\rho}}^{\ }_{\bs{q}^{\ }_2}
\right]=
q^{\mu}_1
q^{\nu}_2
\sum_{\bs{k}\in\mathrm{BZ}}
{F}^{\ }_{\mu\nu,\bs{k}}\,
\left|u^{(\tilde{b})}_{\bs{k}}\right\rangle
\left\langle u^{(\tilde{b})}_{\bs{k}+\bs{q}^{\ }_1+\bs{q}^{\ }_2}\right|
\label{eq: commutation rho q projected}
\end{equation}
to leading order in an expansion in powers of the components
of $\bs{q}^{\ }_{1}$ and $\bs{q}^{\ }_{2}$, where 
\begin{equation}
{F}^{\ }_{\mu\nu,\bs{k}}:=
\partial^{\ }_{\mu}
{A}^{\ }_{\nu,\bs{k}}
-
\partial^{\ }_{\nu}
{A}^{\ }_{\mu,\bs{k}}
\label{eq: Berry field}
\end{equation}
and
\begin{equation}
{A}^{\ }_{\mu,\bs{k}}:=
\sum_{\alpha=1}^{N}
u^{(\tilde{b})*}_{\bs{k},\alpha}\,
\partial^{\ }_{\mu} 
u^{(\tilde{b})}_{\bs{k},\alpha}
\label{eq: Berry connection}
\end{equation}
\end{subequations}
for $\mu,\nu=1,2,3$
are the Abelian Berry curvature and the Abelian Berry connection, 
respectively, and $\partial^{\ }_{\mu}$ is 
understood as the derivative with respect to 
the momentum component $k^{\mu}$.
The operator product expansion%
~\eqref{eq: commutation rho q projected}
closes only if ${F}^{\ }_{\mu\nu,\bs{k}}$ 
is \emph{independent} of $\bs{k}$,
in which case
\begin{subequations}
\begin{equation}
\left[
{\widehat{\rho}}^{\ }_{\bs{q}^{\ }_1},
{\widehat{\rho}}^{\ }_{\bs{q}^{\ }_2}
\right]=
-
\frac{\mathrm{i}}{2\pi}
\left(\bs{q}^{\ }_1\wedge \bs{q}^{\ }_2\right) 
\cdot
\mathbf{Ch}\
{\widehat{\rho}}^{\ }_{\bs{q}^{\ }_1+\bs{q}^{\ }_2}
\label{eq: closed commutation rho q projected}
\end{equation}
to leading order in an expansion in powers of the components
of $\bs{q}^{\ }_{1}$ and $\bs{q}^{\ }_{2}$, where
\begin{equation}
\mathrm{Ch}^{\lambda}:=
\frac{2\pi \mathrm{i}}{L^3}
\frac{\epsilon^{\mu\nu\lambda}}{2}
\sum_{\bs{k}\in\mathrm{BZ}}
{F}^{\ }_{\mu\nu,\bs{k}}
\label{eq: def Chern numbers}
\end{equation}
\end{subequations}
with $\lambda=1,2,3$ are the components of the vector 
$\mathbf{Ch}$
made of the three first Chern numbers characterizing 
any nondegenerate band in 3D space.%
~\cite{Kohmoto92}
(A summation convention is implied for the repeated indices
$\mu,\nu=1,2,3$.)
In the thermodynamic limit by which the linear size $L$
over which the periodic boundary conditions are imposed
is taken to infinity or, equivalently, the lattice
spacing is taken to zero, each first Chern number
is quantized.

The IQHE is an example in 2D 
for which the condition of constant Berry curvature 
${F}^{\ }_{\mu\nu,\bs{k}}$
is met. In this context, the closed operator product expansion%
~\eqref{eq: closed commutation rho q projected} 
was found by GMP
(in fact, the operator product expansion
for the projected density operators
closes to all orders in $\bs{q}$ in this case,
and thus delivers a closed algebra for the projected
density operators).%
~\cite{Girvin85} 
With the help of this algebra, GMP argue,
within a single-mode approximation, that FQH states are incompressible. 

Recently, it was shown that lattice models with flat bands and nonzero 
Chern number also support incompressible FQH ground states,%
~\cite{Neupert11a,Sheng11,Wang11a,Regnault11}
even though their Berry curvature is not constant over the BZ.
This result suggests to approximate the commutator%
~\eqref{eq: commutation rho q projected}
by the closed algebra%
~\eqref{eq: closed commutation rho q projected}, that is, 
to replace ${F}^{\ }_{\mu\nu,\bs{k}}$ 
with its average value over the BZ.%
~\cite{Parameswaran11,Goerbig12}

\begin{widetext}
We will now consider the 3-bracket 
of three projected density operators,  
and expand it to third order in the momenta
\begin{equation}
\begin{split}
\left[
{\widehat{\rho}}^{\ }_{\bs{q}^{\ }_1},
{\widehat{\rho}}^{\ }_{\bs{q}^{\ }_2},
{\widehat{\rho}}^{\ }_{\bs{q}^{\ }_3}\right]=&\,
\epsilon^{ijk}\frac{1}{2}
\sum_{\bs{k}\in\mathrm{BZ}}
\left\{	
q^\mu_i q^\nu_j
{F}^{\ }_{\mu\nu,\bs{k}}
+
q^\mu_i q^\nu_j q^\lambda_k 
{F}^{\ }_{\mu\nu,\bs{k}}
{A}^{\ }_{\lambda,\bs{k}}
\vphantom{\sum_{\alpha=1}^{N}}
\right.
\\
&\,
\left.
+
q^\mu_i q^\nu_j q^\lambda_j 
\left[
\partial^{\ }_{\mu}
\left(
\sum_{\alpha=1}^{N}
u^{(\tilde{b})*}_{\bs{k},\alpha}\,
\partial^{\ }_{\nu}\partial^{\ }_{\lambda} 
u^{(\tilde{b})}_{\bs{k},\alpha}
\right)
-
\left(\partial^{\ }_{\nu}+2 A^{\ }_{\nu}\right)
\partial^{\ }_{\lambda}
{A}^{\ }_{\mu}
\right]
\right\}
\left|
u^{(\tilde{b})}_{\bs{k}}
\right\rangle
\left\langle 
u^{(\tilde{b})}_{\bs{k}+\bs{q}^{\ }_1+\bs{q}^{\ }_2+\bs{q}^{\ }_3}
\right|,
\end{split}
\label{eq: as 3 density product}
\end{equation}
\end{widetext}
where the summation convention over the repeated indices
$i,j,k=1,2,3$ and $\mu,\nu,\lambda=1,2,3$ is implied.
Equation~\eqref{eq: as 3 density product} is invariant under 
the local gauge transformation~\eqref{eq: Abelian gauge trafo lattice}, 
up to contributions of fourth order in $\bs{q}$.
The term of second order in $\bs{q}$ 
comes multiplied by the Berry curvature, 
i.e., the density associated with the topological invariants 
$\mathrm{Ch}^{\lambda}$ for $\lambda=1,2,3$
defined in Eq.~\eqref{eq: def Chern numbers}.
As for the second term on the right-hand side,
we recognize the integrand of the Abelian Chern-Simons form.

The term that dominates the 3-bracket 
of projected density operators at long wavelength 
is thus not equal to the 3-bracket of the position operator
$\widehat{\bs{X}}^{\ }_{r}$. 
According to Eq.%
~(\ref{eq: desired intensive single-particle expectation value for 3-bracket}), 
the latter was determined by the Chern-Simons 3 form 
and not by the Chern number density.
This stands in contrast to the long wavelength limit of the 2-bracket 
(commutator) 
of projected density operators which coincides with the 2-bracket 
of position operators.
However, the connection between the projected density 
and position operators is recovered on the level of the 3-brackets, 
if one considers the derivative of the density operator 
with respect to momentum instead. This choice is motivated by 
the fact that the Fourier components of the density operator
in momentum space  are the generators of translations in momentum space 
[recall Eq.~(\ref{eq: def Fourier trsf density operator})].  
Indeed, it follows from 
Eq.~\eqref{eq: as 3 density product}
that 
\begin{subequations}
\begin{equation}
\begin{split}
&
\left[\partial^{\ }_{q^{\alpha}_{1}}
{\widehat{\rho}}^{\ }_{\bs{q}^{\ }_1},
\partial^{\ }_{q^{\beta}_{2}}
{\widehat{\rho}}^{\ }_{\bs{q}^{\ }_2},
\partial^{\ }_{q^{\gamma}_{3}}
{\widehat{\rho}}^{\ }_{\bs{q}^{\ }_3}\right]
\\
&\quad
=\,
\frac{\epsilon^{\alpha\beta\gamma}}{2}
\sum_{\bs{k}\in\text{BZ}}
\epsilon^{\mu\nu\lambda}A^{\ }_{\mu,\bs{k}}F^{\ }_{\nu\lambda,\bs{k}}
\left|u^{(\tilde{b})}_{\bs{k}}\right\rangle
\left\langle u^{(\tilde{b})}_{\bs{k}+\bs{q}^{\ }_1+\bs{q}^{\ }_2+\bs{q}^{\ }_3}\right|
\end{split}
\label{3-bracket of derivatives of projected densities}
\end{equation}
holds to lowest order in the momenta $\bs{q}^{\ }_1$, $\bs{q}^{\ }_2$ and $\bs{q}^{\ }_2$ 
and is thus determined by
the Chern-Simons 3-form (the Chern-Simons density in 3D).
We define its average over the BZ to be
\begin{equation}
{\theta}:=
\frac{\pi^2}{L^3} \sum_{\bs{k}\in\mathrm{BZ}}
\epsilon^{\mu\nu\lambda}\, 
{F}^{\ }_{\mu\nu,\bs{k}}\, 
{A}^{\ }_{\lambda,\bs{k}},
\label{eq: CS form}
\end{equation}
which is only invariant under the local gauge transformations%
~\eqref{eq: Abelian gauge trafo lattice a}
that leave the boundary conditions in the BZ unchanged.
If the Chern-Simons density is nearly constant and thus independent 
of $\bs{k}$ in the entire BZ, we may approximate 
Eq.~\eqref{3-bracket of derivatives of projected densities}
by
\begin{equation}
\left[\partial^{\ }_{q^{\alpha}_{1}}
{\widehat{\rho}}^{\ }_{\bs{q}^{\ }_1},
\partial^{\ }_{q^{\beta}_{2}}
{\widehat{\rho}}^{\ }_{\bs{q}^{\ }_2},
\partial^{\ }_{q^{\gamma}_{3}}
{\widehat{\rho}}^{\ }_{\bs{q}^{\ }_3}\right]
\approx\,
\epsilon^{\ }_{\alpha\beta\gamma}
\frac{\mathfrak{a}^3\,\theta}{2\pi^2}
\,{\widehat{\rho}}^{\ }_{\bs{q}^{\ }_1+\bs{q}^{\ }_2+\bs{q}^{\ }_3},
\end{equation}
where $\mathfrak{a}$ is the lattice spacing.
\end{subequations}

Insulators for which the invariants $\mathrm{Ch}^{\lambda}$
with $\lambda=1,2,3$
are nonvanishing can be 
viewed as a 3D extension of an IQHE or a layered system of 2D Chern 
insulators. In this case, $\mathrm{Ch}^{\lambda}$ with $\lambda=1,2,3$ 
parametrizes the quantized off-diagonal part of the conductivity 
tensor.%
~\cite{Kohmoto92}
The physics of such insulators is not intrinsically 3D and they are thus not 
our primary interest here.

Even if the Berry curvature vanishes on average in the BZ 
so that $\mathbf{Ch}=0$, ${\theta}$ 
can be nonzero and may take any real value in general. 
The value of ${\theta}$ 
has measurable consequences as it contributes to the 
magneto-electric coupling in a 3D band insulator.%
~\cite{Coh11}
For 3D band insulators with either spin-orbit coupling
that are time-reversal symmetric (symmetry class AII)
or with chiral symmetry (symmetry class AIII), 
${\theta}$ 
is restricted to integer multiples of $\pi$ and represents 
a topological invariant.%
~\cite{Ryu10}
The 3-bracket~\eqref{3-bracket of derivatives of projected densities} 
shows that for 3D tight-binding Hamiltonians 
within the symmetry classes AII or
AIII, the 3-bracket of the momentum derivatives of
projected density operators is dominated by the value of their 
topological invariant ${\theta}$,
just as the 2-bracket (commutator) of 
the momentum derivatives of projected density operators is 
dominated by the value of the Chern number in 2D tight-binding models
within the symmetry class A.
We will illustrate this statement with the help 
of a microscopic lattice model belonging to the symmetry class AIII
in the Sec.\ \ref{sec: Noninteracting three-band tight-binding model}.

\medskip
\section{
Noninteracting three-band tight-binding model
        }
\label{sec: Noninteracting three-band tight-binding model}

The goal of this section is to define a ``simple'' single-particle Bloch 
Hamiltonian that supports a dispersionless isolated band with nontrivial 
topological character, such that the 
3-bracket of the momentum derivatives of the projected
electronic density operators, Eq.%
~\eqref{3-bracket of derivatives of projected densities},
is nonvanishing and the system displays intrinsically 3D physics,
i.e., $\mathrm{Ch}^{\lambda}=0$ for $\lambda=1,2,3$
and $\theta=\pi$.
Our model belongs to symmetry class AIII and has three bands, 
which is the minimum number required to realize the desired $\theta$-term.%
~\cite{Coh11}
One of the three bands is necessarily dispersionless as a 
consequence of chiral symmetry. Therefore, it can be taken as
the basis for the 
construction of fractional topological states in 3D.

\subsection{
Definition
           }

We consider spinless electrons hopping between the sites
$\bs{r}^{\mathsf{T}}\equiv(r^{\ }_{1},r^{\ }_{2},r^{\ }_{3})$
of a 3D cubic lattice $\Lambda$ and on-site orbitals,
whereby each site $\bs{r}$ can accommodate
three orbital degrees of freedom that we label with the Greek index 
$\alpha=1,2,3$. To accommodate the hybridization between any
of the three orbitals, we need to choose a basis for all
$3\times3$ Hermitian matrices. We denote the unit
$3\times3$ matrix by $\lambda^{\ }_{0}$ which, together
with the eight traceless Gell-Mann Hermitian matrices
$\lambda^{\ }_{n}$ with $n=1,\cdots,8$,
form the desired basis of all $3\times3$ Hermitian matrices.
The second quantized tight-binding Hamiltonian is then defined by
\begin{equation}
\begin{split}
\widehat{H}:=&\, 
\frac{1}{2}
\sum_{\bs{r}\in\Lambda}
\sum_{j=1}^3
\left[
\widehat{c}^{\dag}_{\bs{r}}
\left(
\mathrm{i}\lambda^{\ }_{3+j}
-
\lambda^{\ }_{7}
\right)
\widehat{c}^{\ }_{\bs{r}+\bs{e}^{\ }_{j}}
+
\mathrm{h.c.}
\right]
\\
&\,
+M
\sum_{\bs{r}\in\Lambda}\,
\widehat{c}^{\dag}_{\bs{r}}\,
\lambda^{\ }_{7}\,
\widehat{c}^{\ }_{\bs{r}},
\end{split}
\label{eq: def H position space}
\end{equation}
where we have introduced the 3-component operator
$
\widehat{c}^{\dag}_{\bs{r}}\equiv
\left(
\widehat{c}^{\dag}_{\bs{r};1},
\widehat{c}^{\dag}_{\bs{r};2},
\widehat{c}^{\dag}_{\bs{r};3}
\right)
$
with 
$\widehat{c}^{\dag}_{\bs{r};\alpha}$ 
creating a spinless fermion at site $\bs{r}$ in the orbital $\alpha=1,2,3$
and obeying periodic boundary conditions under the translation
$\bs{r}\to\bs{r}+L\bs{e}^{\ }_{j}$
for any of the three orthonormal unit vectors 
$\bs{e}^{\ }_{1}$,
$\bs{e}^{\ }_{2}$,
and
$\bs{e}^{\ }_{3}$
that span the cubic lattice $\Lambda$.
This single-particle Hamiltonian depends on the real-valued parameter $M$.

Translation invariance allows to diagonalize 
Hamiltonian~(\ref{eq: def H position space}) 
upon performing a Fourier transformation on the 
creation and annihilation fermionic operators.
If we denote with BZ the Brillouin zone of the 
3D cubic lattice and with $\bs{k}$ any Bloch
momentum from the BZ that is compatible with the
periodic boundary conditions, then
\begin{subequations}
\label{eq: def H momentum space}
\begin{equation}
\widehat{H}= 
\sum_{\bs{k}\in\mathrm{BZ}}
\widehat{c}^{\dag}_{\bs{k}}\,
\mathcal{H}^{\   }_{\bs{k}}\,
\widehat{c}^{\   }_{\bs{k}}
\label{eq: H momentum space a}
\end{equation}
with the momentum-resolved single-particle $3\times3$ matrix
\begin{equation}
\mathcal{H}^{\ }_{\bs{k}}=\,
\sum_{j=1}^4
\lambda^{\ }_{3+j}\, 
d^{\ }_{\bs{k},j}
\label{eq: H momentum space b}
\end{equation}
that depends on the 4-component real-valued row vector
\begin{equation}
\begin{split}
d^{\mathsf{T}}_{\bs{k}}\equiv&\,
\left(
d^{\ }_{\bs{k},1},
d^{\ }_{\bs{k},2},
d^{\ }_{\bs{k},3},
d^{\ }_{\bs{k},4}
\right)
\\
:=&\,  
\left(
\sin k^{\ }_1,
\sin k^{\ }_2,
\sin k^{\ }_3,
M-\sum_{i=1}^3 \cos k^{\ }_i
\right).
\end{split}
\label{eq: def H momentum space c}
\end{equation}
\end{subequations}

\begin{figure}
\includegraphics[angle=0,scale=0.9]{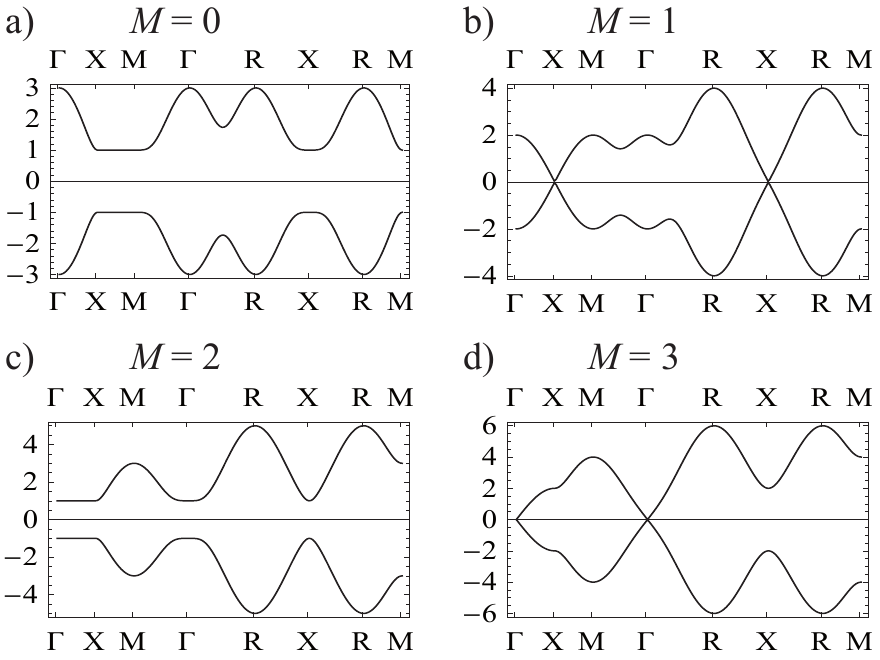}
\caption{\label{fig: energy spectrum of lattice model}
Energy spectrum of the lattice model defined in 
Eq.~\eqref{eq: def H momentum space}
for different values of the parameter $M$.
Panels b) and d) show the gap-closing topological transitions. 
Note the dispersionless band at zero energy in each spectrum.
The spectrum is plotted along the straight path connecting the 
following points in the BZ: 
$\Gamma=(0,0,0)$,
$X=(0,\pi,0)$,
$M=(\pi,\pi,0)$, and
$R=(\pi,\pi,\pi)$.
        }
\end{figure}

With the help of the explicit representation of the
eight Gell-Mann matrices from Appendix%
~\ref{appsec: Gell-Mann matrices}, 
one verifies that
\begin{subequations}
\begin{equation}
\mathcal{C}\,\mathcal{H}^{\ }_{\bs{k}}\, \mathcal{C}^{-1}
=
-\mathcal{H}^{\ }_{\bs{k}}
,\qquad 
\forall\bs{k}\in\mathrm{BZ},
\label{eq: chiral symmetry}
\end{equation}
if and only if the $3\times3$ matrix $\mathcal{C}$ is given by
\begin{equation}
\mathcal{C}:=
\mathrm{diag}(1,1,-1).
\end{equation}
\end{subequations}
The fact that $\mathcal{H}^{\ }_{\bs{k}}$ anticommutes
with $\mathcal{C}$ implies that any pair of eigenstates
$u^{(+)}_{\bs{k}}$ and $u^{(-)}_{\bs{k}}$
of $\mathcal{H}^{\ }_{\bs{k}}$ 
with nonvanishing eigenvalues
is associated with the opposite single-particle eigenenergies
$\varepsilon^{(+)}_{\bs{k}}=-\varepsilon^{(-)}_{\bs{k}}$,
respectively.
Since $\mathcal{H}^{\ }_{\bs{k}}$ is a $3\times3$ Hermitian matrix
for any momentum $\bs{k}$ from the BZ, it then
follows that at least one eigenstate
$u^{(0)}_{\bs{k}}$ 
must have the vanishing eigenvalue
\begin{subequations} 
\label{eq: 3 dispersions}
\begin{equation}
\varepsilon^{(0)}_{\bs{k}}=0,
\end{equation}
irrespective of the Bloch momentum $\bs{k}$ in the BZ.
For any Bloch momentum $\bs{k}$ in the BZ,
the values taken by the nonvanishing eigenvalues
\begin{equation}
\varepsilon^{(+)}_{\bs{k}}=
-\varepsilon^{(-)}_{\bs{k}}=
\left|d^{\ }_{\bs{k}}\right|
\end{equation}
\end{subequations}
follow immediately from the fact that the
four Gell-Mann matrices 
$\lambda^{\ }_{4}$, $\lambda^{\ }_{5}$,
$\lambda^{\ }_{6}$, and $\lambda^{\ }_{7}$,
anticommute pairwise while any one of these 4 Gell-Mann matrices
squares to either 
$\mathrm{diag}(1,0,1)$ or $\mathrm{diag}(0,1,1)$.
The minimum value reached by the
magnitude $\left|d^{\ }_{\bs{k}}\right|$
over the BZ thus determines the energy gap between
the dispersionless band of zero modes and the pair of bands
related by the chiral transformation $\mathcal{C}$.
This energy gap depends parametrically on $M$ and is nonvanishing
if and only if $|M|\neq 1,3$.
In turn, the corresponding Bloch states are derived as follows.
One observes that the 2-component complex-valued row vector
\begin{subequations}
\label{eq: Bloch eigenstates mathcal H}
\begin{equation}
q^{\dag}_{\bs{k}}:= 
|d^{\ }_{\bs{k}}|^{-1}
\left(
d^{\ }_{\bs{k};1}+\mathrm{i}d^{\ }_{\bs{k};2},
d^{\ }_{\bs{k};3}+\mathrm{i}d^{\ }_{\bs{k};4}
\right)
\label{eq: Bloch eigenstates mathcal H a}
\end{equation}
of unit length ($q^{\dag}_{\bs{k}}\,q^{\   }_{\bs{k}}=1$)
enters $\mathcal{H}^{\ }_{\bs{k}}$ according to 
\begin{equation}
\begin{split}
\mathcal{H}^{\ }_{\bs{k}}=&\,
|d^{\ }_{\bs{k}}|
\begin{pmatrix}
0
&
0
& 
q^{\ }_{\bs{k},1}
\\
0
&
0
&
q^{\ }_{\bs{k},2}
\\
q^{*}_{\bs{k},1}
& 
q^{*}_{\bs{k},2}
&
0
\end{pmatrix}
\\
\equiv&\,
|d^{\ }_{\bs{k}}|
\begin{pmatrix}
0^{\   }_{2\times2}
&
q^{\   }_{\bs{k}}
\\
q^{\dag}_{\bs{k}}
&
0
\end{pmatrix}.
\end{split}
\label{eq: Bloch eigenstates mathcal H b}
\end{equation}
One then verifies that
\begin{equation}
u^{(\pm)}_{\bs{k}}=
\frac{1}{\sqrt{2}}
\begin{pmatrix}
+q^{\ }_{\bs{k},1}
\\
+q^{\ }_{\bs{k},2}
\\ 
\pm1
\end{pmatrix},
\qquad
u^{(0)}_{\bs{k}}=
\begin{pmatrix}
+q^{*}_{\bs{k};2}
\\ 
-q^{*}_{\bs{k};1}
\\ 
0
\end{pmatrix}
\label{eq: Bloch eigenstates mathcal H c}
\end{equation}
\end{subequations}
are orthonormal Bloch states of $\mathcal{H}^{\ }_{\bs{k}}$
for any Bloch momentum $\bs{k}$ from the BZ. 
For any value of the parameter $M$ entering
the single-particle Hamiltonian $\widehat{H}$, 
Eqs. 
(\ref{eq: Bloch eigenstates mathcal H c}) 
and
(\ref{eq: 3 dispersions})
define globally over the entire BZ
the desired Bloch states with their dispersions. 
For generic values of $M$, i.e., whenever
$|d^{\ }_{\bs{k}}|$ is nonvanishing over the entire BZ,
there are two dispersive bands whose Bloch states 
$u^{(+)}_{\bs{k}}=\mathcal{C}\,u^{(-)}_{\bs{k}}$
are related
by the chiral transformation
and one dispersionless band
$u^{(0)}_{\bs{k}}=\mathcal{C}\,u^{(0)}_{\bs{k}}$
of zero modes.

Hamiltonian~(\ref{eq: def H momentum space}) 
breaks time-reversal symmetry, 
for the first three components of
$d^{\ }_{\bs{k}}$ are odd while the fourth
component is even under $\bs{k}\to-\bs{k}$ for any value of $M$.
This leaves no room for a particle-hole symmetry by which
Hamiltonian~(\ref{eq: def H momentum space}) 
would anticommute with an antiunitary operator.
Adding to Hamiltonian~(\ref{eq: def H momentum space}) 
any linear combination of the remaining Gell-Mann matrices
$\lambda^{\ }_{1}$, $\lambda^{\ }_{2}$, $\lambda^{\ }_{3}$,
$\lambda^{\ }_{8}$,
and the unit $3\times3$ matrix $\lambda^{\ }_{0}$
breaks the chiral symmetry.
Such perturbations change the symmetry class
of Hamiltonian~(\ref{eq: def H momentum space}) from AIII to A.
Although a chemical potential
(a nonvanishing constant term proportional 
to the unit matrix $\lambda^{\ }_{0}$) 
does break the chiral symmetry,
it does so by moving rigidly the entire energy spectrum up or down
in energy while leaving the Bloch states unchanged. The topological
attributes of the three Bloch bands are thus untouched by the
addition of a chemical potential.

\subsection{
Topological invariants
           }

We shall take the thermodynamic limit $L\to\infty$
with $L$ the linear extend over which periodic boundary conditions
are imposed. In this limit sums over momenta in the BZ
are replaced by integrals over the BZ while the index
$\bs{k}$ becomes the argument of functions.
From now on, we shall identify the BZ with $T^{3}$.
We can then distinguish two related topological invariants
associated to the family of single particle
$3\times3$ matrices $\mathcal{H}(\bs{k})$
labeled by the momentum $\bs{k}$ from a BZ with the topology
of the 3-torus $T^{3}$ owing to the periodic boundary conditions.

The first topological attribute characterizes the bundle of
Hamiltonians $\mathcal{H}(\bs{k})$ over the BZ $T^{3}$.
For any momentum $\bs{k}\in T^{3}$,
there is a one-to-one correspondence between
the $3\times3$ Hermitian matrices $\mathcal{H}(\bs{k})$
and the vector $d(\bs{k})\in\mathbb{R}^{4}$.
For any momentum $\bs{k}\in T^{3}$,
the magnitude $|d(\bs{k})|$ measures the momentum-resolved
energy separation between the zero mode $u^{(0)}(\bs{k})$
and the lower and upper modes  
$u^{(-)}(\bs{k})$ and $u^{(+)}(\bs{k})$,
respectively. The eigenstates $u^{(0)}(\bs{k})$,
$u^{(-)}(\bs{k})$, and $u^{(+)}(\bs{k})$ are independent of
the magnitude of $|d(\bs{k})|$, i.e., they only depend
on the coordinate defined by the unit
$3$-vector $d(\bs{k})/|d(\bs{k})|$
on the 3-sphere $S^{3}$ . It follows
that the topological attributes of the three Bloch bands
of Hamiltonian (\ref{eq: def H momentum space}) 
are determined by the homotopy group $\mathbb{Z}$
of the map defined by
\begin{equation}
\bs{k}\in T^{3}\to d(\bs{k})/|d(\bs{k})|\in S^{3}
\label{eq: map T3 to S3}
\end{equation}
between the BZ $T^{3}$ and the 3-sphere $S^{3}$.
For each parameter $M\neq\pm1,\pm3$ entering in 
Hamiltonian (\ref{eq: def H momentum space}),
the integer value taken by the topological invariant
\begin{subequations}
\label{eq: values taken by nu as M changes}
\begin{equation}
\nu(M):=
\frac{1}{12\pi^{2}}
\int\limits^{\ }_{T^{3}}
\mathrm{d}^{3}\bs{k}\,
\epsilon^{ijkl}\,
\epsilon^{\mu\nu\lambda}\,
\frac{1}{|d|^{4}}
d^{\ }_{i}\,
\partial^{\ }_{\mu}d^{\ }_{j}\,
\partial^{\ }_{\nu}d^{\ }_{k}\,
\partial^{\ }_{\lambda}d^{\ }_{l},
\label{eq: values taken by nu as M changes a}
\end{equation}
determines which homotopy class the map~(\ref{eq: map T3 to S3})
belongs to. Here, we are using the short-hand notation
$\partial^{\ }_{\mu}d^{\ }_{j}\equiv\partial d^{\ }_{j}/d k^{\mu}$,
with $\mu,\nu,\lambda$ labeling the three coordinates of the momentum $\bs{k}$
and $i,j,k,l$ labeling the four components of the vector field $d$,
and the convention for summation over repeated indices. 
Explicit computation of $\nu$ as a function of $M$ delivers
\begin{equation}
\nu(M)=
\begin{cases}
+2,\qquad &|M|<1,\\
-1,\qquad &1<|M|<3,\\
 0,\qquad &3<|M|.
\end{cases}
\label{eq: values taken by nu as M changes b}
\end{equation}
\end{subequations}

Whenever $|M|=1,3$, the gap over the BZ closes
at the discrete points (the lattice spacing is unity)
\begin{equation}
\bs{k}^{\mathsf{T}}_{lmn}:=
\pi\,
(l,m,n),
\qquad
l,m,n=0,1.
\label{eq: def k m n l}
\end{equation}
These eight momenta change by a reciprocal wave vector
under the operation of time reversal,
under which $\bs{k}\to-\bs{k}$. In this sense, they
are time-reversal invariant. The touching of the upper and lower dispersions
at the momenta (\ref{eq: def k m n l}) occurs at zero energy
and delivers a Dirac dispersion in their close vicinity when $|M|=1,3$.
Hence, we call the momenta (\ref{eq: def k m n l})
Dirac points when $|M|=1,3$. 
For small deviations 
away from
$|M|=1,3$, a spectral gap opens up at the momenta (\ref{eq: def k m n l})
that can be associated with a Dirac mass.
Remarkably, the number of Dirac points that change the sign 
of their mass across a transition tuned by changing $M$
through any one of the values $|M|=1,3$
is equal to the change in the topological invariants%
~(\ref{eq: values taken by nu as M changes}).
To see this, observe that the momentum resolved Dirac masses are given by
\begin{equation}
\begin{split}
&
d^{\ }_{\bs{k}^{\ }_{000};4}=
M-3,
\\
&
d^{\ }_{\bs{k}^{\ }_{001};4}=
d^{\ }_{\bs{k}^{\ }_{010};4}=
d^{\ }_{\bs{k}^{\ }_{100};4}=
M-1,
\\
&
d^{\ }_{\bs{k}^{\ }_{110};4}=
d^{\ }_{\bs{k}^{\ }_{101};4}=
d^{\ }_{\bs{k}^{\ }_{011};4}=
M+1,
\\
&
d^{\ }_{\bs{k}^{\ }_{111};4}=
M+3.
\end{split}
\label{eq: Dirac masses }
\end{equation}
With the help of these 8 integers, we define the integer
\begin{equation}
\nu^{\ }_{\mathrm{D}}(M):=
\frac{1}{2} 
\sum_{m,n,l=0,1}
(-1)^{m+n+l}\,
\mathrm{sign}\,d^{\ }_{\bs{k}^{\ }_{mnl};4}.
\label{eq: def Dirac invariant}
\end{equation}
The factor $(-1)^{m+n+l}$ 
assures that the mass sign is taken relative to the chirality 
of the kinetic piece of the Dirac operator. 
One verifies that (see also Appendix~\ref{app: equivalence invariants})
\begin{equation}
\nu^{\ }_{\mathrm{D}}(M)=\nu(M)
\end{equation}
for any $|M|\neq1,3$.

The second topological attribute characterizes the bundle of
Bloch states $u^{(\tilde{a})}(\bs{k})$ over the BZ $T^{3}$
for any of the three bands $\tilde{a}=-,0,+$. 
Whenever $|M|\neq1,3$, it is nothing but
the triplet of Berry phases~\cite{Moore08}
\begin{subequations}
\label{eq: def Abelian Berry phases}
\begin{equation}
\theta^{(\tilde{a})}(M):=
\frac{1}{4\pi}
\int\limits^{\ }_{T^{3}}
\mathrm{d}^{3}\bs{k}\,
\epsilon^{\mu\nu\lambda}\,
A^{(\tilde{a})}_{\mu}\,
\partial^{\ }_{\nu}\,
A^{(\tilde{a})}_{\lambda},
\label{eq: def Abelian Berry phases a}
\end{equation}
where we have introduced the Abelian Berry connection
\begin{equation}
A^{(\tilde{a})}_{\mu}(\bs{k}):=
\left(
u^{(\tilde{a})\dag}
\frac{
\partial
     }
     {
\partial k^{\mu}
     }
u^{(\tilde{a})}
\right)
(\bs{k})
\label{eq: def Abelian Berry phases b}
\end{equation}
\end{subequations}
for any of the three bands $\tilde{a}=-,0,+$.
With the help of Eq.~(\ref{eq: Bloch eigenstates mathcal H c}),
one deduces that
\begin{subequations}
\begin{equation}
\theta^{(0)}(M)=
\frac{1}{4\pi}
\int\limits^{\ }_{T^{3}}
\mathrm{d}^{3}\bs{k}\,
\epsilon^{\mu\nu\lambda}\,
\left(
q^{\dag}\,
\partial^{\ }_{\mu}q\,
\partial^{\ }_{\nu}q^{\dag}\,
\partial^{\ }_{\lambda}q
\right)
(\bs{k})
\end{equation}
and
\begin{equation}
\theta^{(-)}(M)=
\theta^{(+)}(M)=
\frac{1}{4}
\theta^{(0)}(M)
\end{equation}
\end{subequations}
when $|M|\neq1,3$.
Explicit evaluations of the Berry phase of any of the three bands then yields
\begin{equation}
\theta^{(-)}(M)=
\theta^{(+)}(M)=
\frac{1}{4}
\theta^{(0)}(M)=
\frac{\pi}{4}
\nu(M)
\end{equation}
when $|M|\neq1,3$ (see Appendix~\ref{app: density algebra}).

With this computation of the topological invariant $\theta$, 
we have also established that the projected electronic density
in any of the bands of 3-orbital model obeys 
the noncommutative 3-bracket defined in 
Eq.~\eqref{3-bracket of derivatives of projected densities} that is 
dominated by the value of $\theta$. 
Upon partial filling, the flat middle band thus provides a manifold 
of many-body noninteracting ground states 
with macroscopic ground state degeneracy, 
similar to the case of a partially filled Landau level. 
Henceforth, one may expect interesting many-body ground states to 
appear once electron-electron interactions are added to the model.
In that regard, we observe that any many-body Hamiltonian that includes 
an interaction build out of density operators projected to the middle band
is invariant under the chiral transformation%
~\eqref{eq: chiral symmetry}, 
since the projected density operators
themselves are invariant under the chiral transformation%
~\eqref{eq: chiral symmetry}.

\subsection{
Surface states
           }

We shall here provide an interpretation of the topological invariant%
~\eqref{eq: def Dirac invariant} as a manifestation
of the surface states associated with a spatially dependent mass parameter $M$ 
in the Hamiltonian~(\ref{eq: def H momentum space}).
This observation applies when considering the surface states 
that connect bands separated by a bulk gap.
Such surface states, connecting the upper and lower band, 
appear only when the periodic boundary conditions are replaced 
by open boundary conditions that implement
a slab geometry with the surface normal parallel 
to the $r^{\ }_{3}$ direction. 

In order to study the surface modes,
we consider the low energy description of the Hamiltonian%
~(\ref{eq: def H momentum space}) 
by linearizing it around each of the 
$8$ nodal points in the Brillouin zone 
$\bs{k}^{\mathsf{T}}_{lmn}=\pi(l,m, n)$, 
with $l,m,n = 0,1$.
The Hamiltonian~(\ref{eq: def H momentum space}) 
in the linearized approximation factorizes according to
\begin{equation}
\mathcal{H} = \bigotimes_{l,m,n = 0,1} \mathcal{H}^{\ }_{lmn}.
\end{equation}
For example, the expansion $\mathcal{H}$ around 
$\bs{k}^{\mathsf{T}}_{000}$ produces
\begin{equation}
\mathcal{H}^{\ }_{000}=
\begin{pmatrix}
0 & 0 & \hat{k}^{\ }_{-}
\\
0 & 0 & \hat{k}^{\ }_3 - \mathrm{i}\,M^{\ }_{000}
\\
\hat{k}^{\ }_{+} & \hat{k}^{\ }_3 + \mathrm{i}\,M^{\ }_{000}  & 0
\end{pmatrix},
\label{eq:H000}
\end{equation}	
where 
$\hat{k}^{\ }_{\pm}\equiv 
\hat{k}^{\ }_{1}\pm\mathrm{i}\hat{k}^{\ }_{2}$,
$\hat{k}^{\ }_{j}\equiv  
-\mathrm{i}\partial^{\ }_{r^{\ }_j}$, 
for $j=1,2,3$ 
and $M^{\ }_{000}=M-3$.
For a uniform mass $M^{\ }_{000}$, 
the spectrum breaks into three low energy bands
with eigenvalues $0$ and $\pm \sqrt{\bs{k}^2 + |M^{\ }_{000}|^2}$.

We now regard $M^{\ }_{000}$ 
as a domain wall configuration along the $r^{\ }_3$-direction, 
which we choose to parametrize as 
\begin{equation}
\label{eq: domain wall 1}
M^{\ }_{000}(r^{\ }_3)= 
\overline{M}^{\ }_{000} 
\left[ 
\Theta(r^{\ }_3) 
- 
\Theta(-r^{\ }_3) 
\right],
\end{equation}
where $\Theta$ is the Heaviside function.
The choice of a sharp domain wall in (\ref{eq: domain wall 1}) 
facilitates the analytic treatment of
the eigenmode equations and does not affect the generality 
of the following discussion.
Due to the translational invariance on the 
$\bs{e}^{\ }_1$-$\bs{e}^{\ }_2$ plane, we seek solutions of
\begin{equation}
\mathcal{H}^{\ }_{000}\,\psi^{\ }_{000,\bs{\kappa}}= 
\varepsilon^{\ }_{\bs{\kappa}}\,\psi^{\ }_{000,\bs{\kappa}},
\end{equation}
with 
$\psi^{\ }_{000,\bs{\kappa}}(\bs{\rho},r^{\ }_3)= 
e^{\mathrm{i}\bs{\kappa}\cdot\bs{\rho}}\, 
\phi^{\ }_{000,\bs{k}}(r^{\ }_3)$,
whereby $\bs{\rho} = (r^{\ }_1,r^{\ }_2)$ 
and $\bs{\kappa}=(k^{\ }_1, k^{\ }_2)$ 
are, respectively, the coordinates and momenta 
projected on the $\bs{e}^{\ }_1$-$\bs{e}^{\ }_2$ plane. 
The components of the spinor wavefunction
\begin{equation}
\phi^{\ }_{000,\bs{\kappa}}(r^{\ }_3) = 
\big( 
f^{\ }_{\bs{\kappa}}(r^{\ }_3), 
g^{\ }_{\bs{\kappa}}(r^{\ }_3), 
h^{\ }_{\bs{\kappa}}(r^{\ }_3) 
\big)^{\mathsf{T}}
\end{equation}
satisfy
\begin{subequations}
\begin{equation}
\label{eq: diff eqn for f}
f^{\ }_{\bs{\kappa}}(r^{\ }_3)= 
\frac{
k^{\ }_{-}
     }
     {
\varepsilon^{\ }_{\bs{\kappa}}
     } 
h^{\ }_{\bs{\kappa}}(r^{\ }_3),
\end{equation}
\begin{equation}
\label{eq: diff eqn for g}
g^{\ }_{\bs{\kappa}}(r^{\ }_3)= 
\frac{1}{\varepsilon^{\ }_{\bs{\kappa}}} 
\left[ 
-
\mathrm{i}\partial^{\ }_{r^{\ }_3} 
-
\mathrm{i} M^{\ }_{000}(r^{\ }_3) 
\right] 
h^{\ }_{\bs{\kappa}}(r^{\ }_3),
\end{equation}
and
\begin{equation}
\label{eq: diff eqn for h}
\begin{split}
&\,
\left[
-
\partial^{2} _{r^{\ }_3}
+
M^{2}_{000}(r^{\ }_3) 
-
2 \overline{M}^{\ }_{000}\delta(r^{\ }_3) 
\right]  
h^{\ }_{\bs{\kappa}}(r^{\ }_3)
\\
&\qquad\qquad= 
(\varepsilon^{2}_{\bs{\kappa}} -\bs{\kappa}^2) 
h^{\ }_{\bs{\kappa}}(r^{\ }_3).
\end{split}
\end{equation}
\end{subequations}
At $r^{\ }_3 \neq 0$, the solution of Eq.~(\ref{eq: diff eqn for h}) yields
\begin{equation}
\label{eq: sol diff eqn for h}
h^{\ }_{\bs{\kappa}}(r^{\ }_3) = h^{\ }_0\, e\,^{-|r^{\ }_3|/\lambda},
\end{equation}
where $h^{\ }_0$ is a normalization constant and 
$\lambda^{-1}:= 
\sqrt{
\overline{M}^{2}_{000} 
+ 
\bs{\kappa}^2 
- 
\varepsilon^{2}_{\bs{\kappa}}
     } 
> 0$,
while the delta function discontinuity at $r^{\ }_3=0$ 
imposes the condition $\lambda^{-1} = \overline{M}^{\ }_{000}$. 
Therefore, the domain wall configuration~(\ref{eq: domain wall 1}) 
bounds surface states with dispersion 
\begin{equation}
\label{eq: gapless dispersion}
\varepsilon^{\ }_{\pm,\bs{\kappa}}= 
\pm 
|\bs{\kappa}|
\end{equation}
provided $\overline{M}^{\ }_{000}>0$. 
Evaluating the solution (\ref{eq: sol diff eqn for h}) in 
(\ref{eq: diff eqn for f}) and (\ref{eq: diff eqn for g}) 
yields the spinor wavefunction, which,
up to a normalization constant $\mathcal{N}$, reads
\begin{subequations}
\begin{equation}
\psi^{\ }_{000,\pm,\bs{\kappa}} (\bs{\rho},r^{\ }_3) = 
\mathcal{N}\, 
\varphi^{\ }_{000,\pm,\bs{\kappa}}\,
e^{\mathrm{i}\bs{\kappa}\cdot\bs{\rho}}\,
e^{-\overline{M}^{\ }_{000} r^{\ }_3},
\end{equation}
\begin{equation}
\varphi^{\ }_{000,\pm,\bs{\kappa}}= 
2^{-1/2}
\left( 
\pm e^{-\mathrm{i}\alpha^{\ }_{\bs{\kappa}}}, 0, 1 
\right)^{\mathsf{T}}, 
\quad
\frac{k^{\ }_{\pm}}{|\bs{\kappa}|}\equiv  
e^{
\pm\mathrm{i}\alpha^{\ }_{\bs{\kappa}}
  }.
\end{equation}
\end{subequations}

The discussion of the boundary states of the low energy Hamiltonians with 
$n=0$, $\mathcal{H}^{\ }_{l m 0}$, 
is very similar
to that of $\mathcal{H}^{\ }_{000}$. 
In this case, the existence of gapless surface states with dispersion as in 
Eq.~(\ref{eq: gapless dispersion})
for sharp domain wall configurations 
\begin{equation}
M^{\ }_{l m 0}(z)= 
\overline{M}^{\ }_{l m 0} 
\left[ 
\theta(r^{\ }_3) 
- 
\theta(-r^{\ }_3) 
\right]
\end{equation}
requires $ \overline{M}^{\ }_{l m 0} > 0$. 
The explicit form of the eigenspinors
(omitting the $r^{\ }_{3}$ dependent part) is
\begin{eqnarray}
\label{eq: eigenspinors 0}
&&
\varphi^{\ }_{00n,\pm,\bs{\kappa}}= 
2^{-1/2} 
\left( 
\pm e^{-\mathrm{i}\alpha^{\ }_{\bs{k}}}, 0, 1 \right)^{\mathsf{T}},
\nonumber\\
&&
\varphi^{\ }_{10n,\pm,\bs{\kappa}}= 
2^{-1/2}
\left( 
\mp e^{+\mathrm{i}\alpha^{\ }_{\bs{k}}}, 0, 1 
\right)^{\mathsf{T}}, 
\nonumber\\
&&
\varphi^{\ }_{11n,\pm,\bs{\kappa}}= 
2^{-1/2}
\left( 
\mp e^{-\mathrm{i}\alpha^{\ }_{\bs{k}}}, 0, 1 
\right)^{\mathsf{T}}, 
\nonumber\\
&&
\varphi^{\ }_{01n,\pm,\bs{\kappa}}= 
2^{-1/2} 
\left( 
\pm e^{+\mathrm{i}\alpha^{\ }_{\bs{k}}}, 0, 1 
\right)^{\mathsf{T}},
\end{eqnarray}
where $n=0$.
For the boundary states of the low energy Hamiltonians with $n=1$, 
$\mathcal{H}^{\ }_{l m 1}$, the extra minus sign
coming from the Taylor expansion around $k^{\ }_3=\pi$  
implies that the gapless surface states exist for domain wall configurations
\begin{equation}
M^{\ }_{l m 1}(r^{\ }_3)= 
\overline{M}^{\ }_{l m 1} 
\left[ 
\theta(r^{\ }_3) 
- 
\theta(-r^{\ }_3) 
\right],
\end{equation}
provided $\overline{M}^{\ }_{l m 1} < 0$. 
The eigenspinors in this case 
are given by Eq.~\eqref{eq: eigenspinors 0}
with $n=1$.

%read
%\begin{eqnarray}
%\label{eq: eigenspinors pi}
%&&
%\varphi^{\ }_{001,\pm,\bs{\kappa}}= 
%2^{-1/2} 
%\left( 
%\pm e^{-\mathrm{i}\alpha^{\ }_{\bs{\kappa}}}, 0, 1 
%\right)^{\mathsf{T}}, 
%\nonumber\\
%&&
%\varphi^{\ }_{101,\pm,\bs{\kappa}}=
%2^{-1/2} 
%\left( 
%\mp e^{+\mathrm{i}\alpha^{\ }_{\bs{\kappa}}}, 0, 1 
%\right)^{\mathsf{T}}, 
%\nonumber\\
%&&
%\varphi^{\ }_{011,\pm,\bs{\kappa}}=
%2^{-1/2} 
%\left( 
%\pm e^{+\mathrm{i}\alpha^{\ }_{\bs{\kappa}}}, 0, 1 
%\right)^{\mathsf{T}},
%\nonumber\\
%&&
%\varphi^{\ }_{111,\pm,\bs{\kappa}}= 
%2^{-1/2}
%\left( 
%\mp e^{-\mathrm{i}\alpha^{\ }_{\bs{\kappa}}}, 0, 1 
%\right)^{\mathsf{T}}.
%\end{eqnarray}

In order to account for all the possible surface modes 
in a finite size configuration, 
we now take, for the sake of concreteness, our system to be a slab, 
infinite in the $\bs{e}^{\ }_1$-$\bs{e}^{\ }_2$ plane and confined
in the $r^{\ }_{3}$-direction by 
$r^{\mathrm{top}}_3\leq 
r^{\ }_3\leq r^{\mathrm{bottom}}_3$, 
with $r^{\mathrm{top}}_3-r^{\mathrm{bottom}}_3$
assumed to be much larger than any other length scale 
so as to regard the two surfaces as completely decoupled from each other.
Moreover, let us adopt the convention that the vacuum is characterized 
by a positive value of the gap parameter 
($M^{\ }_{\textrm{vac}} > 0$), 
which then changes to negative values for 
$r^{\mathrm{top}}_3 < r^{\ }_3 < r^{\mathrm{bottom}}_3$.
For this particular configuration, the discussion above 
implies the presence of gapless surface states associated with 
$\mathcal{H}^{\ }_{l m 0}$ ($\mathcal{H}^{\ }_{l m 1}$) 
at the surface 
$r^{\ }_3=r^{\mathrm{top}}_3$ ($r^{\ }_3=r^{\mathrm{bottom}}_3$)
for $\overline{M}^{\ }_{l m 0} < 0$ 
($\overline{M}^{\ }_{l m 1} < 0$).

In order to make a connection with the topological invariant%
~\eqref{eq: def Dirac invariant} 
we now compute the winding number of the eigenspinors as
\begin{subequations}
\begin{equation}
\label{eq: winding 0}
\nu^{\ }_{l m 0}= 
+ \frac{1}{\pi\,\mathrm{i}}\,
\oint \mathrm{d}\bs{\kappa}\cdot\,
\left( 
\varphi^{\dag}_{l m 0,\pm,\bs{\kappa}}
\bs{\nabla}^{\ }_{\bs{\kappa}}\,\varphi^{\ }_{l m 0,\pm,\bs{\kappa}} 
\right),
\end{equation}
\begin{equation}
\label{eq: winding pi}
\nu^{\ }_{l m 1}= 
- 
\frac{1}{\pi\,\mathrm{i}}\,
\oint\mathrm{d}\bs{\kappa}\cdot\,
\left( 
\varphi^{\dag}_{l m 1,\pm,\bs{\kappa}}
\bs{\nabla}^{\ }_{\bs{\kappa}}\,\varphi^{\ }_{l m 1,\pm,\bs{\kappa}}) 
\right), 
\end{equation}
\end{subequations} 
where the explicit overall sign difference between (\ref{eq: winding 0}) 
and (\ref{eq: winding pi})
reflects the opposite orientation of the outward normal vectors 
$+\bs{e}^{\ }_3$ and $-\bs{e}^{\ }_3$ at the surfaces
$r^{\ }_3 = r^{\mathrm{top}}_3$ and 
$r^{\ }_3 = r^{\mathrm{bottom}}_3$, 
respectively. 
Direct computation using Eq.%
~(\ref{eq: eigenspinors 0})
gives
\begin{subequations}
\begin{eqnarray}
&&
\nu^{\ }_{000}=\nu^{\ }_{110}=\nu^{\ }_{101}=\nu^{\ }_{011}=-1, 
\\
&&
\nu^{\ }_{100}=\nu^{\ }_{010}=\nu^{\ }_{001}=\nu^{\ }_{111}=+1.
\end{eqnarray}
\end{subequations}

The total winding number of the surface states is encoded in the quantity
\begin{equation}
\tilde{\nu} \equiv \sum_{\overline{M}^{\ }_{l m n} < 0}\, \nu^{\ }_{l m n},
\end{equation}
which acquires the following values:
\begin{equation}
\label{eq: values of nu}
\tilde{\nu}=
\begin{cases}
+2,\qquad &|M|<1,\\
-1,\qquad &1<|M|<3,\\
 0,\qquad &3<|M|.
\end{cases}
\end{equation}
Comparison between Eq.~(\ref{eq: values of nu}) and 
Eq.~\eqref{eq: values taken by nu as M changes b} 
thus establishes a direct relationship between the topological 
index~\eqref{eq: def Dirac invariant} and the total winding number of
the surface states $\tilde{\nu}$. 
Similar analysis of the finite size system spectrum 
for domain wall configurations of the gap parameter 
along either the $x$ or the $y$ directions reveals
the nonexistence of surface states.

\medskip
\section{
Interactions within the single-mode approximation
        }
\label{sec: Interactions within the single-mode approximation}

We begin by reviewing the single-mode approximation (SMA) 
to the FQHE from  Ref.%
~\onlinecite{Girvin85}.

In the IQHE, the external magnetic field organizes the 
single-particle spectrum into degenerate Landau levels,
whereby two consecutive Landau levels 
are separated by the energy gap 
$\hbar\omega^{\ }_{c}$. 
The cyclotron frequency 
$\omega^{\ }_{c}=\hbar/(m^{\ }_{\mathrm{e}}\ell^2_B)$
is proportional to the magnitude $B$ of the uniform magnetic field.

We consider the limit of very strong magnetic fields relative to
the characteristic energy scale $V$ of the electron-electron interactions,
i.e., $\hbar \omega^{\ }_{c}\gg V$.
Moreover, we consider a filling fraction $\nu\equiv\Phi/\Phi^{\ }_{0}<1$
($\Phi$ the magnetic flux and $\Phi^{\ }_{0}$ the flux quantum)
such that the exact many-body ground state
$
\left|\Psi^{\ }_{0}\right\rangle
$
does not break spontaneously any symmetry. The translation invariant
interacting Hamiltonian $\widehat{H}$
describing a nonvanishing density 
of spinless fermions moving in a plane perpendicular
to an external magnetic field of uniform magnitude $B$ and interacting
pairwise with a (screened) Coulomb interaction is then well approximated,
as far as low energy properties go, by its projection 
$\widehat{H}^{\ }_{\mathrm{LLL}}$
onto the vector space spanned by the lowest Landau single-particle levels.

Upon imposing periodic boundary conditions in an area of linear size $L$,
$\widehat{H}^{\ }_{\mathrm{LLL}}$
is given by
\begin{subequations}
\label{eq: H SMA}
\begin{equation}
\widehat{H}^{\ }_{\mathrm{LLL}}= 
\sum_{\bs{q}}\, 
v^{\ }_{\bs{q}}\, 
\delta\widehat{\rho}^{\ }_{-\bs{q}}\,
\delta\widehat{\rho}^{\ }_{+\bs{q}},
\end{equation}
where 
\begin{equation}
v^{\ }_{\bs{q}} = v^{*}_{\bs{q}} = v^{\ }_{-\bs{q}}
\end{equation}
is the Fourier transform of the screened Coulomb interaction, 
while
\begin{equation}
\delta\widehat{\rho}^{\ }_{\bs{q}}:=
\widehat{\rho}^{\ }_{\bs{q}}
- 
\left\langle\,
\Psi^{\ }_{0} 
\left|
\widehat{\rho}^{\ }_{\bs{q}}\, 
\right|
\Psi^{\ }_{0} 
\right\rangle 
\end{equation}
\end{subequations}
is the Fourier component of the
fermion density operator after projection into the LLL 
measured relative to its expectation value
in the exact many-body ground state
$
\left|\Psi^{\ }_{0}\right\rangle
$.

Inspired by the early work of Feynman and Bijl 
in their study of excitations in $^4$He,%
~\cite{Feynman72} GMP
in Ref.~\onlinecite{Girvin85}
consider the variational state
\begin{equation}
\label{eq: SMA state}
|\phi^{\ }_{\bs{k}}\rangle :=
\delta\widehat{\rho}^{\ }_{\bs{k}}\,
|\Psi^{\ }_{0}\rangle,
\end{equation}
whose energy expectation value $\Delta^{\ }_{\bs{k}}$, 
measured relative to the exact ground state energy $E^{\ }_{0}$, 
sets a variational upper bound on the low excitation spectrum of 
the LLL-projected Hamiltonian%
~(\ref{eq: H SMA}).

Assuming the inversion symmetry 
\begin{subequations}
\label{eq: SMA equations}
\begin{equation}
\Delta^{\ }_{+\bs{k}} = \Delta^{\ }_{-\bs{k}},
\label{eq: SMA equations a}
\end{equation}
a direct calculation using Eqs.~(\ref{eq: H SMA}) and%
~(\ref{eq: SMA state}) leads to 
\begin{equation}
\Delta^{\ }_{\bs{k}}= 
\frac{f^{\ }_{\bs{k}}}{s^{\ }_{\bs{k}}},
\label{eq: SMA equations b}
\end{equation}
where
\begin{equation}
f^{\ }_{\bs{k}}= 
\frac{1}{2}\,
\left\langle\Psi^{\ }_{0}\left| \,
\left[
\delta\widehat{\rho}^{\ }_{-\bs{k}}, 
\left[
\widehat{H}^{\ }_{\mathrm{LLL}},
\delta\widehat{\rho}^{\ }_{+\bs{k}} 
\right] 
\right] 
\right|\Psi^{\ }_{0}\right\rangle
\label{eq: SMA equations c}
\end{equation}
and
\begin{equation}
s^{\ }_{\bs{k}}= 
\left\langle\Psi^{\ }_{0}\left| 
\delta\widehat{\rho}^{\ }_{-\bs{k}}\,
\delta\widehat{\rho}^{\ }_{+\bs{k}}  
\right|\Psi^{\ }_{0}\right\rangle.
\label{eq: SMA equations d}
\end{equation}
\end{subequations}
One recognizes on the right-hand side of Eq.%
~(\ref{eq: SMA equations d}) 
the static structure factor.
The insight of GMP in Ref.~\onlinecite{Girvin85} was to realize that 
the density operators projected onto the lowest Landau level 
close the exact algebra
\begin{equation}
\left[
\widehat{\rho}^{\ }_{\bs{q}}, 
\widehat{\rho}^{\ }_{\bs{k}}
\right]=
2\mathrm{i}\,
\sin
\left(
\frac{1}{2}
\left(
\bs{q}\times\bs{k}
\right)
\cdot 
\bs{e}^{\ }_3\,
\ell^{2}_B
\right)
\widehat{\rho}^{\ }_{\bs{q}+\bs{k}}
\label{eq: GMP algebra}
\end{equation}
($\ell^{\ }_B$ is the magnetic length). In turn,
the algebra~(\ref{eq: GMP algebra})
implies that 
\begin{equation}
f^{\ }_{\bs{k}}\sim 
|\bs{k}|^{4} 
\end{equation}
in the small $|\bs{k}|$ limit.
Hence, in the FQHE, a necessary (but not sufficient) condition 
for the existence of a finite gap in the thermodynamic limit is 
to have 
\begin{equation}
s^{\ }_{\bs{k}} \sim |\bs{k}|^{4}
\label{eq: s bs{k} scales like |bs{k}| to 4}
\end{equation}
also hold in the small $|\bs{k}|$ limit. In fact,
Eq.~(\ref{eq: s bs{k} scales like |bs{k}| to 4})
was shown in Ref.~\onlinecite{Girvin85} to be satisfied when 
$\left|\Psi^{\ }_{0}\right\rangle$ is chosen to be any
Laughlin state with filling fraction $\nu=1/m$, 
where $m$ is an odd integer.

In the spirit of GMP,
our starting point is a single-particle Hamiltonian 
defined on a $d$-dimensional Bravais lattice 
and sharing its point group symmetry.
We also assume that there exists
at least one band that is independent of the lattice momentum,
i.e., a flat band, and, furthermore, that is separated from the
other bands by a single-particle gap $\Delta$. We constructed a
3D example thereof in 
Sec.~\ref{sec: Noninteracting three-band tight-binding model}.
We then imagine switching on adiabatically a pairwise interaction that 
preserves the Bravais lattice point-group symmetry, 
say a (screened) Coulomb interaction.
We shall denote with $V$ the corresponding characteristic interaction
energy scale. In the regime for which $\Delta\gg V$, 
Hamiltonian%
~(\ref{eq: H SMA}) 
can be reinterpreted as the interacting Hamiltonian 
projected onto this flat band, provided we identify
$v^{\ }_{\bs{q}}$ 
with the Fourier transform at the lattice momentum $\bs{q}$
of the pairwise fermion interaction,
$
\delta\widehat{\rho}^{\ }_{\bs{q}}
$
with the Fourier transform at lattice momentum $\bs{q}$
of the projected operator describing density fluctuation measured relative
to the fermion density with lattice momentum $\bs{q}$ of the
exact many-body ground state
$\left|\Psi^{\ }_{0}\right\rangle$,
whereby we assume that $\left|\Psi^{\ }_{0}\right\rangle$
does not break spontaneously any point-group symmetry of the lattice.

The projected density operator on a flat band reads
\begin{equation}
\begin{split}
\widehat{\rho}^{\ }_{\bs{k}}=&\,
\sum_{\bs{p}}\, 
u^{\dag}_{\bs{p}}\, 
\cdot\,
u^{\ }_{\bs{p}+\bs{k}}\,
\widehat{\chi}^{\dag}_{\bs{p}}\, 
\widehat{\chi}^{\ }_{\bs{p}+\bs{k}}
\\
\equiv&\,
\sum_{\bs{p}}\, 
M^{\ }_{\bs{p},\bs{k}}\,
\widehat{\chi}^{\dag}_{\bs{p}}\, 
\widehat{\chi}^{\ }_{\bs{p}+\bs{k}},
\end{split}
\label{eq: projected lattice operator}
\end{equation}
where 
$u^{\ }_{\bs{k}}\in\mathbb{C}^{N}$ 
is vector-valued
(its components range over the number 
$N$ of orbitals per site of the Bravais lattice),
while 
$\widehat{\chi}^{\ }_{\bs{k}}$ 
and
$\widehat{\chi}^{\dag}_{\bs{k}}$ 
are the annihilation and creation operators, respectively, 
of single-particle fermionic eigenstates on the isolated flat band 
with lattice momentum $\bs{k}$.
Hence, they satisfy the canonical fermionic 
anticommutation relations
\begin{equation}
\label{eq: fermionic anticommutation}
\left\{
\widehat{\chi}^{\ }_{\bs{k}},  
\widehat{\chi}^{\vphantom{\dag}}_{\bs{k}'} 
\right\}= 
\left\{
\widehat{\chi}^{\dag}_{\bs{k}},  
\widehat{\chi}^{\dag}_{\bs{k}'}
\right\}=0,
\quad 
\left\{
\widehat{\chi}^{\ }_{\bs{k}},  
\widehat{\chi}^{\dag}_{\bs{k}'}
\right\}= 
\delta^{\ }_{\bs{k},\bs{k}'}
\end{equation}
for any pair $\bs{k}$ and $\bs{k}'$ of lattice momenta.
In carrying out the program laid out in 
Eq.~(\ref{eq: SMA equations}) for a general
lattice Hamiltonian with a flat band, 
one notices two immediate obstacles.

The first one arises from the fact that the commutator of two 
(projected) density operators  does not satisfy 
the algebra~(\ref{eq: GMP algebra}) found by GMP for the
FQHE in a uniform magnetic field. However,
it was noticed in Ref.%
~\onlinecite{Parameswaran11} 
that, 
in the limit of small lattice momenta
$
\bs{k}
$ 
and
$
\bs{k}'
$, 
the commutation relation between two projected density operators reads
\begin{subequations}
\label{eq: lattice density algebra}
\begin{equation}
\begin{split}
[\, 
\widehat{\rho}(\bs{k}), 
\widehat{\rho}(\bs{k}') 
\,]=&\,
\int\limits_{\bs{p}}
\Big[
\mathrm{i}\,\left(
\bs{k}
\wedge
\bs{k}'
\right)
\cdot
\bs{\mathcal{B}}(\bs{p})
+
\cdots
\Big]
\\
&\,\times
\widehat{\chi}^{\dag}(\bs{p})\,
\widehat{\chi}(\bs{p}+\bs{k}+\bs{k}')
\end{split}
\label{eq: lattice density algebra a}
\end{equation}
in the thermodynamic limit $L\to\infty$, whereby the short-hand notation 
\begin{equation}
\int\limits_{\bs{p}}\equiv
\int\frac{\mathrm{d}^{d}\bs{p}}{(2\pi/L)^{d}}
\end{equation}
is used,
\begin{equation}
\bs{\mathcal{B}}(\bs{p}):=
- 
\mathrm{i}\,
\left(\bs{\nabla}\wedge\bs{A}\right)(\bs{p})
\label{eq: lattice density algebra b}
\end{equation}
is the (real-valued) Berry field strength of the flat band, and
\begin{equation}
\bs{A}(\bs{p}):=
\left(
u^{\dag}\,
\cdot\,
\bs{\nabla}\,
u
\right)
(\bs{p})
\label{eq: lattice density algebra c}
\end{equation}
\end{subequations}
is the (imaginary-valued) Berry connection of the flat band, 
while
$
\cdots
$
in Eq.%
~(\ref{eq: lattice density algebra a})
accounts for higher order terms in powers of 
$
\bs{k}
$ 
and 
$
\bs{k}'
$.
Consequently, it was proposed in Ref.%
~\onlinecite{Parameswaran11}  
that the numerical observation 
of the FQHE without an external magnetic field in 
2D Chern insulators in Refs.~\onlinecite{Neupert11a,Sheng11,Wang11a,Regnault11}
can be understood on the account that, 
because in a $2$D Chern band insulator the integral of the Berry curvature 
on the Brillouin zone equals the (nonzero) Chern number,
replacing 
$
\bs{\mathcal{B}}(\bs{p})
$
in Eq.%
~(\ref{eq: lattice density algebra})
by its \textit{average},
implies the GMP algebra%
~(\ref{eq: GMP algebra})
in the long-wavelength limit.
However, we would like to stress that, contrary to the 
2D Chern band insulators for which one can associate
the notion of an average Berry curvature due to the
nonzero Chern number, for the $3$D lattice
models studied in Secs.~\ref{sec: Noncommutative geometry}
and~\ref{sec: Noninteracting three-band tight-binding model},
the integral of the Berry curvature vanishes so that
replacing
$
\bs{\mathcal{B}}(\bs{p})
$
by its average is meaningless.
Even for 2D Chern band insulators, the Berry curvature 
is generically nonuniform; a fact that should be reflected in
the exact many-body wavefunction.

The second obstacle to applying the SMA to an interacting lattice model is the
fact that no  good candidate wavefunction is presently known
with which one can compute the static structure factor 
$s^{\ }_{\bs{k}}$
and compare its small 
$\bs{k}$
dependence with that of
$f^{\ }_{\bs{k}}$,
as was done by GMP in Ref.%
~\onlinecite{Girvin85}. Nevertheless,
information about the behavior of
$
f^{\ }_{\bs{k}}
$
for small $\bs{k}$ and the requirement
of a finite gap in the thermodynamic limit, i.e.,
$
\Delta^{\ }_{\bs{k}} 
\rightarrow
\Delta^{\ }_{0}
\neq 0
$
for 
$
\bs{k}
\rightarrow
0
$,
puts a constraint on the static structure factor
for small $\bs{k}$ and, correspondingly, 
on the correlations of the exact many-body wavefunction.

In Appendix%
~\ref{appsec: SMA for a flat band}
we discuss in detail the evaluation of the
function 
$
f^{\ }_{\bs{k}}
$
defined in Eq.%
~(\ref{eq: SMA equations c})
to lowest order in $\bs{k}$.
Our main result is that, 
due to the non-closure of the density algebra for any
$d$-dimensional lattice model, 
the leading contribution to $f(\bs{k})$
reads
\begin{subequations}
\medskip
\begin{widetext}
\begin{equation}
\begin{split}
f(\bs{k})=&\,
\int\limits_{\bs{q}}
\int\limits_{\bs{p}}
\int\limits_{\bs{p}'}
v(\bs{q})
\Big[
\left(
\bs{k}\wedge\bs{q}
\right)
\cdot
\delta\bs{\mathcal{B}}(\bs{p})
\Big]
\Big[
\left(
\bs{k}\wedge\bs{q}
\right)
\cdot
\delta\bs{\mathcal{B}}(\bs{p}')
\Big]
\Big{\langle}
\hat{n}(\bs{p})\,
\hat{n}(\bs{p}')
\Big{\rangle}
\\
&\,
+
\int\limits_{\bs{q}}
\int\limits_{\bs{p}}
v(\bs{q})\,
\frac{\mathrm{i}}{2}
\left(
\bs{k}\wedge\bs{q}
\right)
\cdot
\left(\partial^{\ }_{\mu}\bs{\mathcal{B}}\right)(\bs{p})
k^{\mu}
\Big[
\Big{\langle}\,
\delta\widehat{\rho}(-\bs{q})\, 
\widehat{\chi}^{\dag}(\bs{p})\,
\widehat{\chi}(\bs{p}+\bs{q})
\Big{\rangle}
-
\Big{\langle}\,
\widehat{\chi}^{\dag}(\bs{p}+\bs{q})\,
\widehat{\chi}(\bs{p})\,
\delta\widehat{\rho}(\bs{q})
\Big{\rangle}
\Big]
\end{split}
\label{eq: f_k final expression}
\end{equation}
\end{widetext}
\medskip
where 
$\int\limits_{\bs{q}}\equiv\int\mathrm{d}^{d}\bs{q}/(2\pi/L)^{d}$
and the summation convention 
is implied over the repeated index $\mu=1,\cdots,d$.
In Eq.%
~(\ref{eq: f_k final expression}),
\begin{equation}
\hat{n}(\bs{p}):=
\widehat{\chi}^{\dag}(\bs{p})\,
\widehat{\chi}(\bs{p})
\end{equation}
while 
\begin{equation}
\delta\bs{\mathcal{B}}(\bs{p}):=
\bs{\mathcal{B}}(\bs{p})
-
\overline{\bs{\mathcal{B}}}
\end{equation}
\end{subequations}
denotes the deviations of the Berry curvature 
$
\bs{\mathcal{B}}(\bs{p})
$
away from the uniform background value
$
\overline{\bs{\mathcal{B}}}
$.
This uniform background value is
defined in such a way that, when $d=3$,
\begin{equation}
\begin{split}
\mathrm{Ch}^{\lambda}:=&\,
2\pi 
\times
\frac{1}{2}
\int\limits_{T^{3}} 
\frac{
\mathrm{d}^{3}\bs{p}
     }
     {
(2\pi)^{3}
     }
\mathcal{B}^{\lambda}(\bs{p})
\\
\equiv&\,
\frac{2\pi }{L^{3}}
\times
\frac{1}{2}
\int\limits_{T^{3}} 
\frac{
\mathrm{d}^{3}\bs{p}
     }
     {
(2\pi)^{3}
     }
\overline{\mathcal{B}}^{\lambda}
\end{split}
\end{equation}
with $\lambda=1,2,3$ is compatible with a generalization 
of the $2$D Chern number to layered (quasi-2D) materials.
The result%
~(\ref{eq: f_k final expression})
should be contrasted with the calculation in Ref.%
~\onlinecite{Girvin85}, for which the order $\bs{k}^{2}$ term in 
$
f(\bs{k})
$
vanishes identically as a consequence of the algebra%
~(\ref{eq: GMP algebra}). 
The formula%
~(\ref{eq: f_k final expression}) 
thus establishes a direct relationship, within the SMA,
between the deviations of the Berry field strength away from
a uniform configuration and the order $\bs{k}^{2}$ 
contribution to $f(\bs{k})$
\begin{equation}
f(\bs{k})\sim |\bs{k}|^2\; .
\end{equation}
Such a relation is relevant either
for 2D fractional Chern band insulators for which,
despite a nonzero Chern number,
$\bs{\mathcal{B}}(\bs{p})$ can be 
nonuniform throughout the
Brillouin zone
or for the general classes of 3D lattice models
studied in Secs.~\ref{sec: Noncommutative geometry}
and~\ref{sec: Noninteracting three-band tight-binding model}
for which the integral of 
$\bs{\mathcal{B}}(\bs{p})$
vanishes. 
The result%
~(\ref{eq: f_k final expression})
also indicates that
a prerequisite for the existence of a nonvanishing but finite many-body
gap to excitations above the many-body ground state 
is that the static structure factor $s(\bs{k})$ has also
to vanish as $\bs{k}^{2}$ to allow for the possibility
of a nonzero ratio $\Delta(\bs{k})\equiv f(\bs{k})/s(\bs{k})$
and therefore a nonvanishing SMA gap  
in Eq.% 
~(\ref{eq: SMA equations b}).

\section{
Summary
        }
\label{sec: Discussion}

The noncommutativity of coordinates and density operators 
in a featureless liquid-like electronic state can be a local probe 
of its topological character. In this paper, we have studied how
this fact, which is well-established for quantum Hall fluids in 2D,
carries over to 3D topological states of itinerant electrons.
In the limit of long wavelength, we found that both the 
noncommutative relations
obeyed by projected position and density
operators are characterized by the topological invariant of 
a 3D band structure with chiral symmetry. 
We established a relation between the 
noncommutative relation
of the projected position 
operators and the classical Nambu bracket 
of volume-preserving diffeomorphisms
of 3D fluids, that might bridge
the description of classical
ideal fluids and that of topological incompressible states in 3D.

\medskip
\section*{Acknowledgments}

We thank A. Polychronakos for helpful correspondence.
This work was supported in part by DOE Grant DEFG02-06ER46316
and by the Swiss National Science Foundation.
After posting the original version of this manuscript
on the archives, Ref.~\onlinecite{Estienne12}
appeared. A discussion of the operator product expansion
obeyed by projected density operators in 3D topological insulators
is also given in Ref.~\onlinecite{Estienne12}. 
We thank the authors of Ref.~\onlinecite{Estienne12} 
for insightful communications.

\appendix

\begin{widetext}

\section{
Gauge-invariant regularization of brackets of projected position operators
        }
\label{appsec: Regularization of non-commuting projected position operators}

\subsection{
Definition of the single-particle Hilbert space
           }

Define the three lattices
\begin{subequations}
\begin{equation}
\Lambda^{\star}_{\mathrm{BZ}}:=
\left\{
(k^{\mu})\in\mathbb{R}^{d}
\left|
k^{\mu}=
\frac{2\pi}{\mathfrak{a}}\, n^{\mu},
\quad
n^{\mu}=1,\cdots,\mathcal{N}^{\mu}
\right.
\right\},
\end{equation}
\begin{equation}
\Lambda^{\ }_{r}:=
\left\{
(r^{\mu})\in\mathbb{R}^{d}
\left|
r^{\mu}=
\mathfrak{a}\, n^{\mu},
\quad
n^{\mu}=1,\cdots,\mathcal{N}^{\mu}
\vphantom{\frac{2\pi}{\mathfrak{a}}}
\right.
\right\},
\end{equation}
and
\begin{equation}
\Lambda^{\ }_{R}:=
\left\{
(R^{\mu})\in\mathbb{R}^{d}
\left|
R^{\mu}=
\mathfrak{a}\, n^{\mu},
\quad
n^{\mu}=1,\cdots,\mathcal{N}^{\mu}
\vphantom{\frac{2\pi}{\mathfrak{a}}}
\right.
\right\},
\end{equation}
\end{subequations}
each of which shares the same cardinality
\begin{equation}
\mathcal{N}:=
\prod_{\mu=1}^{d}
\mathcal{N}^{\mu}.
\end{equation}
The lattices $\Lambda^{\ }_{r}$ and $\Lambda^{\ }_{R}$
share the same unit cell of linear extend $\mathfrak{a}$
but they might be shifted by the vector
\begin{equation}
\begin{split}
&
\bs{d}:=
\sum_{\mu=1}^{d}
e^{\ }_{\mu}\,\bs{e}^{\mu},
%\\
%&
\qquad
0\leq e^{\ }_{\mu}<1,
%\\
%&
\qquad
\bs{e}^{\mu}\cdot \bs{e}^{\nu}=\delta^{\mu,\nu},
\qquad
\mu,\nu=1,\cdots,d,
\end{split}
\end{equation}
from the unit cell relative to each other.

The single-particle Hilbert space is defined through a basis
of orthonormal states. We introduce two such bases.

There is the orbital basis
\begin{equation}
\begin{split}
&
\openone=
\left|
\psi^{\alpha}_{\bs{r}}
\right\rangle
\left\langle
\psi^{\alpha}_{\bs{r}}
\right|=
\left|
\psi^{\alpha}_{\bs{k}}
\right\rangle
\left\langle
\psi^{\alpha}_{\bs{k}}
\right|,
%\\
%&
\quad
\langle
\psi^{\alpha}_{\bs{r}}
|
\psi^{\alpha'}_{\bs{r}'}
\rangle=
\delta^{\alpha,\alpha'}\,
\delta^{\ }_{\bs{r},\bs{r}'},
%\\
%&
\quad
\langle
\psi^{\alpha}_{\bs{k}}
|
\psi^{\alpha'}_{\bs{k}'}
\rangle=
\delta^{\alpha,\alpha'}\,
\delta^{\ }_{\bs{k},\bs{k}'},
%\\
%&
\quad
\langle
\psi^{\alpha}_{\bs{r}}
|
\psi^{\alpha'}_{\bs{k}}
\rangle=
\delta^{\alpha,\alpha'}\,
\frac{1}{\sqrt{\mathcal{N}}}\, 
e^{+\mathrm{i}\bs{k}\cdot\bs{r}},
\end{split}
\label{eq: orbital basis}
\end{equation}
with the summation convention implied over repeated indices and
for any pairs $\alpha,\alpha'=1,\cdots,N$ or
$\bs{r},\bs{r}'\in\Lambda^{\ }_{r}$ or
$\bs{k},\bs{k}'\in\Lambda^{\star}_{\hbox{\tiny{BZ}}}$.

There is the band basis
\begin{equation}
\begin{split}
&
\openone=
\left|
W^{a}_{\bs{R}}
\right\rangle
\left\langle
W^{a}_{\bs{R}}
\right|=
\left|
\chi^{a}_{\bs{k}}
\right\rangle
\left\langle
\chi^{a}_{\bs{k}}
\right|,
%\\
%&
\quad
\langle
W^{a}_{\bs{R}}
|
W^{\alpha'}_{\bs{R}'}
\rangle=
\delta^{a,a'}\,
\delta^{\ }_{\bs{R},\bs{R}'},
%\\
%&
\quad
\langle
\chi^{a}_{\bs{k}}
|
\chi^{a'}_{\bs{k}'}
\rangle=
\delta^{a,a'}\,
\delta^{\ }_{\bs{k},\bs{k}'},
%\\
%&
\quad
\langle
W^{a}_{\bs{R}}
|
\chi^{a'}_{\bs{k}}
\rangle=
\delta^{a,a'}\,
\frac{1}{\sqrt{\mathcal{N}}}\, 
e^{+\mathrm{i}\bs{k}\cdot\bs{R}},
\end{split}
\label{eq: band basis}
\end{equation}
with the summation convention implied over repeated indices and
for any pairs $a,a'=1,\cdots,N$ or
$\bs{R},\bs{R}'\in\Lambda^{\ }_{R}$ or
$\bs{k},\bs{k}'\in\Lambda^{\star}_{\hbox{\tiny{BZ}}}$.

The orbital and band basis in momentum space are related
by the momentum resolved
$N\times N$ unitary matrix $U^{\ }_{\bs{k}}$ with the matrix elements
\begin{subequations}
\label{eq: U relates psi to chi}
\begin{equation}
\langle
\psi^{\alpha}_{\bs{k}}
|
\chi^{a}_{\bs{k}}
\rangle=
u^{\alpha a}_{\bs{k}},
\qquad
\alpha,a=1,\cdots,N.
\label{eq: U relates psi to chi a}
\end{equation}
Hence, for any $\bs{k}\in\Lambda^{\star}_{\hbox{\tiny{BZ}}}$,
these matrix elements  
obey the orthonormality conditions
\begin{equation}
u^{\alpha a }_{\bs{k}}\,
u^{\alpha'a*}_{\bs{k}}=
\delta^{\alpha,\alpha'},
\qquad
\alpha,\alpha'=1,\cdots,N,
\label{eq: orthonormality u's on lattice a}
\end{equation}
for row multiplication or
\begin{equation}
u^{\alpha a*}_{\bs{k}}\,
u^{\alpha a'}_{\bs{k}}=
\delta^{a,a'},
\qquad
a,a'=1,\cdots,N,
\label{eq: orthonormality u's on lattice b}
\end{equation}
\end{subequations}
for column multiplication.

The orbital basis in position space
and the band basis in momentum space are related
by the Fourier component
\begin{equation}
\langle
\psi^{\alpha}_{\bs{r}}
|
\chi^{a}_{\bs{k}}
\rangle=
\frac{1}{\sqrt{\mathcal{N}}}\,
u^{\alpha a}_{\bs{k}}\,
e^{+\mathrm{i}\bs{k}\cdot\bs{r}},
\qquad
\alpha,a=1,\cdots,N,
\label{eq: U relates psi to chi c}
\end{equation}
for any $\bs{r}\in\Lambda^{\ }_{r}$ 
and any $\bs{k}\in\Lambda^{\star}_{\hbox{\tiny{BZ}}}$.

The orbital and band basis in position space
are related by the convolution
\begin{equation}
\begin{split}
|
W^{a}_{\bs{R}}
\rangle=&\,
\frac{1}{\sqrt{\mathcal{N}}}\,
e^{-\mathrm{i}\bs{k}\cdot\bs{R}}\,
|
\chi^{a}_{\bs{k}}
\rangle
\\
=&\,
\frac{1}{\sqrt{\mathcal{N}}}\,
e^{-\mathrm{i}\bs{k}\cdot\bs{R}}\,
\Big(
|
\psi^{\alpha}_{\bs{r}}
\rangle
\langle
\psi^{\alpha}_{\bs{r}}
|
\Big)
|
\chi^{a}_{\bs{k}}
\rangle
\\
=&\,
\frac{1}{\sqrt{\mathcal{N}}}\,
e^{-\mathrm{i}\bs{k}\cdot\bs{R}}\,
\Big(
\langle
\psi^{\alpha}_{\bs{r}}
|
\chi^{a}_{\bs{k}}
\rangle
\Big)
|
\psi^{\alpha}_{\bs{r}}
\rangle
\\
\hbox{
\tiny
Eq.~(\ref{eq: U relates psi to chi c})
\qquad
     }
=&\,
\frac{1}{\mathcal{N}}\,
e^{-\mathrm{i}\bs{k}\cdot\left(\bs{R}-\bs{r}\right)}\,
u^{\alpha a}_{\bs{k}}\,
|
\psi^{\alpha}_{\bs{r}}
\rangle
\end{split}
\label{eq: Wannier in terms orbitals}
\end{equation}
for any $\bs{R}\in\Lambda^{\ }_{R}$
with the summation convention over repeated indices on the right-hand side.

\subsection{
Projected lattice position operator
           }

A lattice position operator generates infinitesimal
translations in momentum space.
There is an ambiguity when defining a lattice position operator.
We can either choose to define the position operator
on the lattice $\Lambda^{\ }_{r}$ or on the lattice
$\Lambda^{\ }_{R}$. In the former case, we define
\begin{equation}
\begin{split}
\widehat{\bs{r}}:=&\,
\sum_{\bs{r}\in\Lambda^{\ }_{r}}
\sum_{\alpha=1}^{N}
|
\psi^{\alpha}_{\bs{r}}
\rangle\,
\bs{r}\,
\langle
\psi^{\alpha}_{\bs{r}}
|
\\
\equiv&\,
|
\psi^{\alpha}_{\bs{r}}
\rangle\,
\bs{r}\,
\langle
\psi^{\alpha}_{\bs{r}}
|
\end{split}
\label{eq: spectral decomposition hat r}
\end{equation}
with the summation convention over the repeated indices 
$\alpha=1,\cdots,N$ and $\bs{r}\in\Lambda^{\ }_{r}$ implied
on the second line.
In the latter case, we define
\begin{equation}
\begin{split}
\widehat{\bs{R}}:=&\,
\sum_{\bs{R}\in\Lambda^{\ }_{R}}
\sum_{a=1}^{N}
|
W^{a}_{\bs{R}}
\rangle\,
\bs{R}\,
\langle
W^{a}_{\bs{R}}
|
\\
=&\,
|
W^{a}_{\bs{R}}
\rangle\,
\bs{R}\,
\langle
W^{a}_{\bs{R}}
|
\end{split} 
\label{eq: spectral decomposition hat R}
\end{equation}
with the summation convention over the repeated indices 
$a=1,\cdots,N$ and $\bs{R}\in\Lambda^{\ }_{R}$ implied
on the second line.

We define the projection operator on the first $\widetilde{N}$
occupied bands by
\begin{equation}
\begin{split}
\widehat{p}^{\ }_{\widetilde{N}}:=&\,
\sum_{\bs{k}\in\Lambda^{\star}_{\hbox{\tiny{BZ}}}}
\sum_{\tilde{a}=1}^{\widetilde{N}}
\left|
\chi^{\tilde{a}}_{\bs{k}}
\right\rangle
\left\langle
\chi^{\tilde{a}}_{\bs{k}}
\right|
\\
\equiv&\,
\left|
\chi^{\tilde{a}}_{\bs{k}}
\right\rangle
\left\langle
\chi^{\tilde{a}}_{\bs{k}}
\right|
\end{split}
\label{eq: projection op in band basis}
\end{equation}
with the summation convention over the repeated indices 
$\tilde{a}=1,\cdots,\widetilde{N}$ and 
$\bs{k}\in\Lambda^{\star}_{\hbox{\tiny{BZ}}}$ implied
on the second line.
In the sequel, it will always be understood that
latin indices such as $\tilde{a}$ 
run over the first $\widetilde{N}$ occupied bands.
The projection operator on the first $\widetilde{N}$
occupied bands is represented by
\begin{equation}
\begin{split}
\widehat{p}^{\ }_{\widetilde{N}}=&\,
\frac{1}{\mathcal{N}}
\sum_{\bs{k}\in\Lambda^{\star}_{\hbox{\tiny{BZ}}}}
e^{+\mathrm{i}\bs{k}\cdot\left(\bs{R}-\bs{R}'\right)}
|
W^{\tilde{a}}_{\bs{R}}
\rangle
\langle
W^{\tilde{a}}_{\bs{R}'}
|
\\
=&\,
|
W^{\tilde{a}}_{\bs{R}}
\rangle
\langle
W^{\tilde{a}}_{\bs{R}}
|
\end{split}
\label{eq: projection op in Wannier basis}
\end{equation}
in the Wannier basis
(with the summation convention over the repeated indices
$\tilde{a}=1,\cdots,\widetilde{N}$ and
$\bs{R}\in\Lambda^{\ }_{R}$
on the second line).
The projection operator on the first $\widetilde{N}$
occupied bands is represented by
\begin{equation}
\begin{split}
\widehat{p}^{\ }_{\widetilde{N}}=&\,
u^{\alpha\tilde{a}}_{\bs{k}}\,
u^{\alpha'\tilde{a}*}_{\bs{k}}\,
|
\psi^{\alpha}_{\bs{k}}
\rangle
\langle
\psi^{\alpha'}_{\bs{k}}
|
\end{split}
\end{equation}
in the momentum space orbital basis
(with the summation convention over the repeated indices
$\tilde{a}=1,\cdots,\widetilde{N}$,
$\alpha,\alpha'=1,\cdots,N$,
and
$\bs{k}\in\Lambda^{\star}_{\hbox{\tiny{BZ}}}$). 
It is not diagonal in the orbital indices because of the 
truncation to the occupied band.

The lattice position operator projected on the first $\widetilde{N}$
occupied bands can be either defined by
\begin{equation}
\widehat{\bs{X}}^{\ }_{r}:=
\widehat{p}^{\ }_{\widetilde{N}}\,
\hat{\bs{r}}\,
\widehat{p}^{\ }_{\widetilde{N}}
\label{eq: def X_r}
\end{equation}
or by
\begin{equation}
\widehat{\bs{X}}^{\ }_{R}:=
\widehat{p}^{\ }_{\widetilde{N}}\,
\hat{\bs{R}}\,
\widehat{p}^{\ }_{\widetilde{N}}.
\label{eq: def X_R}
\end{equation}

\subsection{
Lattice discretization
of the single-particle trace over the $1$-bracket 
of the projected position operator
           }
\label{subsec: Lattice regularization of the 1 bracket}

We are first going to show that
\begin{equation}
\begin{split}
\mathrm{Tr}\,
\left(
\widehat{\bs{X}}^{\ }_{r}
-
\widehat{\bs{X}}^{\ }_{R}
\right)=
\sum_{\tilde{a}=1}^{\widetilde{N}}
\left(
\sum_{\bs{r}\in\Lambda^{\ }_{r}}
\bs{r}
-
\sum_{\bs{R}\in\Lambda^{\ }_{R}}
\bs{R}
\right).
\end{split}
\label{eq: Tr X_r-X_R I}
\end{equation}
We are then going to show that
\begin{subequations}
\label{eq: Tr X_r-X_R II}
\begin{equation}
\begin{split}
\mathrm{Tr}\,
\left(
\widehat{\bs{X}}^{\ }_{r}
-
\widehat{\bs{X}}^{\ }_{R}
\right)=
\mathrm{i}
\sum_{\bs{k}\in\Lambda^{\star}_{\hbox{\tiny{BZ}}}}
\mathrm{tr}\,
\bs{A}^{\ }_{\bs{k}}
\end{split}
\label{eq: Tr X_r-X_R II a}
\end{equation}
where, in the thermodynamic limit $\mathcal{N}\to\infty$
and assuming smoothness of the $\bs{k}$ dependence
of the matrix elements~(\ref{eq: U relates psi to chi a}),
$\bs{A}^{\ }_{\bs{k}}$ is the $\widetilde{N}\times\widetilde{N}$
antisymmetric matrix with the components
\begin{equation}
\bs{A}^{\tilde{a}\tilde{b}}_{\bs{k}}:=
u^{\alpha\tilde{a}*}_{\bs{k}}\,
\partial^{\ }_{\bs{k}}\,
u^{\alpha\tilde{b}}_{\bs{k}},
\qquad
\tilde{a},\tilde{b}=1,\cdots,\widetilde{N}.
\label{eq: Tr X_r-X_R II b}
\end{equation}
\end{subequations}
The summation convention over repeated indices is implied.
Comments: (i) Equation~(\ref{eq: Tr X_r-X_R II}) follows from
the identity (the proof of which is postponed to
Sec.~\ref{appsubsec: Lattice discretization of the $1$- and $3$-bracket})
\begin{equation}
\widehat{\bs{X}}^{\ }_{r}=
\widehat{\bs{X}}^{\ }_{R}
+
|
\chi^{\tilde{a}}_{\bs{k}}
\rangle\,
\mathrm{i}\,
\bs{A}^{\tilde{a}\tilde{b}}_{\bs{k}}\,
\langle
\chi^{\tilde{b}}_{\bs{k}}
|.
\label{eq: comment (i) for one bracket}
\end{equation}
(ii) Equation~(\ref{eq: Tr X_r-X_R II})
holds for any choice of the boundary conditions.
(iii)
Equation~(\ref{eq: Tr X_r-X_R I})
is mathematically meaningless in the thermodynamic limit
$\mathcal{N}\to\infty$, for it involves the subtraction of two
ill-conditioned sums. 

\begin{proof}
First, we make two observations.
On the one hand, from the definition~(\ref{eq: def X_r})
\begin{equation}
\begin{split}
\mathrm{Tr}\,
\widehat{\bs{X}}^{\ }_{r}=&\,
\langle
W^{a}_{\bs{R}}
|\,
\widehat{\bs{X}}^{\ }_{r}\,
|
W^{a}_{\bs{R}}
\rangle
\\
\hbox{
\tiny
Eqs.~(\ref{eq: def X_r}) and
(\ref{eq: projection op in Wannier basis})
\qquad
     }
=&\,
\langle
W^{\tilde{a}}_{\bs{R}}
|\,
\hat{\bs{r}}\,
|
W^{\tilde{a}}_{\bs{R}}
\rangle
\\
\hbox{
\tiny
Eq.~(\ref{eq: Wannier in terms orbitals})
\qquad
     }
=&\,
\left[
\frac{1}{\mathcal{N}}\,
e^{+\mathrm{i}\bs{k}\cdot\left(\bs{R}-\bs{r}\right)}\,
u^{\alpha \tilde{a}*}_{\bs{k}}\,
\langle
\psi^{\alpha}_{\bs{r}}
|
\right]\,
\hat{\bs{r}}
\left[
\frac{1}{\mathcal{N}}\,
e^{-\mathrm{i}\bs{k}'\cdot\left(\bs{R}-\bs{r}'\right)}\,
u^{\alpha' \tilde{a}}_{\bs{k}'}\,
|
\psi^{\alpha'}_{\bs{r}'}
\rangle
\right]
\\
\hbox{
\tiny 
Eq.~(\ref{eq: spectral decomposition hat r})
\qquad
     }
=&\,
\left[
\frac{1}{\mathcal{N}}\,
e^{+\mathrm{i}\bs{k}\cdot\left(\bs{R}-\bs{r}\right)}\,
u^{\alpha \tilde{a}*}_{\bs{k}}\,
\langle
\psi^{\alpha}_{\bs{r}}
|
\right]\,
\bs{r}
\left[
\frac{1}{\mathcal{N}}\,
e^{-\mathrm{i}\bs{k}'\cdot\left(\bs{R}-\bs{r}'\right)}\,
u^{\alpha' \tilde{a}}_{\bs{k}'}\,
|
\psi^{\alpha'}_{\bs{r}'}
\rangle
\right]
\\
\hbox{
\tiny
$
\langle
\psi^{\alpha}_{\bs{r}}
|
\psi^{\alpha'}_{\bs{r}'}
\rangle
=
\delta^{\alpha,\alpha'}\,
\delta^{\ }_{\bs{r},\bs{r}'}\,
$\qquad
     }
=&\,
\sum_{\bs{r}\in\Lambda^{\ }_{r}}
\left[
\frac{1}{\mathcal{N}}\,
e^{+\mathrm{i}\bs{k}\cdot\bs{R}}\,
u^{\alpha \tilde{a}*}_{\bs{k}}\,
\right]\,
\bs{r}
\left[
\frac{1}{\mathcal{N}}\,
e^{-\mathrm{i}\bs{k}'\cdot\bs{R}}\,
u^{\alpha \tilde{a}}_{\bs{k}'}\,
\right].
\end{split}
\end{equation}
The implied summation over $\bs{R}$ produces the factor
$\mathcal{N}\,\delta^{\ }_{\bs{k},\bs{k}'}$.
We are left with the implied summations over the orbital
$\alpha=1,\cdots,N$, 
over the occupied bands 
$\tilde{a}=1,\cdots,\widetilde{N}$,
and over the momenta $\bs{k}\in\Lambda^{\star}_{\hbox{\tiny{BZ}}}$,
\begin{equation}
\begin{split}
\mathrm{Tr}\,
\widehat{\bs{X}}^{\ }_{r}=&\,
\frac{1}{\mathcal{N}}\,
\left(
u^{\alpha \tilde{a}*}_{\bs{k}}\,
u^{\alpha \tilde{a}}_{\bs{k}}
\right)
\sum_{\bs{r}\in\Lambda^{\ }_{r}}
\bs{r}
\\
\hbox{
\tiny
Eqs.~(\ref{eq: orthonormality u's on lattice a}) and
(\ref{eq: orthonormality u's on lattice b})
\qquad
     }
=&\,
\sum_{\tilde{a}=1}^{\widetilde{N}}
\sum_{\bs{r}\in\Lambda^{\ }_{r}}
\bs{r}.
\end{split}
\label{eq: Tr X r}
\end{equation}
On the other hand, the definition~(\ref{eq: def X_R})
immediately implies that
\begin{equation}
\begin{split}
\mathrm{Tr}\,
\widehat{\bs{X}}^{\ }_{R}=&\,
\langle
W^{a}_{\bs{R}}
|\,
\widehat{\bs{X}}^{\ }_{R}\,
|
W^{a}_{\bs{R}}
\rangle
\\
\hbox{
\tiny
Eqs.~(\ref{eq: def X_R}) and
(\ref{eq: projection op in Wannier basis})
\qquad
     }
=&\,
\langle
W^{\tilde{a}}_{\bs{R}}
|\,
\hat{\bs{R}}\,
|
W^{\tilde{a}}_{\bs{R}}
\rangle
\\
=&\,
\sum_{\tilde{a}=1}^{\widetilde{N}}
\sum_{\bs{R}\in\Lambda^{\ }_{r}}
\bs{R}.
\end{split}
\label{eq: Tr X R}
\end{equation}
Subtracting Eq.~(\ref{eq: Tr X R})
from Eq.~(\ref{eq: Tr X r})
delivers Eq.~(\ref{eq: Tr X_r-X_R I}).

Second, to prove Eq.~(\ref{eq: Tr X_r-X_R II}),
we start from
Eqs.~(\ref{eq: def X_r}) and
(\ref{eq: projection op in band basis})
to establish that
\begin{equation}
\begin{split}
\mathrm{Tr}\,
\widehat{\bs{X}}^{\ }_{r}=&\,
\langle
\chi^{a}_{\bs{k}}
|\,
\widehat{\bs{X}}^{\ }_{r}\,
|
\chi^{a}_{\bs{k}}
\rangle
\\
\hbox{
\tiny
Eqs.~(\ref{eq: def X_r}) and
(\ref{eq: projection op in band basis})
\qquad
     }
=&\,
\langle
\chi^{\tilde{a}}_{\bs{k}}
|\,
\hat{\bs{r}}\,
|
\chi^{\tilde{a}}_{\bs{k}}
\rangle
\\
=&\,
\langle
\chi^{\tilde{a}}_{\bs{k}}
|\,
\Big(
|
\psi^{\alpha}_{\bs{r}}
\rangle
\langle
\psi^{\alpha}_{\bs{r}}
|
\Big)\,
\hat{\bs{r}}\,
|
\chi^{\tilde{a}}_{\bs{k}}
\rangle
\\
\hbox{
\tiny 
Eq.~(\ref{eq: spectral decomposition hat r})
\qquad
     }
=&\,
\Big(
\langle
\chi^{\tilde{a}}_{\bs{k}}
|
\psi^{\alpha}_{\bs{r}}
\rangle
\Big)
\Big(
\bs{r}\,
\langle
\psi^{\alpha}_{\bs{r}}
|
\chi^{\tilde{a}}_{\bs{k}}
\rangle
\Big)
\\
\hbox{
\tiny
Eq.~(\ref{eq: U relates psi to chi c})
\qquad
     }
=&\,
\Big(
\langle
\chi^{\tilde{a}}_{\bs{k}}
|
\psi^{\alpha}_{\bs{r}}
\rangle
\Big)
\left(
\bs{r}\,
\frac{e^{+\mathrm{i}\bs{k}\cdot\bs{r}}}{\sqrt{\mathcal{N}}}\,
u^{\alpha\tilde{a}}_{\bs{k}}
\right)
\\
=&\,
\Big(
\langle
\chi^{\tilde{a}}_{\bs{k}}
|
\psi^{\alpha}_{\bs{r}}
\rangle
\Big)
\left[
\left(
-\mathrm{i}
\partial^{\ }_{\bs{k}}
\frac{e^{+\mathrm{i}\bs{k}\cdot\bs{r}}}{\sqrt{\mathcal{N}}}
\right)
u^{\alpha\tilde{a}}_{\bs{k}}
\right]
\\
=&\,
\left(
\frac{e^{-\mathrm{i}\bs{k}\cdot\bs{r}}}{\sqrt{\mathcal{N}}}\,
u^{\alpha\tilde{a}*}_{\bs{k}}
\right)
\left[
(-\mathrm{i})
\partial^{\ }_{\bs{k}}
\left(
\frac{e^{+\mathrm{i}\bs{k}\cdot\bs{r}}}{\sqrt{\mathcal{N}}}
u^{\alpha\tilde{a}}_{\bs{k}}
\right)
-
\left(
\frac{e^{+\mathrm{i}\bs{k}\cdot\bs{r}}}{\sqrt{\mathcal{N}}}
(-\mathrm{i})
\partial^{\ }_{\bs{k}}
u^{\alpha\tilde{a}}_{\bs{k}}
\right)
\right]
\\
=&\,
\langle
\chi^{\tilde{a}}_{\bs{k}}
|
\psi^{\alpha}_{\bs{r}}
\rangle
(-\mathrm{i})
\partial^{\ }_{\bs{k}}
\langle
\psi^{\alpha}_{\bs{r}}
|
\chi^{\tilde{a}}_{\bs{k}}
\rangle
+
\mathrm{i}\,
u^{\alpha\tilde{a}*}_{\bs{k}}
\partial^{\ }_{\bs{k}}
u^{\alpha\tilde{a}}_{\bs{k}}.
\label{eq: use product rule dif for Xr a}
\end{split}
\end{equation}
To prove
Eq.~(\ref{eq: Tr X_r-X_R II}),
it suffices to recognize that
\begin{equation}
\mathrm{i}\,
u^{\alpha\tilde{a}*}_{\bs{k}}
\partial^{\ }_{\bs{k}}
u^{\alpha\tilde{a}}_{\bs{k}}
=
\sum_{\bs{k}\in\Lambda^{\star}_{\hbox{\tiny{BZ}}}}
\mathrm{i}\,
\mathrm{tr}\,
\bs{A}^{\ }_{\bs{k}}
\end{equation}
and that, after insertion of the Fourier expansion
within the band basis%
~(\ref{eq: band basis}),
\begin{equation}
\begin{split}
\langle
\chi^{\tilde{a}}_{\bs{k}}
|
\psi^{\alpha}_{\bs{r}}
\rangle
(-\mathrm{i})
\partial^{\ }_{\bs{k}}
\langle
\psi^{\alpha}_{\bs{r}}
|
\chi^{\tilde{a}}_{\bs{k}}
\rangle
=&\,
\left(
\frac{e^{-\mathrm{i}\bs{k}\cdot\bs{R}'}}{\sqrt{\mathcal{N}}}
\langle
W^{\tilde{a}}_{\bs{R}'}
|
\psi^{\alpha}_{\bs{r}}
\rangle
\right)
\left(-\mathrm{i}\right)\partial^{\ }_{\bs{k}}\,
\left(
\frac{e^{+\mathrm{i}\bs{k}\cdot\bs{R}}}{\sqrt{\mathcal{N}}}
\langle
\psi^{\alpha}_{\bs{r}}
|
W^{\tilde{a}}_{\bs{R}}
\rangle
\right)
\\
=&\,
\left(
\frac{e^{-\mathrm{i}\bs{k}\cdot\bs{R}'}}{\sqrt{\mathcal{N}}}
\langle
W^{\tilde{a}}_{\bs{R}'}
|
\psi^{\alpha}_{\bs{r}}
\rangle
\right)
\left(
\bs{R}\,
\frac{e^{+\mathrm{i}\bs{k}\cdot\bs{R}}}{\sqrt{\mathcal{N}}}
\langle
\psi^{\alpha}_{\bs{r}}
|
W^{\tilde{a}}_{\bs{R}}
\rangle
\right)
\\
=&\,
\left(
\sum_{\bs{k}\in\Lambda^{\star}_{\hbox{\tiny{BZ}}}}
\frac{e^{-\mathrm{i}\bs{k}\cdot\left(\bs{R}'-\bs{R}\right)}}{\mathcal{N}}
\right)
\langle
W^{\tilde{a}}_{\bs{R}'}
|
\bs{R}\,
\Big(
|
\psi^{\alpha}_{\bs{r}}
\rangle
\langle
\psi^{\alpha}_{\bs{r}}
\Big)
|
W^{\tilde{a}}_{\bs{R}}
\rangle
\\
=&\,
\langle
W^{\tilde{a}}_{\bs{R}}
|
\bs{R}\,
|
W^{\tilde{a}}_{\bs{R}}
\rangle
\\
=&\,
\mathrm{Tr}\,
\widehat{\bs{X}}^{\ }_{R}.
\end{split}
\label{eq: use product rule dif for Xr b}
\end{equation}

\end{proof}

\subsection{
Lattice discretization
of the $2$-bracket of the projected position operator
           }

We are going to establish that the 2-bracket of the
projected positions operator~(\ref{eq: def X_r}) is
\begin{subequations}
\label{eq: 2 bracket Xr app}
\begin{equation}
\begin{split}
\epsilon^{\ }_{\mu\nu}\,
\widehat{X}^{\mu}_{r}\,
\widehat{X}^{\nu}_{r}
=&\,
|
\chi^{\tilde{a}}_{\bs{k}}
\rangle\,
\left(
-
F^{\tilde{a}\tilde{b}}_{\mu\nu;\bs{k}}
\right)\,
\langle
\chi^{\tilde{b}}_{\bs{k}}
|
\\
=&\,
|
W^{\tilde{a}}_{\bs{R}}
\rangle\,
\left(
-
\frac{e^{+\mathrm{i}\bs{k}\cdot(\bs{R}-\bs{R}')}}{\mathcal{N}}\,
F^{\tilde{a}\tilde{b}}_{\mu\nu;\bs{k}}
\right)\,
\langle
W^{\tilde{b}}_{\bs{R}'}
|
\end{split}
\end{equation}
where
\begin{equation}
F^{\tilde{a}\tilde{b}}_{\mu\nu;\bs{k}}=
\partial^{\ }_\mu\,
A^{\tilde{a}\tilde{b}}_{\bs{k};\nu}
-
\partial^{\ }_\nu\,
A^{\tilde{a}\tilde{b}}_{\bs{k};\mu}
+
\left[
A^{\ }_{\bs{k};\mu},
A^{\ }_{\bs{k};\nu}
\right]^{\tilde{a}\tilde{b}},
\qquad
\tilde{a},\tilde{b}=1,\cdots,\widetilde{N},
\qquad
\bs{k}\in\Lambda^{\star}_{\hbox{\tiny{BZ}}},
\end{equation}
\end{subequations}
in the thermodynamic limit $\mathcal{N}\to\infty$
and assuming smoothness of the $\bs{k}$ dependence
of the matrix elements~(\ref{eq: U relates psi to chi a}).
The summation convention over repeated indices is implied.
In contrast, the 2-bracket of the
projected positions operator~(\ref{eq: def X_R}) 
vanishes
\begin{equation}
\epsilon^{\ }_{\mu\nu}\,
\widehat{X}^{\mu}_{R}\,
\widehat{X}^{\nu}_{R}=
0.
\label{eq: 2 bracket XR app}
\end{equation}
Comments:
(i) No regularization is needed here.
(ii) Equation~(\ref{eq: 2 bracket Xr app})
holds for any choice of the boundary conditions.
\begin{proof}
We begin with the proof of Eq.~(\ref{eq: 2 bracket Xr app})
which we establish by computing the matrix elements of 
$
\widehat{X}^{\mu}_{r}\,
\widehat{X}^{\nu}_{r}
$
in the band basis~(\ref{eq: band basis}) in the Wannier
representation (as opposed to the momentum representation). 
For any triplet of pairs $a,a'=1,\cdots,N$,
$\bs{R},\bs{R}'\in\Lambda^{\ }_{R}$,
and $\mu,\nu=1,\cdots,d$,
we evaluate the matrix element of 
$\widehat{X}^{\mu}_{r}\,\widehat{X}^{\nu}_{r}$
in the Wannier basis given by 
\begin{equation}
\begin{split}
\langle
W^{a}_{\bs{R}}
|
\widehat{X}^{\mu}_{r}\,
\widehat{X}^{\nu}_{r}\,
|
W^{a'}_{\bs{R}'}
\rangle
=&\,
\langle
W^{a}_{\bs{R}}
|
\left(
\widehat{p}^{\ }_{\widetilde{N}}\,
\hat{r}^{\mu}\,
\widehat{p}^{\ }_{\widetilde{N}}
\right)
\,
\left(
\widehat{p}^{\ }_{\widetilde{N}}
\hat{r}^{\nu}\,
\widehat{p}^{\ }_{\widetilde{N}}
\right)\,
|
W^{a'}_{\bs{R}'}
\rangle
\\
=&\,
\delta^{a,\tilde{a}}\times
\delta^{a',\tilde{a}'}\times
\langle
W^{\tilde{a}}_{\bs{R}}
|\,
\hat{r}^{\mu}\,
\,
\widehat{p}^{\ }_{\widetilde{N}}\,
\hat{r}^{\nu}\,
|
W^{\tilde{a}'}_{\bs{R}'}
\rangle.
\end{split}
\end{equation}
With the help of
Eqs.~(\ref{eq: orbital basis})
and (\ref{eq: band basis}),
\begin{scriptsize}
\begin{equation}
\begin{split}
\langle
W^{\tilde{a}}_{\bs{R}}
|\,
\hat{r}^{\mu}\,
\,
\widehat{p}^{\ }_{\widetilde{N}}\,
\hat{r}^{\nu}\,
|
W^{\tilde{a}'}_{\bs{R}'}
\rangle
=&\,
\langle
W^{\tilde{a}}_{\bs{R}}
|\,
\left(\vphantom{\psi^{\alpha'}_{\bs{r}'}}
|
\psi^{\alpha}_{\bs{r}}
\rangle
\langle
\psi^{\alpha}_{\bs{r}}
|
\right)
\hat{r}^{\mu}\,
\,
\left(\vphantom{\psi^{\alpha'}_{\bs{r}'}}
|
\chi^{\tilde{b}}_{\bs{p}}
\rangle
\langle
\chi^{\tilde{b}}_{\bs{p}}
|
\right)
\hat{r}^{\nu}\,
\left(
|
\psi^{\alpha'}_{\bs{r}'}
\rangle
\langle
\psi^{\alpha'}_{\bs{r}'}
|
\right)
|
W^{\tilde{a}'}_{\bs{R}'}
\rangle
\\
\hbox{
\tiny
Eq.~(\ref{eq: spectral decomposition hat r})
\qquad
     }
=&\,
r^{\mu}\,
r^{\prime\nu}\,
\langle
W^{\tilde{a}}_{\bs{R}}
|
\psi^{\alpha}_{\bs{r}}
\rangle
\times
\langle
\psi^{\alpha}_{\bs{r}}
|
\chi^{\tilde{b}}_{\bs{p}}
\rangle
\times
\langle
\chi^{\tilde{b}}_{\bs{p}}
|
\psi^{\alpha'}_{\bs{r}'}
\rangle
\times
\langle
\psi^{\alpha'}_{\bs{r}'}
|
W^{\tilde{a}'}_{\bs{R}'}
\rangle
\\
\hbox{
\tiny
Eqs.%
~(\ref{eq: U relates psi to chi c})+(\ref{eq: Wannier in terms orbitals})
\quad
     }
=&\,
\frac{1}{\mathcal{N}^{3}}
r^{\mu}\,
r^{\prime\nu}\,
\left(
u^{\alpha\tilde{a}*}_{\bs{k}}\,
e^{+\mathrm{i}\bs{k}\cdot\left(\bs{R}-\bs{r}\right)}
\right)
\times
\left(
u^{\alpha\tilde{b}}_{\bs{p}}\,
e^{+\mathrm{i}\bs{p}\cdot\bs{r}}
\right)
\times
\left(
u^{\alpha'\tilde{b}*}_{\bs{p}}\,
e^{-\mathrm{i}\bs{p}\cdot\bs{r}'}
\right)
\times
\left(
u^{\alpha'\tilde{a}'}_{\bs{k}'}\,
e^{-\mathrm{i}\bs{k}'\cdot\left(\bs{R}'-\bs{r}'\right)}
\right)
\\
=&\,
\frac{1}{\mathcal{N}^{3}}
\left(
u^{\alpha\tilde{a}*}_{\bs{k}}\,
e^{+\mathrm{i}\bs{k}\cdot\bs{R}}\,
\frac{\partial}{\partial k^{\mu}}
e^{-\mathrm{i}\bs{k}\cdot\bs{r}}
\right)
\times
\left(
u^{\alpha\tilde{b}}_{\bs{p}}\,
e^{+\mathrm{i}\bs{p}\cdot\bs{r}}
\right)
\times
\left(
u^{\alpha'\tilde{b}*}_{\bs{p}}\,
e^{-\mathrm{i}\bs{p}\cdot\bs{r}'}
\right)
\times
\left(
u^{\alpha'\tilde{a}'}_{\bs{k}'}\,
e^{-\mathrm{i}\bs{k}'\cdot\bs{R}'}\,
\frac{\partial}{\partial k^{\prime\nu}}\,
e^{+\mathrm{i}\bs{k}'\cdot\bs{r}'}
\right).
\end{split}
\end{equation}
\end{scriptsize}

\noindent
We would like to perform the implicit sums over 
$\bs{r}\in\Lambda^{\ }_{r}$ and $\bs{r}'\in\Lambda^{\ }_{r}$.
To this end we use twice
the product rule for differentiation,
\begin{equation}
\begin{split}
f(\partial g) f'(\partial' g')=&\,
\left[
\partial(fg)
-
(\partial f)g
\right]
\left[
\partial'(f'g')
-
(\partial' f')g'
\right]
\\
=&\,
\partial(fg) \partial'(f'g')
-
\partial(fg) (\partial' f')g'
-
(\partial f)g \partial'(f'g')
+
(\partial f)g (\partial' f')g'
\\
=&\,
\partial \partial'(f g f' g')
-
\partial [f g(\partial'f') g']
-
\partial'[(\partial f) g f' g']
+
(\partial f) g (\partial' f) g'
\end{split}
\end{equation}
for any pair of functions $f$ and $g$ of one variable 
and for any  pair of functions $f'$ and $g'$ of another independent variable.
We find
\begin{scriptsize}
\begin{equation}
\begin{split}
\langle
W^{\tilde{a}}_{\bs{R}}
|
\widehat{X}^{\mu}_{r}\,
\widehat{X}^{\nu}_{r}\,
|
W^{\tilde{a}'}_{\bs{R}'}
\rangle
=&\,
+
\frac{1}{\mathcal{N}^{3}}
\frac{\partial}{\partial k^{\mu}}
\frac{\partial}{\partial k^{\prime\nu}}
\left[
\left(
u^{\alpha\tilde{a}*}_{\bs{k}}\,
e^{+\mathrm{i}\bs{k}\cdot\bs{R}}
\right)
\times
\left(
u^{\alpha\tilde{b}}_{\bs{p}}\,
e^{+\mathrm{i}(\bs{p}-\bs{k})\cdot\bs{r}}
\right)
\times
\left(
u^{\alpha'\tilde{b}*}_{\bs{p}}\,
e^{-\mathrm{i}(\bs{p}-\bs{k}')\cdot\bs{r}'}
\right)
\times
\left(
u^{\alpha'\tilde{a}'}_{\bs{k}'}\,
e^{-\mathrm{i}\bs{k}'\cdot\bs{R}'}
\right)
\right]
\\
&\,
-
\frac{1}{\mathcal{N}^{3}}
\frac{\partial}{\partial k^{\mu}}
\left\{
\left(
u^{\alpha\tilde{a}*}_{\bs{k}}\,
e^{+\mathrm{i}\bs{k}\cdot\bs{R}}
\right)
\times
\left(
u^{\alpha\tilde{b}}_{\bs{p}}\,
e^{+\mathrm{i}(\bs{p}-\bs{k})\cdot\bs{r}}
\right)
\times
\left(
u^{\alpha'\tilde{b}*}_{\bs{p}}\,
e^{-\mathrm{i}(\bs{p}-\bs{k}')\cdot\bs{r}'}
\right)
\times
\left[
\partial^{\prime}_{\nu}
\left(
u^{\alpha'\tilde{a}'}_{\bs{k}'}
e^{-\mathrm{i}\bs{k}'\cdot\bs{R}'}
\right)
\right]
\right\}
\\
&\,
-
\frac{1}{\mathcal{N}^{3}}
\frac{\partial}{\partial k^{\prime\nu}}
\left\{
\left[
\partial^{\ }_{\mu}
\left(
u^{\alpha\tilde{a}*}_{\bs{k}}\,
e^{+\mathrm{i}\bs{k}\cdot\bs{R}}
\right)
\right]
\times
\left(
u^{\alpha\tilde{b}}_{\bs{p}}\,
e^{+\mathrm{i}(\bs{p}-\bs{k})\cdot\bs{r}}
\right)
\times
\left(
u^{\alpha'\tilde{b}*}_{\bs{p}}\,
e^{-\mathrm{i}(\bs{p}-\bs{k}')\cdot\bs{r}'}
\right)
\times
\left(
u^{\alpha'\tilde{a}'}_{\bs{k}'}\,
e^{-\mathrm{i}\bs{k}'\cdot\bs{R}'}
\right)
\right\}
\\
&\,
+
\frac{1}{\mathcal{N}^{3}}
\left[
\partial^{\ }_{\mu}
\left(
u^{\alpha\tilde{a}*}_{\bs{k}}\,
e^{+\mathrm{i}\bs{k}\cdot\bs{R}}
\right)
\right]
\times
\left(
u^{\alpha\tilde{b}}_{\bs{p}}\,
e^{+\mathrm{i}(\bs{p}-\bs{k})\cdot\bs{r}}
\right)
\times
\left(
u^{\alpha'\tilde{b}*}_{\bs{p}}\,
e^{-\mathrm{i}(\bs{p}-\bs{k}')\cdot\bs{r}'}
\right)
\times
\left[
\partial^{\prime}_{\nu}
\left(
u^{\alpha'\tilde{a}'}_{\bs{k}'}\,
e^{-\mathrm{i}\bs{k}'\cdot\bs{R}'}
\right)
\right].
\end{split}
\end{equation}
\end{scriptsize}

\noindent
We perform the implicit sum over $\bs{r}\in\Lambda^{\ }_{r}$
on lines 1 and 2.
We perform the implicit sum over $\bs{r}'\in\Lambda^{\ }_{r}$
on line 3. We perform the implicit sum over the pair 
$\bs{r},\bs{r}'\in\Lambda^{\ }_{r}$ on line 4.
The implicit sum over $\bs{r}\in\Lambda^{\ }_{r}$ 
yields the multiplicative
factor $\mathcal{N}\delta^{\ }_{\bs{k},\bs{p}}$,
while the implicit sum over $\bs{r}'\in\Lambda^{\ }_{r}$ 
yields the multiplicative
factor $\mathcal{N}\delta^{\ }_{\bs{k}',\bs{p}}$. 
Thus,
\begin{equation}
\begin{split}
\langle
W^{\tilde{a}}_{\bs{R}}
|
\widehat{X}^{\mu}_{r}\,
\widehat{X}^{\nu}_{r}\,
|
W^{\tilde{a}'}_{\bs{R}'}
\rangle
=&\,
+
\frac{1}{\mathcal{N}^{2}}
\frac{\partial}{\partial k^{\mu}}
\frac{\partial}{\partial k^{\prime\nu}}
\left[
\left(
e^{+\mathrm{i}\bs{k}\cdot\bs{R}}
\delta^{\tilde{a}\tilde{b}}
\right)
\times
\left(
u^{\alpha'\tilde{b}*}_{\bs{k}}\,
e^{-\mathrm{i}(\bs{k}-\bs{k}')\cdot\bs{r}'}
\right)
\times
\left(
u^{\alpha'\tilde{a}'}_{\bs{k}'}\,
e^{-\mathrm{i}\bs{k}'\cdot\bs{R}'}
\right)
\right]
\\
&\,
-
\frac{1}{\mathcal{N}^{2}}
\frac{\partial}{\partial k^{\mu}}
\left\{
\left(
e^{+\mathrm{i}\bs{k}\cdot\bs{R}}
\delta^{\tilde{a}\tilde{b}}
\right)
\times
\left(
u^{\alpha'\tilde{b}*}_{\bs{k}}\,
e^{-\mathrm{i}(\bs{k}-\bs{k}')\cdot\bs{r}'}
\right)
\times
\left[
\partial^{\prime}_{\nu}
\left(
u^{\alpha'\tilde{a}'}_{\bs{k}'}
e^{-\mathrm{i}\bs{k}'\cdot\bs{R}'}
\right)
\right]
\right\}
\\
&\,
-
\frac{1}{\mathcal{N}^{2}}
\frac{\partial}{\partial k^{\prime\nu}}
\left\{
\left[
\partial^{\ }_{\mu}
\left(
u^{\alpha\tilde{a}*}_{\bs{k}}\,
e^{+\mathrm{i}\bs{k}\cdot\bs{R}}
\right)
\right]
\times
\left(
u^{\alpha\tilde{b}}_{\bs{k}'}\,
e^{+\mathrm{i}(\bs{k}'-\bs{k})\cdot\bs{r}}
\right)
\times
\left(
\delta^{\tilde{b}\tilde{a}'}
e^{-\mathrm{i}\bs{k}'\cdot\bs{R}'}
\right)
\right\}
\\
&\,
+
\frac{1}{\mathcal{N}}
\left\{
\left[
\partial^{\ }_{\mu}
\left(
u^{\alpha\tilde{a}*}_{\bs{k}}\,
e^{+\mathrm{i}\bs{k}\cdot\bs{R}}
\right)
\right]
u^{\alpha\tilde{b}}_{\bs{k}}
\right\}
\times
\left\{
\left[
u^{\alpha'\tilde{b}*}_{\bs{k}}\,
\partial^{\ }_{\nu}
\left(
u^{\alpha'\tilde{a}'}_{\bs{k}}\,
e^{-\mathrm{i}\bs{k}\cdot\bs{R}'}
\right)
\right]
\right\}.
\end{split}
\end{equation}
Performing the implicit sum over the projected band index 
$\tilde{b}=1,\cdots,\widetilde{N}$ on lines 1, 2, and 3
gives
\begin{equation}
\begin{split}
\langle
W^{\tilde{a}}_{\bs{R}}
|
\widehat{X}^{\mu}_{r}\,
\widehat{X}^{\nu}_{r}\,
|
W^{\tilde{a}'}_{\bs{R}'}
\rangle
=&\,
+
\frac{1}{\mathcal{N}^{2}}
\frac{\partial}{\partial k^{\mu}}
\frac{\partial}{\partial k^{\prime\nu}}
\left[
u^{\alpha'\tilde{a}*}_{\bs{k}}\,e^{+\mathrm{i}\bs{k}\cdot\bs{R}}
\times
e^{-\mathrm{i}(\bs{k}-\bs{k}')\cdot\bs{r}}
\times
u^{\alpha'\tilde{a}'}_{\bs{k}'}\,e^{-\mathrm{i}\bs{k}'\cdot\bs{R}'}
\right]
\\
&\,
-
\frac{1}{\mathcal{N}^{2}}
\frac{\partial}{\partial k^{\mu}}
\left\{
u^{\alpha'\tilde{a}*}_{\bs{k}}\,e^{+\mathrm{i}\bs{k}\cdot\bs{R}}
\times
e^{-\mathrm{i}(\bs{k}-\bs{k}')\cdot\bs{r}}
\times
\left[
\partial^{\prime}_{\nu}
u^{\alpha'\tilde{a}'}_{\bs{k}'}
e^{-\mathrm{i}\bs{k}'\cdot\bs{R}'}
\right]
\right\}
\\
&\,
-
\frac{1}{\mathcal{N}^{2}}
\frac{\partial}{\partial k^{\prime\nu}}
\left\{
\left[
\partial^{\ }_{\mu}
u^{\alpha\tilde{a}*}_{\bs{k}}\,
e^{+\mathrm{i}\bs{k}\cdot\bs{R}}
\right]
\times
e^{-\mathrm{i}(\bs{k}-\bs{k}')\cdot\bs{r}}
\times
u^{\alpha\tilde{a}'}_{\bs{k}'}\,
e^{-\mathrm{i}\bs{k}'\cdot\bs{R}'}
\right\}
\\
&\,
+
\frac{1}{\mathcal{N}}
\left\{
\left[
\partial^{\ }_{\mu}
\left(
u^{\alpha\tilde{a}*}_{\bs{k}}\,
e^{+\mathrm{i}\bs{k}\cdot\bs{R}}
\right)
\right]
u^{\alpha\tilde{b}}_{\bs{k}}
\right\}
\times
\left\{
\left[
u^{\alpha'\tilde{b}*}_{\bs{k}}\,
\partial^{\ }_{\nu}
\left(
u^{\alpha'\tilde{a}'}_{\bs{k}}\,
e^{-\mathrm{i}\bs{k}\cdot\bs{R}'}
\right)
\right]
\right\}.
\end{split}
\label{eq: Wannier matrix elements 2-bracket manipulation}
\end{equation}
For further simplification, we apply the identity
\begin{equation}
\begin{split}
\underline{
\partial^{\ }_{k}
\partial^{\ }_{k'}
\left[f(k)\, h(k,k')\, g(k')\right]
          }=&\,
+
\underline{
\partial^{\ }_{k}
\left[f(k)\, h(k,k')\, \partial^{\ }_{k'} g(k')\right]
          }
\\
&\,
+
\underline{
\partial^{\ }_{k'}
\left[g(k')\, h(k,k')\, \partial^{\ }_{k} f(k)\right]
          }
\\
&\,
+
f(k)\, 
g(k')\, 
\partial^{\ }_{k} 
\partial^{\ }_{k'} 
h(k,k')
\\
&\,
- 
h(k,k')\, 
\left[
\partial^{\ }_{k} f(k) \vphantom{k'}
\right]
\left[\partial^{\ }_{k} g(k')
\right]
\end{split}
\end{equation}
for the smooth function $f$, $g$, and $h$
to the first three lines of 
Eq.~\eqref{eq: Wannier matrix elements 2-bracket manipulation}.
We find 
\begin{equation}
\begin{split}
\langle
W^{\tilde{a}}_{\bs{R}}
|
\widehat{X}^{\mu}_{r}\,
\widehat{X}^{\nu}_{r}\,
|
W^{\tilde{a}'}_{\bs{R}'}
\rangle
=&\,
+
\frac{1}{\mathcal{N}^{2}}
\left[
\frac{\partial}{\partial k^{\mu}}
\frac{\partial}{\partial k^{\prime\nu}}
e^{-\mathrm{i}(\bs{k}-\bs{k}')\cdot\bs{r}}
\right]
\times
u^{\alpha\tilde{a}*}_{\bs{k}}\,e^{+\mathrm{i}\bs{k}\cdot\bs{R}}
\times
u^{\alpha\tilde{a}'}_{\bs{k}'}\,e^{-\mathrm{i}\bs{k}'\cdot\bs{R}'}
\\
&\,
-
\frac{1}{\mathcal{N}^{2}}
\,
e^{-\mathrm{i}(\bs{k}-\bs{k}')\cdot\bs{r}}
\left[
\partial^{\ }_{\mu}
\left(
u^{\alpha\tilde{a}*}_{\bs{k}}\,
e^{+\mathrm{i}\bs{k}\cdot\bs{R}}
\right)
\right]
\times
\left[
\partial^{\prime}_{\nu}
\left(
u^{\alpha\tilde{a}'}_{\bs{k}'}
e^{-\mathrm{i}\bs{k}'\cdot\bs{R}'}
\right)
\right]
\\
&\,
+
\frac{1}{\mathcal{N}}
\left[
\partial^{\ }_{\mu}
\left(
u^{\alpha\tilde{a}*}_{\bs{k}}\,
e^{+\mathrm{i}\bs{k}\cdot\bs{R}}
\right)
\right]
u^{\alpha\tilde{b}}_{\bs{k}}
\times
\left[
u^{\alpha'\tilde{b}*}_{\bs{k}}\,
\partial^{\ }_{\nu}
\left(
u^{\alpha'\tilde{a}'}_{\bs{k}}\,
e^{-\mathrm{i}\bs{k}\cdot\bs{R}'}
\right)
\right].
\end{split}
\end{equation}
We perform the derivatives on lines 2 and 3 first, 
which we then follow
up with the implicit sum over $\bs{r}$,
\begin{equation}
\begin{split}
\langle
W^{\tilde{a}}_{\bs{R}}
|
\widehat{X}^{\mu}_{r}\,
\widehat{X}^{\nu}_{r}\,
|
W^{\tilde{a}'}_{\bs{R}'}
\rangle
=&\,
+
\frac{1}{\mathcal{N}^{2}}
\left[
\frac{\partial}{\partial k^{\mu}}
\frac{\partial}{\partial k^{\prime\nu}}
e^{-\mathrm{i}(\bs{k}-\bs{k}')\cdot\bs{r}}
\right]
\times
u^{\alpha\tilde{a}*}_{\bs{k}}\,e^{+\mathrm{i}\bs{k}\cdot\bs{R}}
\times
u^{\alpha\tilde{a}'}_{\bs{k}'}\,e^{-\mathrm{i}\bs{k}'\cdot\bs{R}'}
\\
&\,
-
\frac{1}{\mathcal{N}}
\,
e^{+\mathrm{i}\bs{k}\cdot(\bs{R}-\bs{R}')}
\left[
\mathrm{i}R^{\ }_{\mu}
u^{\alpha\tilde{a}*}_{\bs{k}}
+
\partial^{\ }_{\mu}
u^{\alpha\tilde{a}*}_{\bs{k}}\,
\right]
\times
\left[
-\mathrm{i}R_{\nu}'
u^{\alpha\tilde{a}'}_{\bs{k}}
+
\partial^{\prime}_{\nu}
u^{\alpha\tilde{a}'}_{\bs{k}}
\right]
\\
&\,
+
\frac{1}{\mathcal{N}}
e^{+\mathrm{i}\bs{k}\cdot(\bs{R}-\bs{R}')}
\left[
\mathrm{i}R^{\ }_{\mu}
u^{\alpha\tilde{a}*}_{\bs{k}}
+
\partial^{\ }_{\mu}
u^{\alpha\tilde{a}*}_{\bs{k}}
\right]
u^{\alpha\tilde{b}}_{\bs{k}}
\times
u^{\alpha'\tilde{b}*}_{\bs{k}}\,
\left[
-\mathrm{i}R_{\nu}'
u^{\alpha'\tilde{a}'}_{\bs{k}}
+
\partial^{\ }_{\nu}
u^{\alpha'\tilde{a}'}_{\bs{k}}
\right].
\end{split}
\end{equation}
By making use of the orthonormality%
~(\ref{eq: orthonormality u's on lattice a})
and%
~(\ref{eq: orthonormality u's on lattice b})
and the definition~(\ref{eq: Tr X_r-X_R II b})
for the gauge connection,
we can expand the product of the bracketed terms on line 2 and 3
according to
\begin{equation}
\begin{split}
\langle
W^{\tilde{a}}_{\bs{R}}
|
\widehat{X}^{\mu}_{r}\,
\widehat{X}^{\nu}_{r}\,
|
W^{\tilde{a}'}_{\bs{R}'}
\rangle
=&\,
+
\frac{1}{\mathcal{N}^{2}}
\left[
\frac{\partial}{\partial k^{\mu}}
\frac{\partial}{\partial k^{\prime\nu}}
e^{-\mathrm{i}(\bs{k}-\bs{k}')\cdot\bs{r}}
\right]
\times
u^{\alpha\tilde{a}*}_{\bs{k}}\,e^{+\mathrm{i}\bs{k}\cdot\bs{R}}
\times
u^{\alpha\tilde{a}'}_{\bs{k}'}\,e^{-\mathrm{i}\bs{k}'\cdot\bs{R}'}
\\
&\,
-
\frac{e^{\mathrm{i}\bs{k}\cdot(\bs{R}-\bs{R}')}}{\mathcal{N}}
\left[
\underline{
R^{\     }_{\mu}\,
R^{\prime}_{\nu}\,
\delta^{\tilde{a}\tilde{a}'}
+
\mathrm{i}\,
R^{\ }_{\mu}\,
A^{\tilde{a}\tilde{a}'}_{\nu;\bs{k}}
+
\mathrm{i}\,
R^{\prime}_{\nu}\,
A^{\tilde{a}\tilde{a}'}_{\mu;\bs{k}}
          }
+
\left(
\partial^{\ }_{\mu}
u^{\alpha\tilde{a}*}_{\bs{k}}
\right)
\left(
\partial^{\ }_{\nu}
u^{\alpha\tilde{a}'}_{\bs{k}}
\right)
\right]
\\
&\,
+
\frac{e^{\mathrm{i}\bs{k}\cdot(\bs{R}-\bs{R}')}}{\mathcal{N}}
\left(
\mathrm{i}R^{\ }_{\mu}
\delta^{\tilde{a}\tilde{b}}
-
A^{\tilde{a}\tilde{b}}_{\mu;\bs{k}}
\right)
\times
\left(
-\mathrm{i}R_{\nu}'
\delta^{\tilde{b}\tilde{a}'}
+
A^{\tilde{b}\tilde{a}'}_{\nu;\bs{k}}
\right).
\end{split}
\end{equation}
Terms that have been underlined on line 2 cancel with line 3, 
leaving us with
\begin{equation}
\begin{split}
\langle
W^{\tilde{a}}_{\bs{R}}
|
\widehat{X}^{\mu}_{r}\,
\widehat{X}^{\nu}_{r}\,
|
W^{\tilde{a}'}_{\bs{R}'}
\rangle
=&\,
+
\frac{1}{\mathcal{N}^{2}}
\left[
\frac{\partial}{\partial k^{\mu}}
\frac{\partial}{\partial k^{\prime\nu}}
e^{-\mathrm{i}(\bs{k}-\bs{k}')\cdot\bs{r}}
\right]
\times
u^{\alpha\tilde{a}*}_{\bs{k}}\,e^{+\mathrm{i}\bs{k}\cdot\bs{R}}
\times
u^{\alpha\tilde{a}'}_{\bs{k}'}\,e^{-\mathrm{i}\bs{k}'\cdot\bs{R}'}
\\
&\,
-
\frac{e^{\mathrm{i}\bs{k}\cdot(\bs{R}-\bs{R}')}}{\mathcal{N}}\,
\left[
\left(
\partial^{\ }_{\mu}
u^{\alpha\tilde{a}*}_{\bs{k}}
\right)
\left(
\partial^{\ }_{\nu}
u^{\alpha\tilde{a}'}_{\bs{k}}
\right)
+
A^{\tilde{a}\tilde{b} }_{\mu;\bs{k}}\,
A^{\tilde{b}\tilde{a}'}_{\nu;\bs{k}}
\right].
\end{split}
\end{equation}
Here we would have to stop if we do not want to anti-symmetrize the indices 
$\mu=1,\cdots,d$ and $\nu=1,\cdots$; doing so, however, yields
\begin{equation}
\begin{split}
\epsilon_{\mu\nu}\,
\langle
W^{\tilde{a}}_{\bs{R}}
|
\widehat{X}^{\mu}_{r}\,
\widehat{X}^{\nu}_{r}\,
|
W^{\tilde{a}'}_{\bs{R}'}
\rangle
=&\,
-
\frac{e^{\mathrm{i}\bs{k}\cdot(\bs{R}-\bs{R}')}}{\mathcal{N}}
\left[
\epsilon_{\mu\nu}
\epsilon_{\mu\nu}
\partial^{\ }_{\mu}
A^{\tilde{b}\tilde{a}'}_{\nu;\bs{k}}
+
A^{\tilde{a}\tilde{b} }_{\mu;\bs{k}}\,
A^{\tilde{b}\tilde{a}'}_{\nu;\bs{k}}
\right]
\\
=&\,
-
\frac{e^{+\mathrm{i}\bs{k}\cdot(\bs{R}-\bs{R}')}}{\mathcal{N}}\,
F^{\tilde{a}\tilde{a}'}_{\mu\nu;\bs{k}}.
\end{split}
\end{equation}

We continue with the proof of Eq.~(\ref{eq: 2 bracket XR app}),
which we establish by computing the matrix elements of 
$
\widehat{X}^{\mu}_{r}\,
\widehat{X}^{\nu}_{r}
$
in the projected band basis~(\ref{eq: band basis}) in the Wannier
representation (as opposed to the momentum representation). 
For any triplet of pair $a,a'=1,\cdots,N$,
$\bs{R},\bs{R}'\in\Lambda^{\ }_{R}$,
and $\mu,\nu=1,\cdots,d$,
we evaluate the matrix element of 
$\widehat{X}^{\mu}_{R}\,\widehat{X}^{\nu}_{R}$
in the Wannier basis given by
\begin{equation}
\begin{split}
\langle
W^{a}_{\bs{R}}
|
\widehat{X}^{\mu}_{R}\,
\widehat{X}^{\nu}_{R}\,
|
W^{a'}_{\bs{R}'}
\rangle
=&\,
\delta^{a,\tilde{a}}\times
\delta^{a',\tilde{a}'}\times
\langle
W^{a}_{\bs{R}}
|
\left(
\widehat{p}^{\ }_{\widetilde{N}}\,
\hat{R}^{\mu}\,
\widehat{p}^{\ }_{\widetilde{N}}
\right)
\,
\left(
p^{\ }_{\widetilde{N}}
\hat{R}^{\nu}\,
p^{\ }_{\widetilde{N}}
\right)\,
|
W^{a'}_{\bs{R}'}
\rangle
\\
=&\,
\delta^{a,\tilde{a}}\times
\delta^{a',\tilde{a}'}\times
\langle
W^{\tilde{a}}_{\bs{R}}
|\,
\hat{R}^{\mu}\,
\,
p^{\ }_{\widetilde{N}}\,
\hat{R}^{\nu}\,
|
W^{\tilde{a}'}_{\bs{R}'}
\rangle
\\
=&\,
\delta^{a,\tilde{a}}\times
\delta^{a',\tilde{a}'}\times
\langle
W^{\tilde{a}}_{\bs{R}}
|\,
\hat{R}^{\mu}\,
\,
\left(
|
W^{\tilde{a}''}_{\bs{R}''}
\rangle
\langle
W^{\tilde{a}''}_{\bs{R}''}
|
\right)
\hat{R}^{\nu}\,
|
W^{\tilde{a}'}_{\bs{R}'}
\rangle
\\
\hbox{
\tiny 
Eq.~(\ref{eq: spectral decomposition hat R})
\qquad
     }
=&\,
\delta^{a,\tilde{a}}\times
\delta^{a',\tilde{a}'}\times
R^{\mu}\,
R^{\nu}\,
\langle
W^{\tilde{a}}_{\bs{R}}
|
W^{\tilde{a}''}_{\bs{R}''}
\rangle
\langle
W^{\tilde{a}''}_{\bs{R}''}
|
W^{\tilde{a}'}_{\bs{R}'}
\rangle
\\
=&\,
\delta^{a,\tilde{a}}\times
\delta^{a',\tilde{a}'}\times
R^{\mu}\,
R^{\nu}\,
\delta^{\tilde{a},\tilde{a}'}\,
\delta^{\ }_{\bs{R},\bs{R}'}.
\end{split}
\end{equation}
Antisymmetrization yields Eq.~(\ref{eq: 2 bracket XR app})

\end{proof}

\subsection{
Lattice discretization
of the $1$- and $3$-bracket of the projected position operator
           }
\label{appsubsec: Lattice discretization of the $1$- and $3$-bracket}

We are going to show that, 
in the thermodynamic limit $\mathcal{N}\to\infty$
and assuming smoothness of the $\bs{k}$ dependence
of the matrix elements~(\ref{eq: U relates psi to chi a}),
\begin{subequations}
\label{eq: 3 bracket matrix elements}
\begin{equation}
\begin{split}
\widehat{\bs{X}}^{\ }_{r}=&\,
|
\chi^{\tilde{a}}_{\bs{k}}
\rangle\,
\left(
\frac{
e^{-\mathrm{i}(\bs{k}-\bs{k}')\cdot\bs{r}}
     }
     {
\mathcal{N}
     }\,
\bs{r}\,
u^{\alpha\tilde{a}*}_{\bs{k}}\,
u^{\alpha\tilde{a}'}_{\bs{k}'}
\right)
\langle\,
\chi^{\tilde{a}'}_{\bs{k}'}
|
\\
=&\,
|
\chi^{\tilde{a}}_{\bs{k}}
\rangle\,
\left(
\frac{
e^{-\mathrm{i}(\bs{k}-\bs{k}')\cdot\bs{R}}
     }
     {
\mathcal{N}
     }\,
\bs{R}\,
\delta^{\tilde{a}\tilde{a}'}
+
\mathrm{i}\,
\bs{A}^{\tilde{a}\tilde{a}'}_{\bs{k}}\,
\delta^{\ }_{\bs{k},\bs{k}'}
\right)
\langle\,
\chi^{\tilde{a}'}_{\bs{k}'}
|
\\
=&\,
|
\chi^{\tilde{a}}_{\bs{k}}
\rangle\,
\left[
\frac{e^{-\mathrm{i}\bs{k}\cdot\bs{R}'}}{\sqrt{\mathcal{N}}}
\left(
\delta^{\tilde{a},\tilde{a}'}
\bs{R}'
+
\mathrm{i}\,
\bs{A}^{\tilde{a}\tilde{a}'}_{\bs{k}}
\right)
\right]
\langle
W^{\tilde{a}'}_{\bs{R}'}
|
\\
=&\,
|
W^{\tilde{a}}_{\bs{R}}
\rangle\,
\left[
\frac{
e^{+\mathrm{i}\bs{k}\cdot(\bs{R}-\bs{R}')}
     }
     {
\mathcal{N}
     }
\left(
\delta^{\tilde{a},\tilde{a}'}
\bs{R}'
+
\mathrm{i}\,
\bs{A}^{\tilde{a}\tilde{a}'}_{\bs{k}}
\right)
\right]
\langle
W^{\tilde{a}'}_{\bs{R}'}
|
\end{split}
\label{eq: mixed Wannier band matrix of X}
\end{equation}
while
\begin{equation}
\begin{split}
\epsilon^{\ }_{\mu\nu\lambda}\,
\widehat{X}^{\mu}_{r}\,
\widehat{X}^{\nu}_{r}\,
\widehat{X}^{\lambda}_{r}
=&\,
|
W^{\tilde{a}}_{\bs{R} }
\rangle
\left\{
\frac{1}{2}\,
\epsilon^{\mu\nu\lambda}\,
\frac{
e^{+\mathrm{i}\bs{k}\cdot(\bs{R}-\bs{R}')}
     }
     {
\mathcal{N}
     }
\left[
(-)
F^{\tilde{a}\tilde{b}}_{\mu\nu;\bs{k}}\,
\left(
\delta^{\tilde{b},\tilde{a}'}\,
R^{\prime}_{\lambda}
+
\mathrm{i}\,
A^{\tilde{b}\tilde{a}'}_{\lambda;\bs{k}}
\right)
\right]
\right\}
\langle
W^{\tilde{a}'}_{\bs{R}'}
|.
\end{split}
\label{eq: 3 bracket matrix elements b}
\end{equation}
\end{subequations}
The summation convention over repeated indices is implied.
Comments:
(i) A regularization is needed to dispose of the explicit $\bs{R}$
dependence in the position representation of the covariant derivative.
(ii) Equation~(\ref{eq: 3 bracket matrix elements})
holds for any choice of the boundary conditions.
The equality between the first and second right-hand side of
Eq.~(\ref{eq: mixed Wannier band matrix of X})
implies that we can do the identification
\begin{equation}
-\mathrm{i}\,
\partial^{\ }_{\bs{k}'}
\left(
\sum_{\bs{r}\in\Lambda^{\ }_{r}}
\frac{
e^{-\mathrm{i}(\bs{k}-\bs{k}')\cdot\bs{r}}
     }
     {
\mathcal{N}
     }\,
u^{\alpha\tilde{a}*}_{\bs{k}}\,
u^{\alpha\tilde{a}'}_{\bs{k}'}
\right)
+
\left(
\sum_{\bs{r}\in\Lambda^{\ }_{r}}
\frac{
e^{-\mathrm{i}(\bs{k}-\bs{k}')\cdot\bs{r}}
     }
     {
\mathcal{N}
     }\,
u^{\alpha\tilde{a}*}_{\bs{k}}\,
u^{\alpha\tilde{a}'}_{\bs{k}'}
\right)\,
\mathrm{i}\,
\partial^{\ }_{\bs{k}'}
\longleftrightarrow
\frac{
e^{-\mathrm{i}(\bs{k}-\bs{k}')\cdot\bs{R}}
     }
     {
\mathcal{N}
     }\,
\bs{R}\,
\delta^{\tilde{a}\tilde{a}'}
\end{equation}
which will become handy to go back to a formulation in
the continuum for both position and momentum that does not
assume the vanishing of boundary terms.
(iii) Had we chosen to represent the 3-bracket in the Bloch basis,
we could have either written 
\begin{subequations}
\label{eq: 3 bracket matrix elements in Bloch basis}
\begin{equation}
\begin{split}
\epsilon^{\ }_{\mu\nu\lambda}\,
\widehat{X}^{\mu}_{r}\,
\widehat{X}^{\nu}_{r}\,
\widehat{X}^{\lambda}_{r}
=&\,
|
\chi^{\tilde{a}}_{\bs{k}}
\rangle
\left(
\frac{1}{2}\,
\epsilon^{\mu\nu\lambda}\,
(-)
F^{\tilde{a}\tilde{b}}_{\mu\nu;\bs{k}}\,
\frac{
e^{-\mathrm{i}(\bs{k}-\bs{k}')\cdot\bs{r}}
     }
     {
\mathcal{N}
     }\,
r^{\ }_{\lambda}\,
u^{\alpha\tilde{b}*}_{\bs{k}}\,
u^{\alpha\tilde{a}'}_{\bs{k}'}
\right)
\langle
\chi^{\tilde{a}'}_{\bs{k}'}
|
\end{split}
\label{eq: 3 bracket matrix elements in Bloch basis a}
\end{equation}
had we opted not to use the product rule for differentiation
or
\begin{equation}
\begin{split}
\epsilon^{\ }_{\mu\nu\lambda}\,
\widehat{X}^{\mu}_{r}\,
\widehat{X}^{\nu}_{r}\,
\widehat{X}^{\lambda}_{r}
=&\,
|
\chi^{\tilde{a}}_{\bs{k}}
\rangle
\left[
\frac{1}{2}\,
\epsilon^{\mu\nu\lambda}\,
(-)
F^{\tilde{a}\tilde{b}}_{\mu\nu;\bs{k}}\,
\left(
\frac{
e^{-\mathrm{i}(\bs{k}-\bs{k}')\cdot\bs{R}}
     }
     {
\mathcal{N}
     }\,
R^{\ }_{\lambda}\,
\delta^{\tilde{a}\tilde{a}'}
+
\mathrm{i}\,
A^{\tilde{a}\tilde{a}'}_{\bs{k};\lambda}\,
\delta^{\ }_{\bs{k},\bs{k}'}
\right)
\right]
\langle
\chi^{\tilde{a}'}_{\bs{k}'}
|
\end{split}
\label{eq: 3 bracket matrix elements in Bloch basis b}
\end{equation}
\end{subequations}
had we opted to use the product rule for differentiation.
However, the representation on the first line of Eq.%
~(\ref{eq: mixed Wannier band matrix of X})
as well as Eq.%
~(\ref{eq: 3 bracket matrix elements in Bloch basis a})
are meaningless in the thermodynamic limit $\mathcal{N}\to\infty$. 
They fail to separate a finite and
physically meaningful contribution to the trace
of $n$-brackets.

\begin{proof}
Needed is
\begin{equation}
\begin{split}
\langle
W^{a}_{\bs{R} }
|
\left(
\epsilon^{\ }_{\mu\nu\lambda}\,
\widehat{X}^{\mu}_{r}\,
\widehat{X}^{\nu}_{r}\,
\widehat{X}^{\lambda}_{r}
\right)
|
W^{a'}_{\bs{R}'}
\rangle=&
\frac{1}{2}\,
\epsilon^{\ }_{\mu\nu\lambda}\,
\langle
W^{a}_{\bs{R} }
|
\left(
\left[
\widehat{X}^{\mu}_{r},
\widehat{X}^{\nu}_{r}
\right]\,
\widehat{X}^{\lambda}_{r}
\right)
|
W^{a'}_{\bs{R}'}
\rangle
\end{split}
\end{equation}
for any pair $a,a'=1,\cdots,N$ and
any pair $\bs{R},\bs{R}'\in\Lambda^{\ }_{R}$.
With the Fourier expansion within the band basis%
~(\ref{eq: band basis})
and the matrix elements%
~(\ref{eq: 2 bracket Xr app}),
there follows
\begin{equation}
\begin{split}
\langle
W^{a}_{\bs{R} }
|
\left[
\widehat{X}^{\mu}_{r}\,,
\widehat{X}^{\nu}_{r}\,
\right]\,
\widehat{X}^{\lambda}_{r}\,
|
W^{a'}_{\bs{R}'}
\rangle
=&\,
\langle
W^{a}_{\bs{R} }
|
\left[
|
\chi^{\tilde{b}}_{\bs{k}}
\rangle\,
\left(
-
F^{\tilde{b}\tilde{b}'}_{\mu\nu;\bs{k}}\
\right)\,
\langle
\chi^{\tilde{b}'}_{\bs{k}}
|
\right]\,
\widehat{X}^{\lambda}_{r}\,
|
W^{a'}_{\bs{R}'}
\rangle
\\
=&\,
\delta^{a,\tilde{a}}\times
\delta^{a',\tilde{a}'}\times
\langle
W^{\tilde{a}}_{\bs{R} }
|
\chi^{\tilde{a}}_{\bs{k}}
\rangle
\times
\left(
-
F^{\tilde{a}\tilde{b}}_{\mu\nu;\bs{k}}\
\right)
\times
\langle
\chi^{\tilde{b}}_{\bs{k}}
|
\widehat{X}^{\lambda}_{r}\,
|
W^{\tilde{a}'}_{\bs{R}'}
\rangle
\end{split}
\label{eq: step 1 proof for 3 bracket}
\end{equation}
for any pair $a,a'=1,\cdots,N$ and 
for any pair $\bs{R},\bs{R}'\in\Lambda^{\ }_{R}$.
With the Fourier expansion within the band basis%
~(\ref{eq: band basis}),
\begin{equation}
\langle
W^{a}_{\bs{R} }
|
\chi^{\tilde{a}}_{\bs{k}}
\rangle=
\delta^{a,\tilde{a}}\,
\frac{e^{+\mathrm{i}\bs{k}\cdot\bs{R}}}{\sqrt{\mathcal{N}}}.
\label{eq: step 2 proof for 3 bracket}
\end{equation}
Equations%
~(\ref{eq: mixed Wannier band matrix of X}),
(\ref{eq: step 1 proof for 3 bracket}),
and
(\ref{eq: step 2 proof for 3 bracket})
imply 
Eq.~(\ref{eq: 3 bracket matrix elements}).

The proof of
Eq.~(\ref{eq: mixed Wannier band matrix of X})
is done along the same lines as in
Sec.~\ref{subsec: Lattice regularization of the 1 bracket}.
We choose the pair $a,a'=1,\cdots,N$ and
the pair $\bs{k}\in\Lambda^{\star}_{\hbox{\tiny{BZ}}}$,
$\bs{R}\in\Lambda^{\ }_{R}$.
With the help of  Eqs.~(\ref{eq: def X_r}) and
(\ref{eq: projection op in band basis})
\begin{equation}
\begin{split}
\langle
\chi^{a}_{\bs{k}}
|\,
\widehat{\bs{X}}^{\ }_{r}\,
|
W^{a'}_{\bs{R}}
\rangle
=&\,
\delta^{a,\tilde{a}}\times
\delta^{a',\tilde{a}'}\times
\langle
\chi^{\tilde{a}}_{\bs{k}}
|\,
\hat{\bs{r}}\,
|
W^{\tilde{a}'}_{\bs{R}}
\rangle.
\end{split}
\end{equation}
In turn, for any pair 
$\tilde{a},\tilde{a}'=1,\cdots,\widetilde{N}$,
\begin{equation}
\begin{split}
\langle
\chi^{\tilde{a}}_{\bs{k}}
|\,
\hat{\bs{r}}\,
|
W^{\tilde{a}'}_{\bs{R}}
\rangle
=&\,
\langle
\chi^{\tilde{a}}_{\bs{k}}
|\,
\Big(
|
\psi^{\alpha}_{\bs{r}}
\rangle
\langle
\psi^{\alpha}_{\bs{r}}
|
\Big)\,
\hat{\bs{r}}\,
|
W^{\tilde{a}'}_{\bs{R}}
\rangle
\\
\hbox{
\tiny 
Eq.~(\ref{eq: spectral decomposition hat r})
\qquad
     }
=&\,
\Big(
\bs{r}\,
\langle
\chi^{\tilde{a}}_{\bs{k}}
|
\psi^{\alpha}_{\bs{r}}
\rangle
\Big)
\Big(
\langle
\psi^{\alpha}_{\bs{r}}
|
W^{\tilde{a}'}_{\bs{R}}
\rangle
\Big)
\\
\hbox{
\tiny
Eq.~(\ref{eq: U relates psi to chi c})
and
Eq.~(\ref{eq: Wannier in terms orbitals})
\qquad
     }
=&\,
\Big(
\bs{r}\,
\frac{e^{-\mathrm{i}\bs{k}\cdot\bs{r}}}{\sqrt{\mathcal{N}}}
u^{\alpha\tilde{a}*}_{\bs{k}}
\Big)
\left(
\frac{
e^{-\mathrm{i}\bs{k}'\cdot(\bs{R}-\bs{r})}
     }
     {
\mathcal{N}
     }\,
u^{\alpha\tilde{a}'}_{\bs{k}'}
\right)
\\
=&\,
\left[
+\mathrm{i}\,
\partial^{\ }_{\bs{k}}
\Big(
\frac{
e^{-\mathrm{i}\bs{k}\cdot\bs{r}}
     }
     {
\sqrt{\mathcal{N}}
     }
u^{\alpha\tilde{a}*}_{\bs{k}}
\Big)
+
\frac{
e^{-\mathrm{i}\bs{k}\cdot\bs{r}}
     }
     {
\sqrt{\mathcal{N}}
     }
\left(
-\mathrm{i}\,
\partial^{\ }_{\bs{k}}
u^{\alpha\tilde{a}*}_{\bs{k}}
\right)
\right]
\left(
\frac{
e^{-\mathrm{i}\bs{k}'\cdot(\bs{R}-\bs{r})}
     }
     {
\mathcal{N}
     }\,
u^{\alpha\tilde{a}'}_{\bs{k}'}
\right).
\end{split}
\end{equation}
To proceed, we reexpress the first term on the right-hand side
as a product of two overlaps to be differentiated with respect
to momentum, while we perform the implicit sum over
$\bs{r}\in\Lambda^{\ }_{r}$
on the second term on the right-hand side. This 
implicit sum over $\bs{r}\in\Lambda^{\ }_{r}$
produces the multiplicative factor
$\mathcal{N}\times\delta^{\ }_{\bs{k},\bs{k}'}$.
Thus,
\begin{equation}
\begin{split}
\langle
\chi^{\tilde{a}}_{\bs{k}}
|\,
\hat{\bs{r}}\,
|
W^{\tilde{a}'}_{\bs{R}}
\rangle
=&\,
\left[
+\mathrm{i}\,
\partial^{\ }_{\bs{k}}
\Big(
\langle
\chi^{\tilde{a}}_{\bs{k}}
|
\psi^{\alpha}_{\bs{r}}
\rangle
\langle
\psi^{\alpha}_{\bs{r}}
|
W^{\tilde{a}'}_{\bs{R}}
\rangle
\Big)
+
\frac{
e^{-\mathrm{i}\bs{k}\cdot\bs{R}}
     }
     {
\sqrt{\mathcal{N}}
     }
\left(
-\mathrm{i}\,
\partial^{\ }_{\bs{k}}
u^{\alpha\tilde{a}*}_{\bs{k}}
\right)\,
u^{\alpha\tilde{a}'}_{\bs{k}}
\right].
\end{split}
\end{equation}
The implicit sums over 
$\bs{r}\in\Lambda^{\ }_{r}$
and $\alpha=1,\cdots,N$
in the first term on the right-hand side delivers
the resolution of the identity, while we
can use the orthonormality%
~(\ref{eq: orthonormality u's on lattice a})
and%
~(\ref{eq: orthonormality u's on lattice b})
to move the momentum gradient 
in the second term on the right-hand side.
This manipulation gives
\begin{equation}
\begin{split}
\langle
\chi^{\tilde{a}}_{\bs{k}}
|\,
\hat{\bs{r}}\,
|
W^{\tilde{a}'}_{\bs{R}}
\rangle
=&\,
\left[
+\mathrm{i}\,
\partial^{\ }_{\bs{k}}
\Big(
\langle
\chi^{\tilde{a}}_{\bs{k}}
|
W^{\tilde{a}'}_{\bs{R}}
\rangle
\Big)
+
\frac{
e^{-\mathrm{i}\bs{k}\cdot\bs{R}}
     }
     {
\sqrt{\mathcal{N}}
     }\,
u^{\alpha\tilde{a}*}_{\bs{k}}\,
\left(
\mathrm{i}\,
\partial^{\ }_{\bs{k}}
u^{\alpha\tilde{a}'}_{\bs{k}}
\right)
\right].
\end{split}
\end{equation}
Equations~(\ref{eq: U relates psi to chi c})
and (\ref{eq: Tr X_r-X_R II b}) deliver
\begin{equation}
\begin{split}
\langle
\chi^{\tilde{a}}_{\bs{k}}
|\,
\hat{\bs{r}}\,
|
W^{\tilde{a}'}_{\bs{R}}
\rangle
=&\,
\left[
+\mathrm{i}\,
\partial^{\ }_{\bs{k}}
\left(
\frac{
e^{-\mathrm{i}\bs{k}\cdot\bs{R}}
     }
     {
\sqrt{\mathcal{N}}
     }\,
\right)\,
\delta^{\tilde{a},\tilde{a}'}\,
+
\frac{
e^{-\mathrm{i}\bs{k}\cdot\bs{R}}
     }
     {
\sqrt{\mathcal{N}}
     }\,
\mathrm{i}\,
\bs{A}^{\tilde{a}\tilde{a}'}_{\bs{k}}
\right].
\end{split}
\end{equation}
We conclude with
\begin{equation}
\begin{split}
\langle
\chi^{\tilde{a}}_{\bs{k}}
|\,
\hat{\bs{r}}\,
|
W^{\tilde{a}'}_{\bs{R}}
\rangle
=&\,
\frac{
e^{-\mathrm{i}\bs{k}\cdot\bs{R}}
     }
     {
\sqrt{\mathcal{N}}
     }\,
\left(
\bs{R}\,
\delta^{\tilde{a},\tilde{a}'}\,
+
\mathrm{i}\,
\bs{A}^{\tilde{a}\tilde{a}'}_{\bs{k}}
\right).
\end{split}
\end{equation}

The proof of
Eq.~(\ref{eq: 3 bracket matrix elements in Bloch basis})
starts from suitably modifying
Eq.~(\ref{eq: step 1 proof for 3 bracket})
according to
\begin{equation}
\begin{split}
\langle
\chi^{a}_{\bs{k} }
|
\left[
\widehat{X}^{\mu}_{r}\,,
\widehat{X}^{\nu}_{r}\,
\right]\,
\widehat{X}^{\lambda}_{r}\,
|
\chi^{a'}_{\bs{k}'}
\rangle
=&\,
\langle
\chi^{a}_{\bs{k} }
|
\left[
|
\chi^{\tilde{b}}_{\bs{p}}
\rangle\,
\left(
-
F^{\tilde{b}\tilde{b}'}_{\mu\nu;\bs{p}}\
\right)\,
\langle
\chi^{\tilde{b}'}_{\bs{p}}
|
\right]\,
\widehat{X}^{\lambda}_{r}\,
|
\chi^{a'}_{\bs{k}'}
\rangle
\\
=&\,
\delta^{a,\tilde{b}}\times
\delta^{a',\tilde{a}'}\times
\left(
-
F^{\tilde{a}\tilde{b}}_{\mu\nu;\bs{k}}\
\right)
\times
\langle
\chi^{\tilde{b}}_{\bs{k}}
|
\widehat{X}^{\lambda}_{r}\,
|
\chi^{\tilde{a}'}_{\bs{k}'}
\rangle
\end{split}
\end{equation}
where we can either choose the representation
\begin{equation}
\langle
\chi^{\tilde{a}}_{\bs{k}}
|
\widehat{\bs{X}}^{\ }_{r}\,
|
\chi^{\tilde{a}'}_{\bs{k}'}
\rangle=
\frac{
e^{-\mathrm{i}(\bs{k}-\bs{k}')\cdot\bs{r}}
     }
     {
\mathcal{N}
     }\,
\bs{r}\,
u^{\alpha\tilde{a}*}_{\bs{k}}\,
u^{\alpha\tilde{a}'}_{\bs{k}'}
\end{equation}
if we opt not to use the product rule for differentiation
or
\begin{equation}
\langle
\chi^{\tilde{a}}_{\bs{k}}
|
\widehat{\bs{X}}^{\ }_{r}\,
|
\chi^{\tilde{a}'}_{\bs{k}'}
\rangle=
\left(
\frac{
e^{-\mathrm{i}(\bs{k}-\bs{k}')\cdot\bs{R}}
     }
     {
\mathcal{N}
     }\,
\bs{R}\,
\delta^{\tilde{a}\tilde{a}'}
+
\mathrm{i}\,
\bs{A}^{\tilde{a}\tilde{a}'}_{\bs{k}}\,
\delta^{\ }_{\bs{k},\bs{k}'}
\right)
\end{equation}
if we opt to use the product rule for differentiation
as we did in Eqs.~(\ref{eq: use product rule dif for Xr a})
and~(\ref{eq: use product rule dif for Xr b}).

\end{proof}

\subsection{
Gauge invariant regularization of the trace of the 1 and 3 brackets
           }

Equation~(\ref{eq: 3 bracket matrix elements})
is the main result that we need to draw a connection between
the expectation value of the 3-bracket in the noninteracting
filled Fermi sea and the $U(N)$ Chern-Simons action in 3-dimensional
space.

We have shown in Sec.%
~\ref{subsec: Algebra of the position operators}
the ``symbolic''
gauge invariance of the expectation value of the
1-, 2-, and 3-bracket of the projected many-body
position operator in the Fermi sea filling up 
$\widetilde{N}$ Bloch bands.
The qualifier ``symbolic'' must be used since this symmetry
presumes the existence of the expectation value.
There is no ambiguity for the 2-bracket.
The 1- and 3-brackets are however ill defined.
They need to be regularized, i.e., made finite.

It is well known in quantum field theory that regularizations
can break a classical symmetry. Regularizations know about
quantum mechanics, for they involve expectation values of
operators made of additive pieces that do not commute.
In a path integral formalism, quantum mechanics is traded for
coherent states at the price of a measure that requires
a regularization. Here, we need to trace over an operator
that can be decomposed into two additive operators that do 
not commute. The resulting quantum fluctuations require
a regulation of ill-conditioned sums.

However, in the process of regularization the symbolic
gauge invariance can disappear. The question thus becomes
the following. Is it possible to regulate the 1- and 3-bracket 
in a gauge invariant way whereby the gauge invariance only 
applies to pure gauge transformation since large gauge transformations
change the boundary conditions and thus the very nature of the Hilbert space
over which the trace is to be performed?

Our answer is positive and relies on the observation
that we already made in
Eq.~(\ref{eq: comment (i) for one bracket})
and follows from 
Eq.~(\ref{eq: 3 bracket matrix elements})
namely that
\begin{subequations}
\label{eq: def single particle A operator}
\begin{equation}
\widehat{\bs{X}}^{\ }_{r}
-
\widehat{\bs{X}}^{\ }_{R}=
\mathrm{i}\,
\widehat{\bs{A}}
\end{equation}
where we have introduced the operator
\begin{equation}
\widehat{\bs{A}}:=
|
\chi^{\tilde{a}}_{\bs{k}}
\rangle\,
\bs{A}^{\tilde{a}\tilde{b}}_{\bs{k}}\,
\langle
\chi^{\tilde{b}}_{\bs{k}}
|
\end{equation}
\end{subequations}
through its spectral decomposition.

One verifies by direct computation with the help
of Eqs.~(\ref{eq: 3 bracket matrix elements}) 
and (\ref{eq: def single particle A operator}) 
that
\begin{equation}
\begin{split}
F^{(3)}_{\mathrm{finite}}[\bs{A}]:=&\,
\frac{1}{\mathcal{N}}
\langle
W^{a}_{\bs{R}}
|
\left[
\epsilon^{\ }_{\mu\nu\lambda}\,
\widehat{X}^{\mu}_{r}\,
\widehat{X}^{\nu}_{r}\,
\left(
\widehat{X}^{\lambda}_{r}
-
\widehat{X}^{\lambda}_{R}
\right)
\right]
|
W^{a}_{\bs{R}}
\rangle
\\
=&\,
-
\frac{1}{\mathcal{N}}
\times
\frac{\mathrm{i}}{2}
\sum_{\bs{k}\in\Lambda^{\star}_{\hbox{\tiny{BZ}}}}
\epsilon^{\mu\nu\lambda}\,
F^{\tilde{a}\tilde{b}}_{\mu\nu;\bs{k}}\,
A^{\tilde{b}}_{\lambda;\bs{k}}
\end{split}
\label{eq: one out of three regularized but gauge breaking}
\end{equation}
breaks $SU(\widetilde{N})$ pure gauge symmetry.
This regularization is thus not the one we seek.
(Summation convention over repeated indices is implied.)

However, we immediately see that there is an ambiguity
when choosing the space index for which we will do the
replacement
$\widehat{\bs{X}}^{\mu}_{r}\to
 \widehat{\bs{X}}^{\mu}_{r}- 
 \widehat{\bs{X}}^{\mu}_{R}$.
There are three possible choices that would have all
lead to the same right-hand side%
~(\ref{eq: one out of three regularized but gauge breaking}),
namely
\begin{equation}
\begin{split}
F^{(3)}_{\mathrm{finite}}[\bs{A}]=&\,
\frac{1}{\mathcal{N}}
\langle
W^{a}_{\bs{R}}
|
\left[
\epsilon^{\ }_{\mu\nu\lambda}\,
\widehat{X}^{\mu}_{r}\,
\widehat{X}^{\nu}_{r}\,
\left(
\widehat{X}^{\lambda}_{r}
-
\widehat{X}^{\lambda}_{R}
\right)
\right]
|
W^{a}_{\bs{R}}
\rangle
\\
=&\,
\frac{1}{\mathcal{N}}
\langle
W^{a}_{\bs{R}}
|
\left[
\epsilon^{\ }_{\mu\nu\lambda}\,
\widehat{X}^{\mu}_{r}\,
\left(
\widehat{X}^{\nu}_{r}
-
\widehat{X}^{\nu}_{R}
\right)\,
\widehat{X}^{\lambda}_{r}\,
\right]
|
W^{a}_{\bs{R}}
\rangle
\\
=&\,
\frac{1}{\mathcal{N}}
\langle
W^{a}_{\bs{R}}
|
\left[
\epsilon^{\ }_{\mu\nu\lambda}\,
\left(
\widehat{X}^{\mu}_{r}
-
\widehat{X}^{\mu}_{R}
\right)\,
\widehat{X}^{\nu}_{r}\,
\widehat{X}^{\lambda}_{r}
\right]
|
W^{a}_{\bs{R}}
\rangle.
\end{split}
\label{eq: where to place X_r-X_R}
\end{equation}
(Summation convention over repeated indices is implied.)
\begin{proof}
We can first insert and then remove
the resolution of the identity as 
\begin{equation}
\begin{split}
\langle
W^{a}_{\bs{R}}
|
\left[
\widehat{X}^{\mu}_{r}\,
\left(
\widehat{X}^{\nu}_{r}
-
\widehat{X}^{\nu}_{R}
\right)\,
\widehat{X}^{\lambda}_{r}\,
\right]
|
W^{a}_{\bs{R}}
\rangle
=&\,
\langle
W^{a}_{\bs{R}}
|
\widehat{X}^{\mu}_{r}\,
\left(
\widehat{X}^{\nu}_{r}
-
\widehat{X}^{\nu}_{R}
\right)\,
|
W^{a'}_{\bs{R}'}
\rangle
\langle
W^{a'}_{\bs{R}'}
|
\widehat{X}^{\lambda}_{r}\,
|
W^{a}_{\bs{R}}
\rangle
\\
=&\,
\langle
W^{a'}_{\bs{R}'}
|
\widehat{X}^{\lambda}_{r}\,
|
W^{a}_{\bs{R}}
\rangle
\langle
W^{a}_{\bs{R}}
|
\widehat{X}^{\mu}_{r}\,
\left(
\widehat{X}^{\nu}_{r}
-
\widehat{X}^{\nu}_{R}
\right)\,
|
W^{a'}_{\bs{R}'}
\rangle
\\
=&\,
\langle
W^{a'}_{\bs{R}'}
|
\widehat{X}^{\lambda}_{r}\,
\widehat{X}^{\mu}_{r}\,
\left(
\widehat{X}^{\nu}_{r}
-
\widehat{X}^{\nu}_{R}
\right)\,
|
W^{a'}_{\bs{R}'}
\rangle
\end{split}
\end{equation}
for the second line
of Eq.~(\ref{eq: where to place X_r-X_R})
and
\begin{equation}
\begin{split}
\langle
W^{a}_{\bs{R}}
|
\left[
\left(
\widehat{X}^{\mu}_{r}
-
\widehat{X}^{\mu}_{R}
\right)\,
\widehat{X}^{\nu}_{r}\,
\widehat{X}^{\lambda}_{r}\,
\right]
|
W^{a}_{\bs{R}}
\rangle
=&\,
\langle
W^{a}_{\bs{R}}
|
\left(
\widehat{X}^{\mu}_{r}
-
\widehat{X}^{\mu}_{R}
\right)\,
|
W^{a'}_{\bs{R}'}
\rangle
\langle
W^{a'}_{\bs{R}'}
|
\widehat{X}^{\nu}_{r}\,
\widehat{X}^{\lambda}_{r}\,
|
W^{a}_{\bs{R}}
\rangle
\\
=&\,
\langle
W^{a'}_{\bs{R}'}
|
\widehat{X}^{\nu}_{r}\,
\widehat{X}^{\lambda}_{r}\,
|
W^{a}_{\bs{R}}
\rangle
\langle
W^{a}_{\bs{R}}
|
\left(
\widehat{X}^{\mu}_{r}
-
\widehat{X}^{\mu}_{R}
\right)\,
|
W^{a'}_{\bs{R}'}
\rangle
\\
=&\,
\langle
W^{a'}_{\bs{R}'}
|
\widehat{X}^{\nu}_{r}\,
\widehat{X}^{\lambda}_{r}\,
\left(
\widehat{X}^{\mu}_{r}
-
\widehat{X}^{\mu}_{R}
\right)\,
|
W^{a'}_{\bs{R}'}
\rangle
\end{split}
\end{equation}
for the third line
of Eq.~(\ref{eq: where to place X_r-X_R}). 
The space labels $\mu,\nu,\lambda=1,\cdots,d$
have been reordered in cyclic fashion so that contraction
with $\epsilon^{\mu\nu\lambda}$ delivers
Eq.~(\ref{eq: where to place X_r-X_R}).
\end{proof}

The subtraction that we performed in 
Eq.~(\ref{eq: one out of three regularized but gauge breaking}) 
does regulate the expectation value of the 3-bracket
but not in a gauge invariant way. Instead of
Eq.~(\ref{eq: one out of three regularized but gauge breaking}),
we use the more symmetric definition
\begin{equation}
\begin{split}
F^{(3)}_{\mathrm{gauge\ invariant}}[\bs{A}]:=&\,
+
\frac{1}{\mathcal{N}}
\langle
W^{a}_{\bs{R}}
|
\left[
\epsilon^{\ }_{\mu\nu\lambda}\,
\widehat{X}^{\mu}_{r}\,
\widehat{X}^{\nu}_{r}\,
\left(
\widehat{X}^{\lambda}_{r}
-
\widehat{X}^{\lambda}_{R}
\right)
\right]
|
W^{a}_{\bs{R}}
\rangle
\\
&\,
+
\frac{1}{\mathcal{N}}
\langle
W^{a}_{\bs{R}}
|
\left[
\epsilon^{\ }_{\mu\nu\lambda}\,
\widehat{X}^{\mu}_{r}\,
\left(
\widehat{X}^{\nu}_{r}
-
\widehat{X}^{\nu}_{R}
\right)\,
\widehat{X}^{\lambda}_{r}\,
\right]
|
W^{a}_{\bs{R}}
\rangle
\\
&\,
+
\frac{1}{\mathcal{N}}
\langle
W^{a}_{\bs{R}}
|
\left[
\epsilon^{\ }_{\mu\nu\lambda}\,
\left(
\widehat{X}^{\mu}_{r}
-
\widehat{X}^{\mu}_{R}
\right)\,
\widehat{X}^{\nu}_{r}\,
\widehat{X}^{\lambda}_{r}\,
\right]
|
W^{a}_{\bs{R}}
\rangle
\\
&\,
-
\frac{1}{\mathcal{N}}
\langle
W^{a}_{\bs{R}}
|
\left[
\epsilon^{\ }_{\mu\nu\lambda}\,
\left(
\widehat{X}^{\mu}_{r}
-
\widehat{X}^{\mu}_{R}
\right)\,
\left(
\widehat{X}^{\nu}_{r}
-
\widehat{X}^{\nu}_{R}
\right)\,
\left(
\widehat{X}^{\lambda}_{r}
-
\widehat{X}^{\lambda}_{R}
\right)
\right]
|
W^{a}_{\bs{R}}
\rangle.
\end{split}
\label{eq: expectation value that gives F3 gauge invariant}
\end{equation}
One verifies by direct computation with the help
of Eqs.~(\ref{eq: 3 bracket matrix elements}) 
and (\ref{eq: def single particle A operator}) that
\begin{equation}
F^{(3)}_{\mathrm{gauge\ invariant}}[\bs{A}]=
-\mathrm{i}\,
\frac{3}{2}\,
\frac{1}{\mathcal{N}}
\sum_{\bs{k}\in\Lambda^{\star}_{\hbox{\tiny{BZ}}}}
\epsilon^{\mu\nu\lambda}\,
\mathrm{tr}\,
\left(
F^{\ }_{\mu\nu;\bs{k}}\,
A^{\ }_{\lambda;\bs{k}}
-
\frac{2}{3}
A^{\ }_{\mu;\bs{k}}\,
A^{\ }_{\nu;\bs{k}}\,
A^{\ }_{\lambda;\bs{k}}
\right)
\end{equation}
is proportional to the integral over the Brillouin zone
of the Chern-Simons 3 form.

The operator over which the trace is taken on 
the right-hand side of Eq.%
~(\ref{eq: expectation value that gives F3 gauge invariant})
can be rewritten in a way that brings it to a linear combination
of 3-brackets, thereby justifying the upper index $(3)$
for the functional
$F^{(3)}_{\mathrm{gauge\ invariant}}[\bs{A}]$
over the manifold of $su(\widetilde{N})$ gauge fields.
Indeed, we are allowed to reorder the $3\times6=18$ operators
over which the trace is taken on 
the first three lines of the right-hand side of Eq.%
~(\ref{eq: expectation value that gives F3 gauge invariant})
as follows,
\begin{equation}
\begin{split}
&
\epsilon^{\ }_{IJK}
\left[
\widehat{X}^{I}_{r}\,
\widehat{X}^{J}_{r}\,
\left(
\widehat{X}^{K}_{r}
-
\widehat{X}^{K}_{R}
\right)
+
\widehat{X}^{I}_{r}\,
\left(
\widehat{X}^{J}_{r}
-
\widehat{X}^{J}_{R}
\right)\,
\widehat{X}^{K}_{r}
+
\left(
\widehat{X}^{I}_{r}
-
\widehat{X}^{I}_{R}
\right)\,
\widehat{X}^{J}_{r}\,
\widehat{X}^{K}_{r}
\right]=
\\
&
\qquad\qquad\qquad\qquad\qquad\qquad
\left[
\widehat{X}^{\mu}_{r},
\widehat{X}^{\nu}_{r},
\left(
\widehat{X}^{\lambda}_{r}
-
\widehat{X}^{\lambda}_{R}
\right)
\right]
+
\left[
\widehat{X}^{\mu}_{r},
\left(
\widehat{X}^{\nu}_{r}
-
\widehat{X}^{\nu}_{R}
\right),
\widehat{X}^{\lambda}_{r}
\right]
+
\left[
\left(
\widehat{X}^{\mu}_{r}
-
\widehat{X}^{\mu}_{R}
\right),
\widehat{X}^{\nu}_{r},
\widehat{X}^{\lambda}_{r}
\right]
\end{split}
\end{equation}
where $I,J,K=\mu,\nu,\lambda$.
One also verifies that
\begin{equation}
\left[
\left(
\widehat{X}^{\mu}_{r}
-
\widehat{X}^{\mu}_{R}
\right),
\left(
\widehat{X}^{\nu}_{r}
-
\widehat{X}^{\nu}_{R}
\right),
\left(
\widehat{X}^{\lambda}_{r}
-
\widehat{X}^{\lambda}_{R}
\right)
\right]=
\epsilon^{\ }_{IJK}\,
\left(
\widehat{X}^{I}_{r}
-
\widehat{X}^{I}_{R}
\right)\,
\left(
\widehat{X}^{J}_{r}
-
\widehat{X}^{J}_{R}
\right)\,
\left(
\widehat{X}^{K}_{r}
-
\widehat{X}^{K}_{R}
\right)
\end{equation}
where $I,J,K=\mu,\nu,\lambda$.
We may then define the regularized 3-bracket to be the linear combination
\begin{equation}
\begin{split}
\left[
\widehat{X}^{\mu}_{r},
\widehat{X}^{\nu}_{r},
\widehat{X}^{\lambda}_{r}
\right]^{\ }_{\mathrm{reg}}:=
\frac{1}{2}
\Bigg\{&
\left[
\widehat{X}^{\mu}_{r},
\widehat{X}^{\nu}_{r},
\left(
\widehat{X}^{\lambda}_{r}
-
\widehat{X}^{\lambda}_{R}
\right)
\right]
+
\left[
\widehat{X}^{\mu}_{r},
\left(
\widehat{X}^{\nu}_{r}
-
\widehat{X}^{\nu}_{R}
\right),
\widehat{X}^{\lambda}_{r}
\right]
+
\left[
\left(
\widehat{X}^{\mu}_{r}
-
\widehat{X}^{\mu}_{R}
\right),
\widehat{X}^{\nu}_{r},
\widehat{X}^{\lambda}_{r}
\right]
\\
&\,
-
\left[
\left(
\widehat{X}^{\mu}_{r}
-
\widehat{X}^{\mu}_{R}
\right),
\left(
\widehat{X}^{\nu}_{r}
-
\widehat{X}^{\nu}_{R}
\right),
\left(
\widehat{X}^{\lambda}_{r}
-
\widehat{X}^{\lambda}_{R}
\right)
\right]
\Bigg\}.
\end{split}
\label{eq: appdef of reg 3 bracket}
\end{equation}
Here, we have multiplied the curly braces by the normalization
$1/2$ as we demand that the regularization preserves the number of
3-brackets to be regularized. To regularize a single 3-bracket,
we added three 3-brackets and subtracted one 3-bracket.
the number $3-1=2$ is thus the integer by which we choose to divide
the curly bracket on the right-hand side of Eq.%
~(\ref{eq: appdef of reg 3 bracket}). Because the
regularized 3-bracket is a linear superposition of 3-brackets,
it remains odd under the exchange of any pair of its consecutive arguments,
\begin{equation}
\left[
\widehat{X}^{\sigma(\mu)}_{r},
\widehat{X}^{\sigma(\nu)}_{r},
\widehat{X}^{\sigma(\lambda)}_{r}
\right]^{\ }_{\mathrm{reg}}=
(-)^{\mathrm{sgn}(\sigma)}
\left[
\widehat{X}^{\mu}_{r},
\widehat{X}^{\nu}_{r},
\widehat{X}^{\lambda}_{r}
\right]^{\ }_{\mathrm{reg}}
\end{equation}
with $\sigma$ denoting any permutation of 3 objects and
$\mathrm{sgn}(\sigma)=0,1$ with $0$ if the permutation is even
and $1$ if the permutation is odd.
The regularized 3-bracket also vanishes whenever two of its
arguments are equal,
\begin{equation}
\left[
\widehat{X}^{\mu}_{r},
\widehat{X}^{\nu}_{r},
\widehat{X}^{\lambda}_{r}
\right]^{\ }_{\mathrm{reg}}
=
0
\end{equation}
if $\mu=\nu$ or $\nu=\lambda$ or $\mu=\lambda$.
Finally, the regularized 3-bracket is invariant under
pure gauge transformations of the form%
~(\ref{eq: gauge invariance local density k space bis b})
since
\begin{equation}
\left[
\widehat{X}^{\mu}_{r},
\widehat{X}^{\nu}_{r},
\widehat{X}^{\lambda}_{r}
\right]^{\ }_{\mathrm{reg}}
=
-\mathrm{i}\,
\frac{3}{4}\,
\frac{1}{\mathcal{N}}
\sum_{\bs{k}\in\Lambda^{\star}_{\hbox{\tiny{BZ}}}}
\epsilon^{IJK}\,
\mathrm{tr}\,
\left(
F^{\ }_{IJ;\bs{k}}\,
A^{\ }_{K;\bs{k}}
-
\frac{2}{3}
A^{\ }_{I;\bs{k}}\,
A^{\ }_{J;\bs{k}}\,
A^{\ }_{K;\bs{k}}
\right)
\end{equation}
where $I,J,K=\mu,\nu,\lambda$.

\subsection{
Regularized 3-bracket and the Nambu bracket
           }
\label{appsubsec: Regularized 3-bracket and the Nambu bracket}

We are going to prove
Eq.~(\ref{eq: master formula 3 bracket f1 X f2 X f3 X})
to which we refer the reader for the notation and definitions.

To perform a Taylor expansion on 
$
\mathrm{Tr}\,
\left[
f^{\ }_{1}(\widehat{\bs{X}}^{\ }_{r}),
f^{\ }_{2}(\widehat{\bs{X}}^{\ }_{r}),
f^{\ }_{3}(\widehat{\bs{X}}^{\ }_{r})
\right]^{\ }_{\mathrm{reg}}
$,
we need to start with a Taylor expansion on
\begin{equation}
\begin{split}
\left[
f^{\ }_{1}(\widehat{\bs{X}}^{\ }_{r}),
f^{\ }_{2}(\widehat{\bs{X}}^{\ }_{r}),
f^{\ }_{3}(\widehat{\bs{X}}^{\ }_{r})
\right]^{\ }_{\mathrm{reg}}
:=&\,
+
\left[
f^{\ }_{1}(\widehat{\bs{X}}^{\ }_{r}),
f^{\ }_{2}(\widehat{\bs{X}}^{\ }_{r}),
f^{\ }_{3}(\widehat{\bs{X}}^{\ }_{r}-\widehat{\bs{X}}^{\ }_{R})
\right]
\\
&\,
+
\left[
f^{\ }_{1}(\widehat{\bs{X}}^{\ }_{r}),
f^{\ }_{2}(\widehat{\bs{X}}^{\ }_{r}-\widehat{\bs{X}}^{\ }_{R}),
f^{\ }_{3}(\widehat{\bs{X}}^{\ }_{r})
\right]
\\
&\,
+
\left[
f^{\ }_{1}(\widehat{\bs{X}}^{\ }_{r}-\widehat{\bs{X}}^{\ }_{R}),
f^{\ }_{2}(\widehat{\bs{X}}^{\ }_{r}),
f^{\ }_{3}(\widehat{\bs{X}}^{\ }_{r})
\right]
\\
&\,
-
\left[
f^{\ }_{1}(\widehat{\bs{X}}^{\ }_{r}-\widehat{\bs{X}}^{\ }_{R}),
f^{\ }_{2}(\widehat{\bs{X}}^{\ }_{r}-\widehat{\bs{X}}^{\ }_{R}),
f^{\ }_{3}(\widehat{\bs{X}}^{\ }_{r}-\widehat{\bs{X}}^{\ }_{R})
\right].
\end{split}
\end{equation}
To this end, we recall that
\begin{equation}
[A,B,C]=
A[B,C]
+
B[C,A]
+
C[A,B].
\end{equation}
Thus,
\begin{equation}
\begin{split}
\left[
f^{\ }_{1}(\widehat{\bs{X}}^{\ }_{r}),
f^{\ }_{2}(\widehat{\bs{X}}^{\ }_{r}),
f^{\ }_{3}(\widehat{\bs{X}}^{\ }_{r})
\right]^{\ }_{\mathrm{reg}}
=&\,
+
f^{\ }_{1}(\widehat{\bs{X}}^{\ }_{r})
\left[
f^{\ }_{2}(\widehat{\bs{X}}^{\ }_{r}),
f^{\ }_{3}(\widehat{\bs{X}}^{\ }_{r}-\widehat{\bs{X}}^{\ }_{R})
\right]
+
\hbox{cyclic permutations of 1,2,3}
\\
&\,
+
f^{\ }_{1}(\widehat{\bs{X}}^{\ }_{r})
\left[
f^{\ }_{2}(\widehat{\bs{X}}^{\ }_{r}-\widehat{\bs{X}}^{\ }_{R}),
f^{\ }_{3}(\widehat{\bs{X}}^{\ }_{r})
\right]
+
\hbox{cyclic permutations of 1,2,3}
\\
&\,
+
f^{\ }_{1}(\widehat{\bs{X}}^{\ }_{r}-\widehat{\bs{X}}^{\ }_{R})
\left[
f^{\ }_{2}(\widehat{\bs{X}}^{\ }_{r}),
f^{\ }_{3}(\widehat{\bs{X}}^{\ }_{r})
\right]
+
\hbox{cyclic permutations of 1,2,3}
\\
&\,
-
f^{\ }_{1}(\widehat{\bs{X}}^{\ }_{r}-\widehat{\bs{X}}^{\ }_{R})
\left[
f^{\ }_{2}(\widehat{\bs{X}}^{\ }_{r}-\widehat{\bs{X}}^{\ }_{R}),
f^{\ }_{3}(\widehat{\bs{X}}^{\ }_{r}-\widehat{\bs{X}}^{\ }_{R})
\right]
-
\hbox{cyclic permutations of 1,2,3}.
\label{appeq: reg 3-bracket in terms 2 brackets}
\end{split}
\end{equation}
We are now ready to insert the Taylor expansions
\begin{equation}
f^{\ }_{i}(\bs{x})=
f^{\ }_{i}(\bs{0})
+
\sum_{\mu=1}^{3}
\left(
\partial^{\ }_{\mu}\,
f^{\ }_{i}
\right)(\bs{0})\,
x^{\mu}
+
\cdots
\label{appeq: Taylor expansion fi}
\end{equation}
for $i=1,2,3$
after substituting $\bs{x}$ with the corresponding projected
position operator. Because
$f^{\ }_{i}(\bs{0})$ with $i=1,2,3$
are $\mathbb{C}$ numbers,
the commutators in 
Eq.~(\ref{appeq: reg 3-bracket in terms 2 brackets})
must necessarily be of second order 
in the projected position operators if they are to be
nonvanishing. This means that the insertion of
Eq.~(\ref{appeq: Taylor expansion fi})
into
Eq.~(\ref{appeq: reg 3-bracket in terms 2 brackets})
can be organized into the expansion
\begin{subequations}
\begin{equation}
\left[
f^{\ }_{1}(\widehat{\bs{X}}^{\ }_{r}),
f^{\ }_{2}(\widehat{\bs{X}}^{\ }_{r}),
f^{\ }_{3}(\widehat{\bs{X}}^{\ }_{r})
\right]^{\ }_{\mathrm{reg}}
=
\left[
f^{\ }_{1}(\widehat{\bs{X}}^{\ }_{r}),
f^{\ }_{2}(\widehat{\bs{X}}^{\ }_{r}),
f^{\ }_{3}(\widehat{\bs{X}}^{\ }_{r})
\right]^{(2)}_{\mathrm{reg}}
+
\left[
f^{\ }_{1}(\widehat{\bs{X}}^{\ }_{r}),
f^{\ }_{2}(\widehat{\bs{X}}^{\ }_{r}),
f^{\ }_{3}(\widehat{\bs{X}}^{\ }_{r})
\right]^{(3)}_{\mathrm{reg}}
+
\cdots
\end{equation}
where
\begin{equation}
\begin{split}
\left[
f^{\ }_{1}(\widehat{\bs{X}}^{\ }_{r}),
f^{\ }_{2}(\widehat{\bs{X}}^{\ }_{r}),
f^{\ }_{3}(\widehat{\bs{X}}^{\ }_{r})
\right]^{(2)}_{\mathrm{reg}}
=&\,
+
\left[
f^{\ }_{1}\,
\left(
\partial^{\ }_{\mu}\,
f^{\ }_{2}
\right)
\left(
\partial^{\ }_{\nu}\,
f^{\ }_{3}
\right)
+
\hbox{cyclic permutations of 1,2,3}
\right]
(\bs{0})\,
\left[
\widehat{X}^{\mu}_{r},
\widehat{X}^{\nu}_{r}-\widehat{X}^{\nu}_{R}
\right]
\\
&\,
+
\left[
f^{\ }_{2}\,
\left(
\partial^{\ }_{\mu}\,
f^{\ }_{3}
\right)
\left(
\partial^{\ }_{\nu}\,
f^{\ }_{1}
\right)
+
\hbox{cyclic permutations of 1,2,3}
\right]
(\bs{0})\,
\left[
\widehat{X}^{\mu}_{r}-\widehat{X}^{\mu}_{R},
\widehat{X}^{\nu}_{r}
\right]
\\
&\,
+
\left[
f^{\ }_{3}\,
\left(
\partial^{\ }_{\mu}\,
f^{\ }_{1}
\right)
\left(
\partial^{\ }_{\nu}\,
f^{\ }_{2}
\right)
+
\hbox{cyclic permutations of 1,2,3}
\right]
(\bs{0})\,
\left[
\widehat{X}^{\mu}_{r},
\widehat{X}^{\nu}_{r}
\right]
\\
&\,
-
\left[
f^{\ }_{1}\,
\left(
\partial^{\ }_{\mu}\,
f^{\ }_{2}
\right)
\left(
\partial^{\ }_{\nu}\,
f^{\ }_{3}
\right)
+
\hbox{cyclic permutations of 1,2,3}
\right](\bs{0})
\left[
\widehat{X}^{\mu}_{r}-\widehat{X}^{\nu}_{R},
\widehat{X}^{\nu}_{r}-\widehat{X}^{\nu}_{R}
\right]
\end{split}
\end{equation}
while
\begin{equation}
\left[
f^{\ }_{1}(\widehat{\bs{X}}^{\ }_{r}),
f^{\ }_{2}(\widehat{\bs{X}}^{\ }_{r}),
f^{\ }_{3}(\widehat{\bs{X}}^{\ }_{r})
\right]^{(3)}_{\mathrm{reg}}
=
\left(
\partial^{\ }_{\mu}\,
f^{\ }_{1}
\right)
\left(
\partial^{\ }_{\nu}\,
f^{\ }_{2}
\right)
\left(
\partial^{\ }_{\lambda}\,
f^{\ }_{3}
\right)(\bs{0})\,
\left[
\widehat{X}^{\mu}_{r},
\widehat{X}^{\nu}_{r},
\widehat{X}^{\lambda}_{r}
\right]^{(3)}_{\mathrm{reg}}.
\end{equation}
\end{subequations}
The summation convention over the repeated indices
$\mu,\nu,\lambda=1,2,3$ is understood.
If we take advantage of the fact that
\begin{equation}
\left[
\widehat{X}^{\mu}_{R},
\widehat{X}^{\nu}_{R}
\right]
=0,
\qquad \mu,\nu=1,2,3,
\end{equation}
we find the remarkable simplification
\begin{equation}
\begin{split}
\left[
f^{\ }_{1}(\widehat{\bs{X}}^{\ }_{r}),
f^{\ }_{2}(\widehat{\bs{X}}^{\ }_{r}),
f^{\ }_{3}(\widehat{\bs{X}}^{\ }_{r})
\right]^{(2)}_{\mathrm{reg}}
=&\,
+
2
\left[
f^{\ }_{1}\,
\left(
\partial^{\ }_{\mu}\,
f^{\ }_{2}
\right)
\left(
\partial^{\ }_{\nu}\,
f^{\ }_{3}
\right)
+
\hbox{cyclic permutations of 1,2,3}
\right]
(\bs{0})\,
\left[
\widehat{X}^{\mu}_{r},
\widehat{X}^{\nu}_{r}
\right]
\\
=&\,
\epsilon^{ijk}\,
f^{\ }_{i}
\left\{
f^{\ }_{j},
f^{\ }_{k}
\right\}^{\mu\nu}_{\mathrm{P}}(\bs{0})\,
\left[
\widehat{X}^{\mu}_{r},
\widehat{X}^{\nu}_{r}
\right].
\end{split}
\end{equation}
Another simplification due to the full antisymmetry of the 3-bracket
delivers
\begin{equation}
\left[
f^{\ }_{1}(\widehat{\bs{X}}^{\ }_{r}),
f^{\ }_{2}(\widehat{\bs{X}}^{\ }_{r}),
f^{\ }_{3}(\widehat{\bs{X}}^{\ }_{r})
\right]^{(3)}_{\mathrm{reg}}
=
\left\{
f^{\ }_{1},
f^{\ }_{2},
f^{\ }_{3}
\right\}^{\ }_{\mathrm{N}}(\bs{0})\,
\left[
\widehat{X}^{1}_{r},
\widehat{X}^{2}_{r},
\widehat{X}^{3}_{r}
\right]^{\ }_{\mathrm{reg}}.
\end{equation}
We thus arrive at the operator identity
\begin{equation}
\left[
f^{\ }_{1}(\widehat{\bs{X}}^{\ }_{r}),
f^{\ }_{2}(\widehat{\bs{X}}^{\ }_{r}),
f^{\ }_{3}(\widehat{\bs{X}}^{\ }_{r})
\right]^{\ }_{\mathrm{reg}}
=
\epsilon^{ijk}\,
f^{\ }_{i}
\left\{
f^{\ }_{j},
f^{\ }_{k}
\right\}^{\mu\nu}_{\mathrm{P}}(\bs{0})\,
\left[
\widehat{X}^{\mu}_{r},
\widehat{X}^{\nu}_{r}
\right]
+
\left\{
f^{\ }_{1},
f^{\ }_{2},
f^{\ }_{3}
\right\}^{\ }_{\mathrm{N}}(\bs{0})\,
\left[
\widehat{X}^{1}_{r},
\widehat{X}^{2}_{r},
\widehat{X}^{3}_{r}
\right]^{\ }_{\mathrm{reg}}
+
\cdots.
\label{appeq: operator identity for reg 3 bracket expansion}
\end{equation}

\section{
Gell-Mann matrices
        }
\label{appsec: Gell-Mann matrices}

The Gell-Mann matrices are 3$\times$3 Hermitian matrices that 
are a representation of generators of SU(3). They are defined as
\begin{equation}
\begin{split}
&
\lambda^{\ }_1=
\begin{pmatrix}
0&1&0\\1&0&0\\0&0&0
\end{pmatrix},
\qquad
\lambda^{\ }_2=
\begin{pmatrix}
0&-\mathrm{i}&0\\ \mathrm{i}&0&0\\0&0&0
\end{pmatrix},
\qquad
\lambda^{\ }_3=
\begin{pmatrix}
1&0&0\\0&-1&0\\0&0&0
\end{pmatrix},
\qquad
\lambda^{\ }_4=
\begin{pmatrix}
0&0&1\\ 0&0&0\\1&0&0
\end{pmatrix},
\\
&
\lambda^{\ }_5=
\begin{pmatrix}
0&0&-\mathrm{i}\\0&0&0\\ \mathrm{i}&0&0
\end{pmatrix},
\qquad
\lambda^{\ }_6=
\begin{pmatrix}
0&0&0\\ 0&0&1\\0&1&0
\end{pmatrix},
\qquad
\lambda^{\ }_7=
\begin{pmatrix}
0&0&0\\0&0&-\mathrm{i}\\ 0&\mathrm{i}&0
\end{pmatrix},
\qquad
\lambda^{\ }_8=
\frac{1}{\sqrt{3}}
\begin{pmatrix}
1&0&0\\ 0&1&0\\0&0&-2
\end{pmatrix}.
\end{split}
\end{equation}

\end{widetext}

\medskip
\section{
Topological invariants in the 3-orbital model
        }
\label{app: density algebra}

In this Appendix, we evaluate the 
Chern numbers
\begin{equation}
\mathrm{Ch}^{\lambda}=
\frac{\mathrm{i}}{(2\pi)^{2}}
\frac{\epsilon^{\mu\nu\lambda}}{2}
\int\limits_{\mathrm{BZ}} \mathrm{d}^{3}\bs{k}\,
F^{\ }_{\mu\nu}(\bs{k})
\in\mathbb{Z},
\label{eq: def Chern numbers in TL app}
\end{equation}
for $\lambda=1,2,3$ and
the Chern-Simons-invariant
\begin{equation}
\theta:=
\frac{\epsilon^{\mu\nu\lambda}}{8\pi} 
\int\limits_{\mathrm{BZ}} \mathrm{d}^{3}\bs{k}\,
F^{\ }_{\mu\nu}(\bs{k})\, 
A^{\ }_{\lambda}(\bs{k}),
\label{eq: CS form app}
\end{equation}
for the projection on the dispersionless middle band 
of the three-orbital model defined by Eq.~\eqref{eq: def H position space}
in the thermodynamic limit.
(We have dropped the symbol ${\vphantom{A}}$
refering to the projection for notational simplicity.)
For the three-orbital model defined in 
Eq.~\eqref{eq: def H position space}, 
the block off-diagonal projector $q(\bs{k})$ defined in 
Eq.~\eqref{eq: Bloch eigenstates mathcal H a}
delivers a natural choice of gauge for the Berry connection of the flat band
\begin{subequations}
\begin{equation}
\begin{split}
\bs{A}(\bs{k})=&\, q^{\dag}(\bs{k})\bs{\nabla}q(\bs{k}).
\end{split}
\end{equation}
In this case, $\bs{A}$ can be decomposed as
\begin{equation}
\begin{split}
\bs{A}(\bs{k})
=&\,
\begin{pmatrix}
A'(k^{\ }_1,k^{\ }_2,k^{\ }_3)+A''(k^{\ }_1,k^{\ }_2,k^{\ }_3)\\
A'(k^{\ }_2,k^{\ }_1,k^{\ }_3)-A''(k^{\ }_2,k^{\ }_1,k^{\ }_3)\\
A^{\ }_3(k^{\ }_1,k^{\ }_2,k^{\ }_3)
\end{pmatrix},
\end{split}
\end{equation}
where
\begin{eqnarray}
A'(\bs{k})&=&
-\mathrm{i}\,
\frac{
\sin k^{\ }_1\, 
\sin k^{\ }_3
     }
     {
G(\bs{k})
     },
\\
A''(\bs{k})&=&
+\mathrm{i}\,
\frac{ 
\cos k^{\ }_1\, 
\sin k^{\ }_2
     }
     {
G(\bs{k})
     },
\\
A^{\ }_3(\bs{k})&=&
-\mathrm{i}
\frac{
1
+ 
\cos k^{\ }_3
\left( 
\cos k^{\ }_1
+ 
\cos k^{\ }_2
-
M
\right)
     }
     {
G(\bs{k})
     },
\end{eqnarray}
and 
\begin{equation}
G(\bs{k})
=3+
\left(M-\sum_{\mu=1}^3 \cos k^{\ }_\mu\right)-\sum_{\mu=1}^3 \cos^2 k^{\ }_\mu.
\end{equation}
\end{subequations}
It follows that
\begin{subequations}
\begin{equation}
\begin{split}
A'(k^{\ }_1,k^{\ }_2,k^{\ }_3)
=&\,-A'(-k^{\ }_1,k^{\ }_2,k^{\ }_3)\\
=&\,+A'(k^{\ }_1,-k^{\ }_2,k^{\ }_3)\\
=&\,-A'(k^{\ }_1,k^{\ }_2,-k^{\ }_3),
\end{split}
\end{equation}
as well as
\begin{equation}
\begin{split}
A''(k^{\ }_1,k^{\ }_2,k^{\ }_3)
=&\,+A''(-k^{\ }_1,k^{\ }_2,k^{\ }_3)\\
=&\,-A''(k^{\ }_1,-k^{\ }_2,k^{\ }_3)\\
=&\,+A''(k^{\ }_1,k^{\ }_2,-k^{\ }_3),
\end{split}
\end{equation}
\label{symmetry rel}
\end{subequations}
while $A^{\ }_3(\bs{k})$ is an even function of 
$k^{\ }_1$, 
$k^{\ }_2$, 
and $k^{\ }_3$.

As a consequence, all terms appearing in 
$F^{\ }_{13}(\bs{k})$ and $F^{\ }_{23}(\bs{k})$ 
are an odd function of either $k^{\ }_1$ or $k^{\ }_2$.
Thus,
\begin{equation}
\mathrm{Ch}^{1}=\mathrm{Ch}^{2}=0.
\end{equation}
Furthermore,
\begin{equation}
\begin{split}
\mathrm{Ch}^{3}\propto&\,
\int\limits_{\mathrm{BZ}} \mathrm{d}^3\bs{k}\, 
F^{\ }_{12}
\\
=&\,
\int\limits_{\mathrm{BZ}} 
\mathrm{d}^3\bs{k} 
\left\{
\partial^{\ }_1
\left[
A'(k^{\ }_2,k^{\ }_1,k^{\ }_3) 
- 
A''(k^{\ }_2,k^{\ }_1,k^{\ }_3)
\right]
\right.\\
&\,-
\left.
\partial^{\ }_2
\left[
A'(k^{\ }_1,k^{\ }_2,k^{\ }_3) 
+ 
A''(k^{\ }_1,k^{\ }_2,k^{\ }_3)
\right]
\right\}
\\
=&\,
-2 \int\limits_{\mathrm{BZ}} \mathrm{d}^3\bs{k} \,
\partial^{\ }_2A''(k^{\ }_2,k^{\ }_1,k^{\ }_3)
\\
=&\,
-2 
\left[
A''(2\pi,k^{\ }_1,k^{\ }_3)
-
A''(0,k^{\ }_1,k^{\ }_3)
\right]
\\
=&\,0,
\end{split}
\end{equation}
since $\partial^{\ }_2A''(k^{\ }_2,k^{\ }_1,k^{\ }_3)$ 
is a continuous function of $k^{\ }_2$ with periodicity $2\pi$.
We conclude that the Chern numbers $\mathbf{Ch}$ defined in 
Eq.~\eqref{eq: def Chern numbers in TL app} vanish identically.

\begin{figure}
\includegraphics[angle=0,scale=0.8]{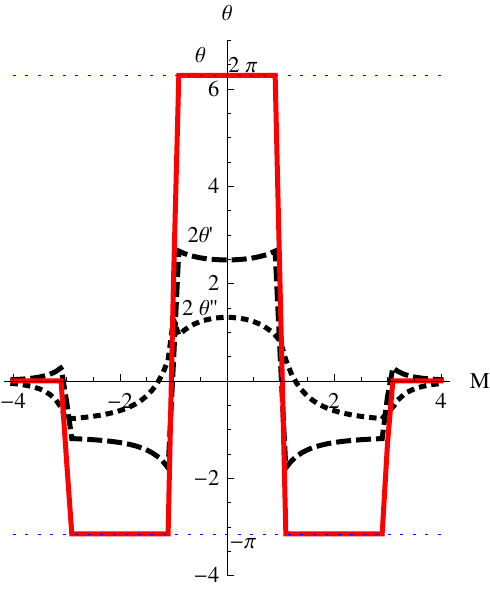}
\caption{\label{fig: plot theta for eta=2}
(Color online)
Numerical evaluation of the topological invariant $\theta^{(0)}=\pi \nu(M)$ 
(solid line)
for the model~\eqref{eq: def H momentum space}
The parameters $\theta'$ and $\theta''$ 
that sum up to the topological invariant $\theta$
are defined in Eqs.~\eqref{eq: theta prime} and~\eqref{eq: theta primeprime},
respectively.
        }
\end{figure}

To calculate $\theta$ defined in Eqs.~\eqref{eq: CS form app} 
we consider integrals of the form
\begin{equation}
\int\limits_{\mathrm{BZ}} 
\mathrm{d}^3\bs{k}\, 
A^{\ }_{\mu}\partial^{\ }_\nu A^{\ }_{\lambda},
\qquad \mu\neq\nu\neq\lambda,
\end{equation}
which are nonvanishing in general.
On one hand, defining
\begin{subequations}
\begin{equation}
\begin{split}
+\theta':=&\,
\frac{1}{8\pi}
\int\limits_{\mathrm{BZ}} 
\mathrm{d}^3\bs{k}\, 
A^{\ }_{1}\partial^{\ }_2 A^{\ }_{3}
\\
=&\,
\frac{1}{8\pi}
\int\limits_{\mathrm{BZ}} 
\mathrm{d}^3\bs{k}\, 
A''(k^{\ }_1,k^{\ }_2,k^{\ }_3)
\partial^{\ }_2 A^{\ }_{3}(k^{\ }_1,k^{\ }_2,k^{\ }_3),
\end{split}
\label{eq: theta prime}
\end{equation}
partial integration delivers
\begin{equation}
-\theta'=\frac{1}{8\pi}
\int\limits_{\mathrm{BZ}} 
\mathrm{d}^3\bs{k}\, A^{\ }_{3}\partial^{\ }_2 A^{\ }_{1},
\end{equation}
and using the identity 
$A^{\ }_{3}(k^{\ }_1,k^{\ }_2,k^{\ }_3)=
A^{\ }_{3}(k^{\ }_2,k^{\ }_1,k^{\ }_3)$
one obtains
\begin{equation}
\begin{split}
+\theta'=&\,\frac{1}{8\pi}
\int\limits_{\mathrm{BZ}} 
\mathrm{d}^3\bs{k}\, A^{\ }_{3}\partial^{\ }_1 A^{\ }_{2},
\\
-\theta'=&\,\frac{1}{8\pi}
\int\limits_{\mathrm{BZ}} 
\mathrm{d}^3\bs{k}\, A^{\ }_{2}\partial^{\ }_1 A^{\ }_{3}.
\end{split}
\end{equation}
On the other hand, defining
\begin{equation}
+\theta'':=
\frac{1}{8\pi}
\int\limits_{\mathrm{BZ}} 
\mathrm{d}^3\bs{k}\, 
A^{\ }_{2}\partial^{\ }_3 A^{\ }_{1}
\label{eq: theta primeprime}
\end{equation}
partial integration delivers
\begin{equation}
-\theta''=\frac{1}{8\pi}
\int\limits_{\mathrm{BZ}} 
\mathrm{d}^3\bs{k}\, A^{\ }_{1}\partial^{\ }_3 A^{\ }_{2}.
\end{equation}
\end{subequations}
Finally, numerical evaluation of
\begin{equation}
\theta=4 \theta'+ 2 \theta''
\end{equation}
reveals that $\theta$ is quantized in units of $\pi$ as announced,
while $\theta'$ and $\theta''$ are not quantized 
and are not equal in general (see Fig.~\ref{fig: plot theta for eta=2}).

\section{
Equivalence of Chern-Simons and Dirac invariants
        }
\label{app: equivalence invariants}

The purpose of this Appendix is to prove that the Abelian Chern-Simons 
invariant, defined by
\begin{equation}
\theta:=
\frac{1}{4\pi}
\int\limits^{\ }_{T^{3}}
\mathrm{d}^{3}\bs{k}\,
\epsilon^{\mu\nu\lambda}\,
A^{\ }_{\mu}\,
\partial^{\ }_{\nu}\,
A^{\ }_{\lambda},
\label{eq: def CS invariant app}
\end{equation}
with the Abelian Berry connection $A^{\ }_{\mu}(\bs{k})$
is equivalent to the Dirac invariant $\nu^{\ }_{\mathrm{D}}$ 
defined in Eq.~\eqref{eq: def Dirac invariant}
for the case of a Bloch Hamiltonian with chiral symmetry and three bands. 
The topological attributes of such a Hamiltonian are characterized by 
its normalized off-diagonal 
part $q(\bs{k})$ from Eq.~\eqref{eq: Bloch eigenstates mathcal H a}
in terms of which the Abelian Berry connection reads
\begin{equation}
A^{\ }_{\mu}(\bs{k})=
q^{-1}(\bs{k})\partial^{\ }_\mu q(\bs{k}).
\end{equation}
Here, $q(\bs{k})$ represents a map from $T^{\ }_3$ (the BZ) 
to $S^{\ }_3$ and $\theta/\pi$ 
is the associated winding number. 
As a member of $S^{\ }_3$, $q(\bs{k})$ can be parametrized 
by three angular coordinates
\begin{equation}
q=:
\begin{pmatrix}
\cos \alpha \,e^{\mathrm{i}\varphi}\\
\sin \alpha \,e^{\mathrm{i}\vartheta}
\end{pmatrix},
\end{equation}
and the Berry connection reads accordingly
\begin{equation}
A^{\ }_{\mu}=
\mathrm{i}\,
\cos^2 \alpha \, \partial^{\ }_\mu \varphi
+
\mathrm{i}\,
\sin^2 \alpha \, \partial^{\ }_\mu \vartheta,
\end{equation}
where we suppress the variable $\bs{k}$ for the moment.
As we shall see, contributions to the winding number%
~\eqref{eq: def CS invariant app}
arise from vortex lines in $\varphi(\bs{k})$ and $\vartheta(\bs{k})$.
Rewriting
\begin{equation}
\begin{split}
\theta=&\,
-
\frac{\epsilon^{\mu\nu\lambda}}{4\pi}
\int\limits^{\ }_{T^{3}}
\mathrm{d}^{3}\bs{k}\,
\sin 2\alpha
\left(\partial^{\ }_\mu \alpha\right)
\left(\partial^{\ }_\nu \vartheta\right)
\partial^{\ }_\lambda \varphi
\\
=&\,
\frac{\epsilon^{\mu\nu\lambda}}{4\pi}
\Biggl\{
\oint
\mathrm{d}^{2}k^{\ }_\mu\,
\cos^{2}\alpha\,
\left(\partial^{\ }_\nu \vartheta\right)
\partial^{\ }_\lambda \varphi
\\
&\,-
\int\limits^{\ }_{T^{3}}
\mathrm{d}^{3}\bs{k}\,
\cos^{2}\alpha\,
\left[
\left(\partial^{\ }_\mu\partial^{\ }_\nu\vartheta\right) 
\partial^{\ }_\lambda \varphi
+
\left(\partial^{\ }_\mu\partial^{\ }_\lambda \varphi\right) 
\partial^{\ }_\nu \vartheta 
\right]
\Biggr\},
\end{split}
\label{eq: manipulation CS invariant app}
\end{equation}
the antisymmetric double derivatives in the last term contribute
a delta-function for $\bs{k}$
on the vortex lines times the winding of the vortex.

Let us now specialize on the model given by 
Eq.~\eqref{eq: def H position space}
in which case
\begin{eqnarray}
\varphi&=&
\mathrm{arg} \left(\sin \, k^{\ }_1+\mathrm{i}\sin\, k^{\ }_2\right),
\\
\vartheta&=&
\mathrm{arg} \left[\sin \, k^{\ }_3+\mathrm{i}
\left(M-\sum^{3}_{i=1}\cos\, k^{\ }_i\right)\right],
\end{eqnarray}
and $\cos\alpha=1$ in the vortex lines of $\varphi$, 
while $\cos\alpha=0$ in the vortex lines of $\vartheta$.
Observe also that the first term in 
Eq.~\eqref{eq: manipulation CS invariant app} vanishes, 
since the either of the partial derivatives 
$\partial^{\ }_\nu \vartheta$ and $\partial^{\ }_\lambda \varphi$
vanishes on each surface with the normal $k^{\ }_\mu$.
The four vortex lines of $\varphi$ are parametrized by
\begin{equation}
\bs{k}^{\mathsf{T}}_{mn}:=(m\pi,n\pi, k^{\ }_3),
\qquad
m,n\in\{0,1\}
\end{equation}
and their winding numbers are $(-1)^{m+n}$ .
Eq.~\eqref{eq: manipulation CS invariant app} then simplifies to
\begin{equation}
\begin{split}
\theta/\pi=&\,
-
\frac{1}{2\pi}
\sum^{1}_{m,n=0}
(-1)^{m+n}
\int\limits^{\ }_{T^{3}}
\mathrm{d}^{3}\bs{k}\,
\delta(\bs{k}-\bs{k}^{\ }_{mn})
\partial^{\ }_3 \vartheta 
\\
=&\,
\frac{1}{2}
\sum^{1}_{m,n,l=0}
(-1)^{m+n+l}
\mathrm{sign}\,d^{\ }_{\bs{k}_{mnl};4}
\end{split}
\label{eq: result CS invariant app}
\end{equation}
where we have written the number of phase windings of $\varphi$ 
in the vortex line of $\vartheta$ as
\begin{equation}
\begin{split}
-\int_0^{2\pi}
\frac{\mathrm{d}k^{\ }_3\,}{2\pi}
\partial^{\ }_3 \vartheta (\bs{k}^{\ }_{mn})
=&\,
\frac{
\mathrm{sign}\,d^{\ }_{\bs{k}_{mn0};4}
-
\mathrm{sign}\,d^{\ }_{\bs{k}_{mn1};4}
}{2}
\\
=&\,
\sum_{l=0}^1
\frac{(-1)^l}{2}
\mathrm{sign}\,d^{\ }_{\bs{k}_{mnl};4}
\end{split}
\end{equation}
and $\bs{k}^{\ }_{mnl}$ is defined as in Eq.~\eqref{eq: def k m n l}.
In writing Eq.~\eqref{eq: result CS invariant app}, 
we have recovered the Dirac invariant~\eqref{eq: def Dirac invariant}.

\medskip
\section{
SMA for a flat band
        }
\label{appsec: SMA for a flat band}

We present some of the intermediate steps needed to
derive Eq.%
~(\ref{eq: f_k final expression}). 
(For ease of presentation, we use Latin instead of Greek indices 
for the momentum components in what follows.
Summation convention over repeated indices is also implied.)

Our aim is to evaluate Eq.%
~(\ref{eq: SMA equations c})
up to order $\bs{q}^{2}\bs{k}^{2}$.
The commutator in Eq.%
~(\ref{eq: SMA equations c})
can be conveniently broken into
four contributions,
\begin{subequations}
\label{eq: terms of f}
\begin{equation}
f^{\ }_{\bs{k}}= 
f^{\ }_{1,\bs{k}} 
+ 
f^{\ }_{2,\bs{k}} 
+ 
f^{\ }_{3,\bs{k}} 
+ 
f^{\ }_{4,\bs{k}},
\end{equation}
each of which read
\begin{equation}
f^{\ }_{1,\bs{k}}:=
\frac{1}{2}\,
\sum_{\bs{q}}\,
v^{\ }_{\bs{q}}\,
\Big{\langle}
\left[
\delta\widehat{\rho}^{\ }_{-\bs{k}},
\delta\widehat{\rho}^{\ }_{-\bs{q}}
\right]
\left[
\delta\widehat{\rho}^{\ }_{+\bs{q}},
\delta\widehat{\rho}^{\ }_{+\bs{k}}
\right]
\Big{\rangle},
\end{equation}
\begin{equation}
f^{\ }_{2,\bs{k}}:=
\frac{1}{2}\,
\sum_{\bs{q}}\,
v^{\ }_{\bs{q}}\,
\Big{\langle}
\left[ 
\delta\widehat{\rho}^{\ }_{-\bs{q}}, 
\delta\widehat{\rho}^{\ }_{+\bs{k}} 
\right]
\left[ 
\delta\widehat{\rho}^{\ }_{-\bs{k}}, 
\delta\widehat{\rho}^{\ }_{+\bs{q}} 
\right]
\Big{\rangle},
\end{equation}
\begin{equation}
f^{\ }_{3,\bs{k}}:=
\frac{1}{2}\,
\sum_{\bs{q}}\,
v^{\ }_{\bs{q}}\,
\Big{\langle}\,
\delta\widehat{\rho}^{\ }_{-\bs{q}}\, 
\left[ 
\delta\widehat{\rho}^{\ }_{-\bs{k}},
\left[
\delta\widehat{\rho}^{\ }_{+\bs{q}}, 
\delta\widehat{\rho}^{\ }_{+\bs{k}} 
\right]
\right]
\Big{\rangle},
\end{equation}
and
\begin{equation}
f^{\ }_{4,\bs{k}}:=
\frac{1}{2}\,
\sum_{\bs{q}}\,
v^{\ }_{\bs{q}}\,
\Big{\langle}\,
\left[
\delta\widehat{\rho}^{\ }_{-\bs{k}},
\left[ 
\delta\widehat{\rho}^{\ }_{-\bs{q}}, 
\delta\widehat{\rho}^{\ }_{+\bs{k}} 
\right]
\right]
\delta\widehat{\rho}^{\ }_{+\bs{q}}
\Big{\rangle}.
\end{equation}
\end{subequations}

The commutator of two projected density operators can be expressed, 
with the aid of Eq.~(\ref{eq: fermionic anticommutation}), 
as
\begin{subequations}
\begin{equation}
\left[
\widehat{\rho}^{\ }_{\bs{q}}, 
\widehat{\rho}^{\ }_{\bs{k}}
\right]=
\sum_{\bs{p}}\,
R^{\ }_{\bs{p},\bs{q},\bs{k}}\,
\widehat{\chi}^{\dag}_{\bs{p}}\, 
\widehat{\chi}^{\ }_{\bs{p}+\bs{q}+\bs{k}},
\end{equation}
where
\begin{equation}
R^{\ }_{\bs{p},\bs{q},\bs{k}}:=
M^{\ }_{\bs{p},\bs{q}}\,
M^{\ }_{\bs{p}+\bs{q},\bs{k}}
-
M^{\ }_{\bs{p}+\bs{k},\bs{q}}\,
M^{\ }_{\bs{p},\bs{k}}.
\end{equation}
\end{subequations}
\medskip

The nested commutators of three projected density operators can be expressed, 
with the aid of Eq.~(\ref{eq: fermionic anticommutation}), 
as
\begin{subequations}
\begin{equation}
\Big{[}\,  
\widehat{\rho}^{\ }_{\bs{k}},
\left[
\widehat{\rho}^{\ }_{\bs{q}}, 
\widehat{\rho}^{\ }_{\bs{k}}
\right]
\Big{]}=
\sum_{\bs{p}}\,
\Lambda^{\ }_{\bs{p},\bs{q},\bs{k}}\,
\widehat{\chi}^{\dag}_{\bs{p}}\, 
\widehat{\chi}^{\ }_{\bs{p}+\bs{q}},
\end{equation}
where
\begin{equation}
\Lambda^{\ }_{\bs{p},\bs{q},\bs{k}}:=
R^{\ }_{\bs{p}-\bs{k},\bs{q},\bs{k}}\,
M^{\ }_{\bs{p},-\bs{k}}
-
R^{\ }_{\bs{p},\bs{q},\bs{k}}\,
M^{\ }_{\bs{p}+\bs{q}+\bs{k},-\bs{k}}.
\end{equation}
\end{subequations}
Observe here that the identity
\begin{equation}
\Big{[}\,  
\widehat{\rho}^{\ }_{\bs{k}},
\left[ 
\widehat{\rho}^{\ }_{\bs{q}}, 
\widehat{\rho}^{\ }_{\bs{k}}
\right]
\Big{]}^{\dag}=
\Big{[}\,  
\widehat{\rho}^{\ }_{-\bs{k}},
\left[ 
\widehat{\rho}^{\ }_{-\bs{q}}, 
\widehat{\rho}^{\ }_{-\bs{k}}
\right]
\Big{]}
\end{equation}
implies that 
\begin{equation}
\label{eq: Lambda condition}
\Lambda^{* }_{\bs{p},\bs{q},\bs{k}}=
\Lambda^{\ }_{\bs{p}+\bs{q},-\bs{q},-\bs{k}}.
\end{equation}

Needed is the expansion of
$R^{\ }_{\bs{p},\bs{q},\bs{k}}$
and 
$\Lambda^{\ }_{\bs{p},\bs{q},\bs{k}}$ 
up to order $\bs{q}^{2}\bs{k}^{2}$.
We start with
\begin{widetext}
\begin{subequations}
\begin{equation}
\label{eq: M(p,q)}
\begin{split}
M^{\ }_{\bs{p},\bs{q}}
=&\, 
u^{\dag}_{\bs{p}} 
\cdot 
u^{\ }_{\bs{p}+\bs{q}}
\\
=&\,
u^{\dag}_{\bs{p}} \cdot 
\left(
u^{\ }_{\bs{p}} 
+ 
q^{i}\,\partial^{\ }_{i} u^{\ }_{\bs{p}} 
+ 
\frac{1}{2}\,
q^{i}q^{j}\,
\partial^{\ }_{i}\partial^{\ }_{j}u^{\ }_{\bs{p}} 
+ 
\cdots
\right)
\\
=&\,
1
+  
q^{i}\, 
u^{\dag}_{\bs{p}} 
\cdot 
\partial^{\ }_{i}u^{\ }_{\bs{p}}
+ 
\frac{1}{2}\,
q^{i}q^{j}\,
u^{\dag}_{\bs{p}} 
\cdot 
\partial^{\ }_{i}\partial^{\ }_{j}u^{\ }_{\bs{p}} 
+ 
\cdots
\\
=&\,
1
+  
q^{i}\,A^{\ }_{i,\bs{p}}
+ 
\frac{1}{2}\,
q^{i}q^{j}\,
u^{\dag}_{\bs{p}} 
\cdot 
\partial^{\ }_{i}\partial^{\ }_{j}u^{\ }_{\bs{p}} 
+ 
\cdots
\end{split}
\end{equation}
where we have introduced the (imaginary-valued) Berry connection
\begin{equation}
A^{\ }_{i,\bs{p}}
\equiv
 u^{\dag}_{\bs{p}} \cdot \partial^{\ }_{i} u^{\ }_{\bs{p}}
\end{equation}
\end{subequations}
and the summation convention over repeated indices 
$i,j=1,\cdots,d$ is implied.
The symbol
$\partial^{\ }_{i}$ 
with $i=1,\cdots,d$
is to be regarded as a derivative with respect to the argument of 
the function on which it acts. Similarly,
\begin{equation}
\label{eq: M(p+q,k)}
\begin{split}
M^{\ }_{\bs{p}+\bs{q},\bs{k}}=&\,
1 
+  
k^{i}\,A^{\ }_{i} 
+ 
q^{i}k^{j}\partial^{\ }_{i}u^{\dag}\cdot\partial^{\ }_{j} u
+
\frac{1}{2}
\left(
k^{i}k^{j}+2 q^{i}k^{j}
\right)\, 
u^{\dag}\cdot\partial^{\ }_{i}\partial^{\ }_{j}u
\\
&\,+
\frac{1}{2}
\left(
q^{i}q^{j}k^{m} 
+ 
k^{i}k^{j}q^{m}
\right)\, 
u^{\dag}
\cdot
\partial^{\ }_{i}\partial^{\ }_{j}\partial^{\ }_{m}u
+
\frac{1}{2}
\left(
k^{i}k^{j}q^{m} 
+ 
2q^{m}q^{i}k^{j}
\right)\, 
\partial^{\ }_{m}u^{\dag}
\cdot
\partial^{\ }_{i}\partial^{\ }_{j}\partial^{\ }_{m}u
+
\frac{1}{2}\, 
q^{i}q^{j}k^{m}\,
\partial^{\ }_{i}
\partial^{\ }_{j} 
u^{\dag}
\cdot
\partial^{\ }_{m} u
\\
&\,+
\frac{1}{4}\, 
q^{i}q^{j}k^{l}k^{m}\, 
u^{\dag}
\cdot
\partial^{\ }_{i}
\partial^{\ }_{j}
\partial^{\ }_{l}
\partial^{\ }_{m} 
u
+
\frac{1}{2}\, 
k^{i}k^{j}q^{l}q^{m}\, 
\partial^{\ }_{l}
u^{\dag}
\cdot
\partial^{\ }_{i}\partial^{\ }_{j}\partial^{\ }_{m}u
+
\frac{1}{2}\, 
q^{i}q^{j}k^{l}k^{m}\, 
\partial^{\ }_{i}\partial^{\ }_{j}
u^{\dag}
\cdot
\partial^{\ }_{l}\partial^{\ }_{m}\partial^{\ }_{m}u
+
\cdots
\end{split}
\end{equation}
where the summation convention over the repeated indices 
$i,j,l,m=1,\cdots,d$ is implied.

We multiply
Eq.~(\ref{eq: M(p,q)}) 
by 
Eq.~(\ref{eq: M(p+q,k)}) 
and antisymmetrize with respect to the interchange
of $\bs{q}$ and $\bs{k}$. We obtain
\begin{subequations}
\begin{equation}
\begin{split}
R(\bs{p},\bs{q},\bs{k})=&\,
q^{i}k^{j}
\left(
T^{(2)}_{i j}
\right)(\bs{p})
+
\left(
k^{i}k^{j}q^{m} 
- 
q^{i}q^{j}k^{m}
\right)
\left(
T^{(3)}_{i j ; m}
\right)(\bs{p})
+
k^{i}k^{j}q^{l}q^{m}
\left(
T^{(4)}_{i j ; l m}
\right)(\bs{p}),
\end{split}
\end{equation}
where the summation convention over the repeated indices 
$i,j,l,m=1,\cdots,d$ is implied and
we have introduced the short-hand notation
\begin{equation}
\left(
T^{(2)}_{i j}
\right)(\bs{p}):=
\left(
F^{\ }_{i j}
\right)(\bs{p})
\equiv
\left(
\partial^{\ }_{i} A^{\ }_{j} - \partial^{\ }_{i} A^{\ }_{j}
\right)(\bs{p}),
\end{equation}
\begin{equation}
\begin{split}
\left(
T^{(3)}_{i j ; m}
\right)(\bs{p}):=
\frac{1}{2}
&\,\Big{(}
\partial^{\ }_{m} u^{\dag} 
\cdot 
\partial^{\ }_{i}\partial^{\ }_{j} u
-
\partial^{\ }_{i}\partial^{\ }_{j} u^{\dag} 
\cdot 
\partial^{\ }_{m} u
-
2 
\partial^{\ }_{j} u^{\dag} 
\cdot 
\partial^{\ }_{i}\partial^{\ }_{m} u
-
2 A^{\ }_{j} \partial^{\ }_{i} A^{\ }_{m}
\Big{)}(\bs{p}),
\end{split}
\end{equation}
and
\begin{equation}
\begin{split}
\left(
T^{(4)}_{i j ; l m}
\right)(\bs{p}):=&\,
\frac{1}{4}
\Big{[}
\partial^{\ }_{l}\partial^{\ }_{m} u^{\dag}
\cdot
\partial^{\ }_{i}\partial^{\ }_{j} u
-
\partial^{\ }_{i}\partial^{\ }_{j} u^{\dag} 
\cdot 
\partial^{\ }_{l}\partial^{\ }_{m} u
+ 
2 A^{\ }_{l} \partial^{\ }_{m}
\left(
u^{\dag} 
\cdot 
\partial^{\ }_{i}\partial^{\ }_{j} u
\right)
-
2 A^{\ }_{i} \partial^{\ }_{j}
\left(
u^{\dag} 
\cdot 
\partial^{\ }_{l}\partial^{\ }_{m} u
\right)
\\
&\,
+
2 \partial^{\ }_{l} u^{\dag} 
\cdot 
\partial^{\ }_{i}\partial^{\ }_{j}\partial^{\ }_{m} u
-
2 \partial^{\ }_{i} u^{\dag} 
\cdot 
\partial^{\ }_{l}\partial^{\ }_{j}\partial^{\ }_{m} u
\Big{]}(\bs{p})
\end{split}
\end{equation}
\end{subequations}
for $i,j,l,m=1,\cdots,d$.
We evaluate
\begin{equation}
\begin{split}
\Lambda(\bs{p},\bs{q},\bs{k})=&\,
R(\bs{p}-\bs{k},\bs{q},\bs{k})\,
M(\bs{p},-\bs{k})
-
R(\bs{p},\bs{q},\bs{k})\,
M(\bs{p}+\bs{q}+\bs{k},-\bs{k})
\\
=&\,
\Big[
R(\bs{p},\bs{q},\bs{k}) 
- 
k^{a} \partial^{\ }_{a} 
R(\bs{p},\bs{q},\bs{k}) 
+ 
\cdots
\Big]
\Big[
1 
- 
k^{b} A^{\ }_{b}(\bs{p}) 
+ 
\cdots
\Big]
-
R(\bs{p},\bs{q},\bs{k})
\Big[
1 
- 
k^{a} A^{\ }_{a}(\bs{p}) 
-  
k^{a} q^{b} \partial^{\ }_{b} A^{\ }_{a}(\bs{p}) 
+ 
\cdots
\Big]
\\
=&\,
R(\bs{p},\bs{q},\bs{k}) 
-  
k^{a} \partial^{\ }_{a} 
R(\bs{p},\bs{q},\bs{k}) 
-  
k^{a} A^{\ }_{a}(\bs{p}) 
R(\bs{p},\bs{q},\bs{k})
+
\cdots
\\
&\,
\quad
- 
R(\bs{p},\bs{q},\bs{k})
+  
k^{a} 
A^{\ }_{a}(\bs{p}) 
R(\bs{p},\bs{q},\bs{k})
+  
k^{a} 
q^{b} 
\partial^{\ }_{b}A^{\ }_{a}(\bs{p}) 
R(\bs{p},\bs{q},\bs{k}) 
+ 
\cdots
\\
=&\,
- 
k^{a} 
\partial^{\ }_{a} 
R(\bs{p},\bs{q},\bs{k}) 
+  
k^{a} 
q^{b} 
\partial^{\ }_{b}A^{\ }_{a}(\bs{p}) 
R(\bs{p},\bs{q},\bs{k})
+ 
\cdots
\\
=&\,
-k^{a}
\Big[
q^{i}k^{j}
\left(
\partial^{\ }_{a}T^{(2)}_{i j}
\right)(\bs{p})
+
- q^{i}q^{j}k^{m}
\left(
\partial^{\ }_{a}T^{(3)}_{i j ; m}
\right)(\bs{p})
+
\cdots
\Big]
+
 k^{a} q^{b} \partial^{\ }_{b}A^{\ }_{a}(\bs{p})
\Big[
q^{i}k^{j}
\left(
T^{(2)}_{i j}
\right)(\bs{p})
+
\cdots
\Big]
\\
=&\,
-
q^{i}k^{j}k^{a}
\left(
\partial^{\ }_{a}T^{(2)}_{i j}
\right)(\bs{p})
+
q^{i}q^{j}k^{m}k^{a}
\left(
\partial^{\ }_{a}T^{(3)}_{i j ; m}
\right)(\bs{p})
+
q^{b}k^{a}q^{i}k^{j}\partial^{\ }_{b}A^{\ }_{a}(\bs{p})
\left(
T^{(2)}_{i j}
\right)(\bs{p})
+
\cdots
\end{split}
\end{equation}
where the summation convention over the repeated indices 
$a,b,i,j,m=1,\cdots,d$ is implied.

At last, we are in a position to evaluate 
the terms contributing to the function $f^{\ }_{\bs{k}}$
in Eq.%
~(\ref{eq: terms of f}). We start with 
\begin{equation}
\label{eq: f1 lattice}
\begin{split}
f^{\ }_{1,\bs{k}}
&\,=
\frac{1}{2}\,\sum_{\bs{q}}\,
v^{\ }_{\bs{q}}\,
\Big{\langle}
[ \delta\widehat{\rho}^{\ }_{-\bs{k}}, \delta\widehat{\rho}^{\ }_{-\bs{q}} ]
[ \delta\widehat{\rho}^{\ }_{\bs{q}}, \delta\widehat{\rho}^{\ }_{\bs{k}} ]
\Big{\rangle}
\\
&\,=
\frac{1}{2}\,
\sum_{\bs{q}}\,
v^{\ }_{\bs{q}}\,
\Big{\langle}
\sum_{\bs{p}}\, 
R(\bs{p},-\bs{k},-\bs{q})\, 
\widehat{\chi}^{\dag}_{\bs{p}}\,
\widehat{\chi}^{\ }_{\bs{p}-\bs{k}-\bs{q}}
\sum_{\bs{p}'}\, 
R(\bs{p}',\bs{q},\bs{k})\, 
\widehat{\chi}^{\dag}_{\bs{p}'}\,
\widehat{\chi}^{\ }_{\bs{p}'+\bs{k}+\bs{q}}
\Big{\rangle}
\\
&\,=
\frac{1}{2}\,
\sum_{\bs{q}}\,v^{\ }_{\bs{q}}\,
\sum_{\bs{p},\bs{p}'}\, 
R(\bs{p},-\bs{k},-\bs{q})\, 
R(\bs{p}',\bs{q},\bs{k})
\Big{\langle}
\widehat{\chi}^{\dag}_{\bs{p}}\,
\widehat{\chi}^{\ }_{\bs{p}-\bs{k}-\bs{q}}\,
\widehat{\chi}^{\dag}_{\bs{p}'}\,
\widehat{\chi}^{\ }_{\bs{p}'+\bs{k}+\bs{q}}
\Big{\rangle}
\\
&\,=
\frac{1}{2}\,
\sum_{\bs{q}}\,
\sum_{\bs{p},\bs{p}'}\, 
v^{\ }_{\bs{q}}
\Big[
(-k^{a})(-q^{b})
\left(
T^{(2)}_{a b}
\right)(\bs{p}) + \cdots
\Big]
\Big[
q^{i}k^{j}
\left(
T^{(2)}_{i j}
\right)(\bs{p}') + \cdots
\Big]
\Big{\langle}
\widehat{\chi}^{\dag}_{\bs{p}}\,
\widehat{\chi}^{\ }_{\bs{p}-\bs{k}-\bs{q}}\,
\widehat{\chi}^{\dag}_{\bs{p}'}\,
\widehat{\chi}^{\ }_{\bs{p}'+\bs{k}+\bs{q}}
\Big{\rangle}
\\
&\,=
-
\frac{1}{2}\,
\sum_{\bs{q}}\,
\sum_{\bs{p},\bs{p}'}\, 
v^{\ }_{\bs{q}}
\Big[
k^{a}q^{b}
\left(
F^{\ }_{a b}
\right)(\bs{p})
\Big]
\Big[
k^{i}q^{j}
\left(
F^{\ }_{i j}
\right)(\bs{p}')
\Big]
\Big{\langle}
\widehat{\chi}^{\dag}_{\bs{p}}\,
\widehat{\chi}^{\ }_{\bs{p}-\bs{k}-\bs{q}}\,
\widehat{\chi}^{\dag}_{\bs{p}'}\,
\widehat{\chi}^{\ }_{\bs{p}'+\bs{k}+\bs{q}}
\Big{\rangle}
+ \cdots
\\
&\,=
-
\frac{1}{2}\,
\sum_{\bs{q}}\,
\sum_{\bs{p},\bs{p}'}\, 
v^{\ }_{\bs{q}}
\Big[
\left(
\bs{k}\wedge\bs{q}
\right)
\cdot
\bs{B}(\bs{p})
\Big]
\Big[
\left(
\bs{k}\wedge\bs{q}
\right)
\cdot
\bs{B}(\bs{p}')
\Big]
\Big{\langle}
\widehat{\chi}^{\dag}_{\bs{p}}\,
\widehat{\chi}^{\ }_{\bs{p}-\bs{k}-\bs{q}}\,
\widehat{\chi}^{\dag}_{\bs{p}'}\,
\widehat{\chi}^{\ }_{\bs{p}'+\bs{k}+\bs{q}}
\Big{\rangle}
+ \cdots,
\end{split}
\end{equation}
where we used that 
$B^{i}=\epsilon^{i j m}\partial^{\ }_{j} A^{\ }_{m}= 
\frac{1}{2} \epsilon^{i j m}F^{\ }_{j m}$ 
or, equivalently, 
$F^{\ }_{i j}=\epsilon^{\ }_{i j m}B^{m}$.
We now break the Berry field strength into two contributions, i.e., 
$
\bs{B}(\bs{p})=
\overline{\bs{B}}
+
\delta\bs{B}(\bs{p})
$.
If so,
\begin{equation}
\label{eq: f1 lattice 2}
\begin{split}
f^{\ }_{1,\bs{k}}&\,=
-
\frac{1}{2}\,
\sum_{\bs{q}}\,
\sum_{\bs{p},\bs{p}'}\, 
v^{\ }_{\bs{q}}
\Big[
\left(
\bs{k}\wedge\bs{q}
\right)
\cdot
\left(
\overline{\bs{B}}
+
\delta\bs{B}(\bs{p})
\right)
\Big]
\Big[
\left(
\bs{k}\wedge\bs{q}
\right)
\cdot
\left(
\overline{\bs{B}}
+
\delta\bs{B}(\bs{p}')
\right)
\Big]
\Big{\langle}
\widehat{\chi}^{\dag}_{\bs{p}}\,
\widehat{\chi}^{\ }_{\bs{p}-\bs{k}-\bs{q}}\,
\widehat{\chi}^{\dag}_{\bs{p}'}\,
\widehat{\chi}^{\ }_{\bs{p}'+\bs{k}+\bs{q}}
\Big{\rangle}
+ \cdots 
\\
&\,=
-
\frac{1}{2}\,
\sum_{\bs{q}}\,
\sum_{\bs{p},\bs{p}'}\, 
v^{\ }_{\bs{q}}
\Big[
\left(
\bs{k}\wedge\bs{q}
\right)
\cdot
\overline{\bs{B}}
\Big]
\Big[
\left(
\bs{k}\wedge\bs{q}
\right)
\cdot
\overline{\bs{B}}
\Big]
\Big{\langle}
\widehat{\chi}^{\dag}_{\bs{p}}\,
\widehat{\chi}^{\ }_{\bs{p}-\bs{k}-\bs{q}}\,
\widehat{\chi}^{\dag}_{\bs{p}'}\,
\widehat{\chi}^{\ }_{\bs{p}'+\bs{k}+\bs{q}}
\Big{\rangle}
\\
&\,
\quad
- 
\frac{1}{2}\,
\sum_{\bs{q}}\,
\sum_{\bs{p},\bs{p}'}\, 
v^{\ }_{\bs{q}}
\Big[
\left(
\bs{k}\wedge\bs{q}
\right)
\cdot
\overline{\bs{B}}
\Big]
\Big[
\left(
\bs{k}\wedge\bs{q}
\right)
\cdot
\delta\bs{B}(\bs{p}')
\Big]
\Big{\langle}
\widehat{\chi}^{\dag}_{\bs{p}}\,
\widehat{\chi}^{\ }_{\bs{p}-\bs{k}-\bs{q}}\,
\widehat{\chi}^{\dag}_{\bs{p}'}\,
\widehat{\chi}^{\ }_{\bs{p}'+\bs{k}+\bs{q}}
\Big{\rangle}
\\
&\,
\quad 
- 
\frac{1}{2}\,
\sum_{\bs{q}}\,
\sum_{\bs{p},\bs{p}'}\, 
v^{\ }_{\bs{q}}
\Big[
\left(
\bs{k}\wedge\bs{q}
\right)
\cdot
\delta\bs{B}(\bs{p})
\Big]
\Big[
\left(
\bs{k}\wedge\bs{q}
\right)
\cdot
\overline{\bs{B}}
\Big]
\Big{\langle}
\widehat{\chi}^{\dag}_{\bs{p}}\,
\widehat{\chi}^{\ }_{\bs{p}-\bs{k}-\bs{q}}\,
\widehat{\chi}^{\dag}_{\bs{p}}\,
\widehat{\chi}^{\ }_{\bs{p}'+\bs{k}+\bs{q}}
\Big{\rangle}
\\
&\,
\quad 
- 
\frac{1}{2}\,
\sum_{\bs{q}}\,
\sum_{\bs{p},\bs{p}'}\, 
v^{\ }_{\bs{q}}
\Big[
\left(
\bs{k}\wedge\bs{q}
\right)
\cdot
\delta\bs{B}(\bs{p})
\Big]
\Big[
\left(
\bs{k}\wedge\bs{q}
\right)
\cdot
\delta\bs{B}(\bs{p}')
\Big]
\Big{\langle}
\widehat{\chi}^{\dag}_{\bs{p}}\,
\widehat{\chi}^{\ }_{\bs{p}-\bs{k}-\bs{q}}\,
\widehat{\chi}^{\dag}_{\bs{p}'}\,
\widehat{\chi}^{\ }_{\bs{p}'+\bs{k}+\bs{q}}
\Big{\rangle}
+ \cdots
\\
&\,=
-
\frac{1}{2}\,
\sum_{\bs{q}}\, 
v^{\ }_{\bs{q}}
\Big[
\left(
\bs{k}\wedge\bs{q}
\right)
\cdot
\overline{\bs{B}}
\Big]
\Big[
\left(
\bs{k}\wedge\bs{q}
\right)
\cdot
\overline{\bs{B}}
\Big]
\Big{\langle}
\widehat{\rho}^{\ }_{-\bs{k}-\bs{q}}\,
\widehat{\rho}^{\ }_{\bs{k}+\bs{q}}
\Big{\rangle}
\\
&\,
\quad
- 
\frac{1}{2}\,
\sum_{\bs{q}}\,
\sum_{\bs{p}'}\, 
v^{\ }_{\bs{q}}
\Big[
\left(
\bs{k}\wedge\bs{q}
\right)
\cdot
\overline{\bs{B}}
\Big]
\Big[
\left(
\bs{k}\wedge\bs{q}
\right)
\cdot
\delta\bs{B}(\bs{p}')
\Big]
\Big{\langle}
\widehat{\rho}^{\ }_{-\bs{k}-\bs{q}}\,
\widehat{\chi}^{\dag}_{\bs{p}'}\,
\widehat{\chi}^{\ }_{\bs{p}'+\bs{k}+\bs{q}}
\Big{\rangle}
\\
&\,
\quad 
- 
\frac{1}{2}\,
\sum_{\bs{q}}\,
\sum_{\bs{p},\bs{p}'}\, 
v^{\ }_{\bs{q}}
\Big[
\left(
\bs{k}\wedge\bs{q}
\right)
\cdot
\delta\bs{B}(\bs{p})
\Big]
\Big[
\left(
\bs{k}\wedge\bs{q}
\right)
\cdot
\overline{\bs{B}}
\Big]
\Big{\langle}
\widehat{\chi}^{\dag}_{\bs{p}}\,
\widehat{\chi}^{\ }_{\bs{p}-\bs{k}-\bs{q}}\,
\widehat{\rho}^{\ }_{\bs{k}+\bs{q}}
\Big{\rangle}
\\
&\,
\quad 
- 
\frac{1}{2}\,
\sum_{\bs{q}}\,
\sum_{\bs{p},\bs{p}'}\, 
v^{\ }_{\bs{q}}
\Big[
\left(
\bs{k}\wedge\bs{q}
\right)
\cdot
\delta\bs{B}(\bs{p})
\Big]
\Big[
\left(
\bs{k}\wedge\bs{q}
\right)
\cdot
\delta\bs{B}(\bs{p}')
\Big]
\Big{\langle}
\widehat{\chi}^{\dag}_{\bs{p}}\,
\widehat{\chi}^{\ }_{\bs{p}-\bs{k}-\bs{q}}\,
\widehat{\chi}^{\dag}_{\bs{p}'}\,
\widehat{\chi}^{\ }_{\bs{p}'+\bs{k}+\bs{q}}
\Big{\rangle}
+ 
\cdots.
\end{split}
\end{equation}
In a uniform liquid-like ground state we have 
$
\langle\, \widehat{\rho}^{\ }_{\bs{k}}\, \rangle 
\propto 
\delta^{\ }_{\bs{k},\bs{0}}
$ and, due
to the relation 
$
k^{a}q^{b}\,
F^{\ }_{ab}
=
\left(
\bs{k}\wedge\bs{q}
\right)\cdot\bs{B}
$,
we can replace
$
\widehat{\rho}^{\ }_{\pm\bs{k}\pm\bs{q}}
$
by 
$
\delta \widehat{\rho}^{\ }_{\pm\bs{k}\pm\bs{q}}
$.
As a consequence, 
we can drop the first three terms
on the last equality of~(\ref{eq: f1 lattice 2})
up to order $\bs{q}^{2}\bs{k}^{2}$.
We are then left with:
\begin{equation}
\label{eq: f1 lattice 3}
f^{\ }_{1,\bs{k}}
=
- \frac{1}{2}\,\sum_{\bs{q}}\,\sum_{\bs{p},\bs{p}'}\, 
v^{\ }_{\bs{q}}
\Big[
\left(
\bs{k}\wedge\bs{q}
\right)
\cdot
\delta\bs{B}(\bs{p})
\Big]
\Big[
\left(
\bs{k}\wedge\bs{q}
\right)
\cdot
\delta\bs{B}(\bs{p}')
\Big]
\Big{\langle}
\hat{n}^{\ }_{\bs{p}}
\hat{n}^{\ }_{\bs{p}'}
\Big{\rangle}
+ \cdots ,
\end{equation} 
where 
$\hat{n}^{\ }_{\bs{p}} \equiv 
\widehat{\chi}^{\dag}_{\bs{p}}\,\widehat{\chi}^{\ }_{\bs{p}}$ 
is the number operator projected on the lowest band. Similarly,
\begin{equation}
\label{eq: f2 lattice}
\begin{split}
f^{\ }_{2,\bs{k}}&\,=
\frac{1}{2}\,\sum_{\bs{q}}\,
v^{\ }_{\bs{q}}\,
\Big{\langle}
[ \delta\widehat{\rho}^{\ }_{-\bs{q}}, \delta\widehat{\rho}^{\ }_{\bs{k}} ]
[ \delta\widehat{\rho}^{\ }_{-\bs{k}}, \delta\widehat{\rho}^{\ }_{\bs{q}} ]
\Big{\rangle}
\\
&\,=
\frac{1}{2}\,\sum_{\bs{q}}\,
v^{\ }_{\bs{q}}\,
\Big{\langle}
[ \delta\widehat{\rho}^{\ }_{\bs{k}}, \delta\widehat{\rho}^{\ }_{-\bs{q}} ]
[ \delta\widehat{\rho}^{\ }_{\bs{q}}, \delta\widehat{\rho}^{\ }_{-\bs{k}} ]
\Big{\rangle}
\\
&\,=
f^{\ }_{1,-\bs{k}}
\\
&\,=
- \frac{1}{2}\,\sum_{\bs{q}}\,\sum_{\bs{p},\bs{p}'}\, 
v^{\ }_{\bs{q}}
\Big[
\left(
\bs{k}\wedge\bs{q}
\right)
\cdot
\delta\bs{B}(\bs{p})
\Big]
\Big[
\left(
\bs{k}\wedge\bs{q}
\right)
\cdot
\delta\bs{B}(\bs{p}')
\Big]
\Big{\langle}
\hat{n}^{\ }_{\bs{p}}
\hat{n}^{\ }_{\bs{p}'}
\Big{\rangle}
+ 
\cdots,
\end{split}
\end{equation}
while
\begin{equation}
\label{eq: f3 lattice}
\begin{split}
f^{\ }_{3,\bs{k}}
&\,=
\frac{1}{2}\,\sum_{\bs{q}}\,
v^{\ }_{\bs{q}}\,
\Big{\langle}\,
\delta\widehat{\rho}^{\ }_{-\bs{q}}\, 
[ 
\delta\widehat{\rho}^{\ }_{-\bs{k}},\,
[ \delta\widehat{\rho}^{\ }_{\bs{q}}, \delta\widehat{\rho}^{\ }_{\bs{k}} ]\,
]
\Big{\rangle}
\\
&\,=
\frac{1}{2}\,\sum_{\bs{q}}\,
v^{\ }_{\bs{q}}\,
\Big{\langle}\,
\delta\widehat{\rho}^{\ }_{-\bs{q}}\, 
\sum_{\bs{p}}\,
\Lambda(\bs{p},\bs{q},\bs{k})\,
\widehat{\chi}^{\dag}_{\bs{p}}\,
\widehat{\chi}^{\ }_{\bs{p}+\bs{q}}
\Big{\rangle}
\\
&\,=
\frac{1}{2}\,\sum_{\bs{q}}\,\sum_{\bs{p}}\,
v^{\ }_{\bs{q}}\,
\Lambda(\bs{p},\bs{q},\bs{k})\,
\Big{\langle}\,
\delta\widehat{\rho}^{\ }_{-\bs{q}}\, 
\widehat{\chi}^{\dag}_{\bs{p}}\,
\widehat{\chi}^{\ }_{\bs{p}+\bs{q}}
\Big{\rangle}.
\end{split}
\end{equation}
The matrix element 
$
\Big{\langle}\,
\delta\widehat{\rho}^{\ }_{-\bs{q}}\, 
\widehat{\chi}^{\dag}_{\bs{p}}\,
\widehat{\chi}^{\ }_{\bs{p}+\bs{q}}
\Big{\rangle}
$
vanishes in the limit 
$
\bs{q} 
\rightarrow
0
$
and, 
therefore, 
the only term that contributes to $f^{\ }_{3,\bs{k}}$ 
up to order $\bs{q}^{2}\bs{k}^{2}$ is
\begin{equation}
\label{eq: f3 lattice 2}
\begin{split}
f^{\ }_{3,\bs{k}}
&\,=
\frac{1}{2}\,\sum_{\bs{q}}\,\sum_{\bs{p}}\,
v^{\ }_{\bs{q}}\,
\Big[
-q^{i}k^{j}k^{a}
\left(
\partial^{\ }_{a} F^{\ }_{i j}
\right)(\bs{p})
\Big]
\Big{\langle}\,
\delta\widehat{\rho}^{\ }_{-\bs{q}}\, 
\widehat{\chi}^{\dag}_{\bs{p}}\,
\widehat{\chi}^{\ }_{\bs{p}+\bs{q}}
\Big{\rangle}
\\
&\,=
\frac{1}{2}\,\sum_{\bs{q}}\,\sum_{\bs{p}}\,
v^{\ }_{\bs{q}}\,
\Big[
\left(
\bs{k}\wedge\bs{q}
\right)
\cdot
\left(
\frac{\partial \bs{B}}{\partial p^{a}}
\right)
(\bs{p})\,
\Big]k^{a}\,
\Big{\langle}\,
\delta\widehat{\rho}^{\ }_{-\bs{q}}\, 
\widehat{\chi}^{\dag}_{\bs{p}}\,
\widehat{\chi}^{\ }_{\bs{p}+\bs{q}}
\Big{\rangle}.
\end{split}
\end{equation}
The condition%
~(\ref{eq: Lambda condition})
implies that 
$
f^{\ }_{4,\bs{k}}=
f^{* }_{3,\bs{k}}
$,
which then delivers
\begin{equation}
\label{eq: f4 lattice}
f^{\ }_{4,\bs{k}}
=
\frac{1}{2}\,\sum_{\bs{q}}\,\sum_{\bs{p}}\,
v^{\ }_{\bs{q}}\,
\Big[-
\left(
\bs{k}\wedge\bs{q}
\right)
\cdot
\left(
\frac{\partial \bs{B}}{\partial p^{a}}
\right)
(\bs{p})\,
\Big]k^{a}\,
\Big{\langle}\,
\widehat{\chi}^{\dag}_{\bs{p}+\bs{q}}\,
\widehat{\chi}^{\ }_{\bs{p}}
\delta\widehat{\rho}^{\ }_{\bs{q}}\,
\Big{\rangle},
\end{equation}
where we have used that 
$
\left(
\bs{B}(\bs{p})
\right)^{*}
=
-
\bs{B}(\bs{p})
$.

Putting together all the contributions, we obtain
\begin{equation}
\begin{split}
f^{\ }_{\bs{k}}=&\,
- 
\sum_{\bs{q}}\,
\sum_{\bs{p},\bs{p}'}\, 
v^{\ }_{\bs{q}}
\Big[
\left(
\bs{k}\wedge\bs{q}
\right)
\cdot
\delta\bs{B}(\bs{p})
\Big]
\Big[
\left(
\bs{k}\wedge\bs{q}
\right)
\cdot
\delta\bs{B}(\bs{p}')
\Big]
\Big{\langle}
\hat{n}^{\ }_{\bs{p}}
\hat{n}^{\ }_{\bs{p}'}
\Big{\rangle}
\\
&\,
+
\frac{k^{a}}{2}\,
\sum_{\bs{q}}\,
\sum_{\bs{p}}\,
v^{\ }_{\bs{q}}\,
\left[
\left(
\bs{k}\wedge\bs{q}
\right)
\cdot
\left(
\frac{\partial \bs{B}}{\partial p^{a}}
\right)
(\bs{p})\,
\left\langle
\delta\widehat{\rho}^{\ }_{-\bs{q}}\, 
\widehat{\chi}^{\dag}_{\bs{p}}\,
\widehat{\chi}^{\ }_{\bs{p}+\bs{q}}
\right\rangle
-
\left(
\bs{k}\wedge\bs{q}
\right)
\cdot
\left(
\frac{\partial \bs{B}}{\partial p^{a}}
\right)
(\bs{p})
\left\langle
\widehat{\chi}^{\dag}_{\bs{p}+\bs{q}}\,
\widehat{\chi}^{\ }_{\bs{p}}\,
\delta\widehat{\rho}^{\ }_{\bs{q}}\,
\right\rangle
\right]
\end{split}
\end{equation}
where the summation convention over the repeated indices 
$a=1,\cdots,d$ is implied.
Finally, the analytical continuation
$\bs{\mathcal{B}}\equiv-\mathrm{i}\bs{B}$
delivers Eq.%
~(\ref{eq: f_k final expression}).

\medskip
\end{widetext}

\vskip 4 true cm


\begin{thebibliography}{99}

\bibitem{Klitzing80}
%New Method for High-Accuracy Determination of 
%the Fine-Structure Constant Based on Quantized Hall Resistance
K. V. Klitzing, G. Dorda, and M. Pepper, 
Phys.\ Rev.\ Lett.\ \textbf{45}, 494 (1980). 

\bibitem{Thouless82}
%Quantized Hall Conductance in a Two-Dimensional Periodic Potential
D. Thouless, M. Kohmoto, M. Nightingale, and M. den Nijs, 
Phys.\ Rev.\ Lett.\ \textbf{49}, 405 (1982).

\bibitem{Avron83} 
%Homotopy and Quantization in Condensed Matter Physics
J.\ E.\ Avron, R.\ Seiler, and B.\ Simon,
Phys.\ Rev.\ Lett.\ \textbf{51}, 51 (1983). 

\bibitem{Simon83}
%Holonomy, the Quantum Adiabatic Theorem, and Berry's Phase
B. Simon,
Phys.\ Rev.\ Lett.\ \textbf{51}, 2167 (1983).

\bibitem{Tsui82}
%Two-Dimensional Magnetotransport in the Extreme Quantum Limit
D. C. Tsui, H. L. Stormer, and A. C. Gossard, 
Phys.\ Rev.\ Lett.\ \textbf{48}, 1559 (1982). 

\bibitem{Laughlin83}
R. B. Laughlin, 
Phys.\ Rev.\ Lett.\ \textbf{50}, 1395 (1983).

\bibitem{Kane05a}
%Quantum Spin Hall Effect in Graphene
C.\ L.\ Kane and E.\ J.\ Mele,
Phys.\ Rev.\ Lett.\ \textbf{95}, 226801 (2005).

\bibitem{Kane05b} 
%Z2 Topological Order and the Quantum Spin Hall Effect
C.\ L.\ Kane and E.\ J.\ Mele,
Phys.\ Rev.\ Lett.\ \textbf{95}, 146802 (2005).

\bibitem{Bernevig06a} 
%Spin quantum Hall effect
B.\ A.\ Bernevig and S.-C.\ Zhang,
Phys.\ Rev.\ Lett.\ \textbf{96}, 106802 (2006).

\bibitem{Bernevig06b} 
%Quantum Spin Hall Effect and 
%Topological Phase Transition in HgTe Quantum Wells
B.\ A.\ Bernevig, T.\ L.\ Hughes, and S.-C.\ Zhang, 
Science \textbf{314}, 1757 (2006).

\bibitem{Konig07} 
%Quantum Spin Hall Insulator State in HgTe Quantum Wells 
M.\ K\"onig, % \textit{et al.}, 
S.\ Wiedmann, C.\ Br\"une, A.\ Roth, 
H.\ Buhmann, L.\ W.\ Molenkamp, X.-L.\ Qi, and S.-C.\ Zhang,
Science \textbf{318}, 766 (2007).

\bibitem{Fu07}
%\textit{``Topological insulators in three dimensions,''}
L. Fu, C. L. Kane, and E. J. Mele, 
Phys.\ Rev.\ Lett.\ \textbf{98}, 106803 (2007).

\bibitem{Moore07}
%\textit{``Topological invariants of time-reversal-invariant 
%band structures,''} 
J. E. Moore, and L. Balents, 
Phys.\ Rev.\ B \textbf{75}, 121306(R) (2007).

\bibitem{Qi08}
%Topological field theory of time-reversal invariant insulators
X. L. Qi, T. L. Hughes, and S. C. Zhang, 
Phys.\ Rev.\ B \textbf{78}, 195424 (2008).

\bibitem{Hsieh08}
%\textit{``A topological Dirac insulator in a quantum spin Hall phase,''}
D. Hsieh, D. Qian, L. Wray, Y. Xia, Y. S. Hor, R. J. Cava, and M. Z. Hasan,
Nature \textbf{452}, 970 (2008).

\bibitem{Hsieh09}
%\textit{``Observation of Unconventional Quantum Spin Textures 
%in Topological Insulators,''}
D. Hsieh, Y. Xia, L. Wray, D. Qian, A. Pal, 
J. H. Dil, J. Osterwalder, F. Meier, 
G. Bihlmayer, C. L. Kane, Y. S. Hor, R. J. Cava, 
and M. Z. Hasan,
Science \textbf{323}, 919 (2009).

\bibitem{Schnyder08}  
%``Classification of topological insulators and superconductors
%in three spatial dimensions,''
A.\ P.\ Schnyder, S.\ Ryu, A.\ Furusaki, and A.\ W.\ W.\ Ludwig,
Phys.\ Rev.\ B \textbf{78}, 195125 (2008).
%[arXiv:0803.2786v3 (cond-mat.mes-hall)].

\bibitem{Kitaev09}
%``Periodic table for topological insulators and superconductors,'' 
A.\ Kitaev,
AIP Conf.\ Proc.\ \textbf{1134}, 22 (2009).
%[arXiv:0901.2686 (cond-mat.mes-hall)].

\bibitem{Ryu10}
%Topological insulators and superconductors: 
%ten-fold way and dimensional hierarchy 
S.\ Ryu, A.\ P.\ Schnyder, A.\ Furusaki, and A.\ W.\ W.\ Ludwig,
New J.\ Phys.\ \textbf{12}, 065010 (2010).

\bibitem{Jain89} 
%Incompressible quantum Hall states
J. K. Jain, 
Phys.\ Rev.\ B \textbf{40}, 8079 (1989).

\bibitem{Wen91}
%NonAbelian statistics in the fractional quantum Hall states
X.-G.\ Wen,
Phys.\ Rev.\ Lett.\ \textbf{66}, 802 (1991).

\bibitem{Wen99} 
%Projective construction of nonAbelian quantum Hall liquids
X.-G.\ Wen,
Phys.\ Rev.\ B \textbf{60}, 8827 (1999).

\bibitem{Levin09}
%Gapless layered three-dimensional fractional quantum Hall states
M.\ Levin and M.\ P.\. A.\ Fisher,
Phys.\ Rev.\ B \textbf{79}, 235315 (2009).

\bibitem{Barkeshli10}
%Effective field theory and projective construction 
%for Zk parafermion fractional quantum Hall states
M.\ Barkeshli and X.-G.\ Wen,
Phys.\ Rev.\ B \textbf{81}, 155302 (2010).

\bibitem{Vaezi11}
%Fractional quantum Hall effect at zero magnetic field
A.\ Vaezi, 
(unpublished)
arXiv:1105.0406.

\bibitem{Lu11} 
%Symmetry protected fractional Chern insulators 
%and fractional topological insulators
Y.-M.\ Lu and Y.\ Ran,
(unpublished)
arXiv:1109.0226.

\bibitem{McGreevy11}
%Fractional Chern Insulators from the nth Root of Bandstructure
J.\ McGreevy, B.\ Swingle, and K.-A.\ Tran,
(unpublished)
arXiv:1109.1569v2.

\bibitem{Maciejko10}
%Fractional Topological Insulators in Three Dimensions
J.\ Maciejko, X.-L.\ Qi, A.\ Karch, and S.-C.\ Zhang, 
Phys.\ Rev.\ Lett.\ \textbf{105}, 246809 (2010).

\bibitem{Swingle11}
%Correlated topological insulators and the fractional magnetoelectric effect
B. Swingle, M. Barkeshli, J. McGreevy, and T. Senthil,
Phys. Rev. B \textbf{83}, 195139 (2011).

\bibitem{Maciejko11} 
%Models of three-dimensional fractional topological insulators
J.\ Maciejko, X.-L.\ Qi, A.\ Karch, and S.-C.\ Zhang,
(unpublished)
arXiv:1111.6816. 

\bibitem{Haldane12}
%Self-duality and long-wavelength behavior of 
%the Landau-level guiding-center structure function, 
%and the shear modulus of fractional quantum Hall fluids
F. D. M. Haldane,
(unpublished)
arXiv:1112.0990.

\bibitem{Haldane11}
%Geometrical Description of the Fractional Quantum Hall Effect
F. D. M. Haldane,
Phys.\ Rev.\ Lett.\ \textbf{107}, 116801 (2011).

\bibitem{Girvin85}
%Collective-Excitation Gap in the Fractional Quantum Hall Effect
S. M. Girvin, A. H. MacDonald, and P. M. Platzman, 
Phys.\ Rev.\ Lett.\ \textbf{54}, 581 (1985).

\bibitem{Feynman72}
R. P. Feynman, 
Statistical Mechanics (Benjamin, Reading, Mass., 1972), Chap. 11.

\bibitem{Iso92}
%Fermions in the lowest Landau level. 
% Bosonization, W�� algebra, droplets, chiral bosons 
S. Iso, D. Karabali, and B. Sakita,
Phys.\ Lett.\ B \textbf{296}, 143 (1992).

\bibitem{Cappelli93}
%Large N limit in the quantum Hall effect, 
A. Cappelli, C. A. Trugenberger, and G. R. Zemba, 
Phys.\ Lett.\ B \textbf{306}, 100 (1993);
%INFINITE SYMMETRY IN THE QUANTUM HALL-EFFECT
Nucl.\ Phys.\ B\textbf{396}, 465 (1993) and
%CONFORMAL SYMMETRY AND UNIVERSAL PROPERTIES OF QUANTUM HALL STATES 
Nucl.\ Phys.\ B\textbf{398}, 531 (1993).

\bibitem{Martinez93}
%Current operators in the lowest Landau level
J. Martinez and M. Stone, 
Int.\ J. Mod.\ Phys. B\textbf{7},  4389 (1993).

\bibitem{Parameswaran11}
%Fractional Chern Insulators and the W-Infinity Algebra 
S. Parameswaran, R. Roy, and S. Sondhi, 
arXiv:1106.4025 (unpublished).

\bibitem{Goerbig12}
%From Fractional Chern Insulators to a Fractional Quantum Spin Hall Effect
M. O. Goerbig,
%arXiv:1107.1986 (unpublished).
The European Physical Journal B \textbf{85}, 14 (2012). 

\bibitem{Bernevig11}
%Emergent Many-Body Translational Symmetries of Abelian and NonAbelian Fractionally Filled Topological Insulators
B.\ A.\ Bernevig and N.\ Regnault,
arXiv:1110.4488 (unpublished).

\bibitem{Neupert11b}
%Fractional topological liquids with time-reversal symmetry 
%and their lattice realization
T.\ Neupert, L.\ Santos, S.\ Ryu, C.\ Chamon, and C.\ Mudry,
Phys.\ Rev.\ B \textbf{84}, 165107 (2011).

\bibitem{Neupert12}
%The topological Hubbard model and its high-temperature quantum Hall effect 
T.\ Neupert, L.\ Santos, S.\ Ryu, C.\ Chamon, and C.\ Mudry,
Phys.\ Rev.\ Lett.\ \textbf{108}, 046806 (2012).

\bibitem{Xiao11} 
%Interface engineering of quantum Hall effects 
%in digital heterostructures of transition-metal oxides
D. Xiao, W. Zhu, Y. Ran, N. Nagaosa, and S. Okamoto,
Nature Communications \textbf{2}, 596 (2011).

%section density algebra

\bibitem{GMP vs W algebra} 
The GMP algebra is an instance of an algebra
also known as the Moyal or $W^{\ }_{\infty}$ algebra,
see Refs.~\onlinecite{Moyal49},
\onlinecite{Fairlie89},
\onlinecite{Bakas89},
and
\onlinecite{Hoppe90}.

\bibitem{Moyal49}
%Quantum mechanics as a statistical theory Journal
%Mathematical Proceedings of the Cambridge Philosophical Society 
J. E. Moyal,
Math.\ Proc.\ Camb.\ Phil,\ Soc.\ \textbf{45}, 99 (1949). 

\bibitem{Fairlie89}
%Trigonometric structure constants for new infinite-dimensional algebras 
D.B. Fairlie, P. Fletcher, and C.K. Zachos,
Phys.\ Lett.\ B\textbf{218}, 203 (1989).

\bibitem{Bakas89}
%The large-N limit of extended conformal symmetries
I. Bakas,
Phys.\ Lett.\ B\textbf{228}, 57 (1989).

\bibitem{Hoppe90}
%Infinitel many versions of U(infty)
J. Hoppe and P Schaller,
Phys.\ Lett.\ B\textbf{237}, 407 (1990).

\bibitem{Nambu73}
%Generalized Hamiltonian Dynamics
Y.\ Nambu,
Phys.\ Rev.\ D \textbf{7}, 2405 (1973).

\bibitem{Neupert11a}
%Fractional Quantum Hall States at Zero Magnetic Field
T.\ Neupert, L.\ Santos, C.\ Chamon, and C.\ Mudry,
Phys.\ Rev.\ Lett.\ \textbf{106}, 236804 (2011).

\bibitem{Sheng11}
%Fractional quantum Hall effect in the absence of Landau levels
D.\ N.\ Sheng, Z.\ Gu, K.\ Sun, and L.\ Sheng,
Nature Communications \textbf{2}, 389 (2011).

\bibitem{Wang11a}
%Fractional Quantum Hall Effect of Hard-Core Bosons in Topological Flat Bands
Y.-F.\ Wang, Z.-C.\ Gu, C.-D.\ Gong, D.\ N.\ Sheng,
Phys.\ Rev.\ Lett.\ \textbf{107}, 146803 (2011).

\bibitem{Regnault11}
%Fractional Chern Insulator 
N.\ Regnault and B.\ A.\ Bernevig,
Phys.\ Rev.\ X \textbf{1}, 021014 (2011). 
%arXiv:1105.4867 (unpublished).

\bibitem{Wang10}
%Topological Order Parameters for Interacting Topological Insulators
Zhong Wang, Xiao-Liang Qi, and Shou-Cheng Zhang, 
Phys.\ Rev.\ Lett.\ \textbf{105}, 256803 (2010).
 
\bibitem{Malashevich10}
%Theory of orbital magnetoelectric response 
A. Malashevich, I. Souza, S. Coh, and D. Vanderbilt,
New J. Phys.\ \textbf{12}, 05032 (2010).

\bibitem{Essin10} 
%Orbital magnetoelectric coupling in band insulators
A. M. Essin, A. M. Turner, J. E. Moore, and D. Vanderbilt, 
Phys.\ Rev.\ B \textbf{81}, 205104 (2010). 

\bibitem{Coh11}
%Chern-Simons orbital magnetoelectric coupling in generic insulators
S.\ Coh, D.\ Vanderbilt, A.\ Malashevich, and I.\ Souza,
Phys.\ Rev.\ B \textbf{83}, 085108 (2011).

\bibitem{Chen11a}
%Topological insulator and the θ vacuum in a system without boundaries
Kuang-Ting Chen and Patrick A. Lee,
Phys.\ Rev.\ B \textbf{83}, 125119 (2011).

\bibitem{Chen11b}
%Unified formalism for calculating polarization, magnetization, 
%and more in a periodic insulator
Kuang-Ting Chen and Patrick A. Lee,
Phys.\ Rev.\ B \textbf{84}, 205137 (2011).

\bibitem{Thouless84}  
%Wannier functions for magnetic sub-bands 
D. J. Thouless, 
J. Phys. C: Solid State Phys. \textbf{17}, L325 (1984). 

\bibitem{Soluyanov11}
%Wannier representation of Z2 topological insulators
Alexey A. Soluyanov and David Vanderbilt,
Phys.\ Rev.\ B \textbf{83}, 035108 (2011).

\bibitem{King-Smith93}
%Theory of polarization of crystalline solids
R.\ D.\ King-Smith and D.\ Vanderbilt, 
Phys.\ Rev.\ B \textbf{47}, 1651 (1993).

\bibitem{footnote reg 3-bracket}
The regularization of the $3$-bracket involving the operators
$\widehat{X}^{\ }_{R}$ defined in the Wannier basis allows for
a representation of $\mathrm{CS}^{(3)}$ in terms of expectation
values of the position operators in the Wannier states, 
similar to that used in Ref.~\onlinecite{Coh11}. 
In Eq. (41) of Ref.~\onlinecite{Estienne12},
a similar regularization procedure for the trace of the $3$-bracket of the
projected density operators was defined so as to deliver an expression
for $\mathrm{CS}^{(3)}$.

\bibitem{Estienne12}
%D-Algebra Structure of Topological Insulators
B.\ Estienne, N.\ Regnault, and B.A.\ Bernevig,
arXiv:1202.5543 (unpublished).

\bibitem{footnote on Stokes thm}
When $d=2$, this quantization is a consequence of
the Gauss-Bonnet theorem that can be understood as follows
if we consider the Abelian case
$\widetilde{N}=1$ when $d=2$. 
If so, $F^{\ }_{12}=-F^{\ }_{21}$ 
is the (two-dimensional) rotation of a scalar field.
Stokes theorem then allows to convert the two-dimensional integral 
into a one-dimensional integral along the boundary, i.e., a winding number.
Any higher-dimensional integral in $d>2$ over a 
$T^{d}$ torus can then be decomposed into an integral over 
the Cartesian product of $T^{2}$ and $T^{d-2}$. We have established
that the integral over $T^{2}$ 
is quantized for every given $\bs{k}\in T^{d-2}$. 
The integrand over $T^{d-2}$ is then necessarily
constant as a function of $\bs{k}\in T^{d-2}$, 
since we can think of $\bs{k}\in T^{d-2}$
as an extra parameter in the Hamiltonian as a function of
which the gap does not close by assumption.

\bibitem{Avron95}
%Viscosity of Quantum Hall Fluids
J.E.\ Avron, R.\ Seiler, and P.\ G.\ Zograf,
Phys.\ Rev.\ Lett.\ \textbf{75}, 697 (1995).

\bibitem{Bachcall91}
S.\ Bachcall and L.\ Susskind,
Int.\ J.\ Mod.\ Phys.\ B \textbf{5}, 2735 (1991);
L.\ Susskind, arxiv:hep-th/0101029 (unpublished).

\bibitem{Jackiw04}
R.\ Jackiw, V.P.\ Nair, S.-Y.\ Pi, and A.P.\ Polychronakos,
J.\ Phys.\ A: Math.\ Gen.\ \textbf{37}, R327 (2004).

\bibitem{Polychronakos07}
A.P.\ Polychronakos, \textit{Seminaire Poincare X}, Institut Henri Poincare, Paris, 2007;
arXiv:0706.1095.

\bibitem{footnote on ideal 3D classical fluids}
L. Santos, T. Neupert, S. Ryu, C. Chamon, and C. Mudry,
unpublished.

\bibitem{Cho11}
%Topological BF field theory description of topological insulators
G.Y.\ Cho and J.E.\ Moore,
Annals Phys.\ \textbf{326}, 1515 (2011).

\bibitem{Kohmoto92}
M.\ Kohmoto, B.\ I.\ Halperin, and Y.-S.\ Wu, 
Phys.\ Rev.\ B \textbf{45}, 13488 (1992).

%section noninteracting model

\bibitem{Moore08}
%Topological surface states in three-dimensional magnetic insulators 
J.\ E.\ Moore, Y.\ Ran, and X.-G.\ Wen,
Phys. Rev. Lett. \textbf{101}, 186805 (2008).

\bibitem{Gell-Mann62}  
%Symmetries of Baryons and Mesons
M.\ Gell-Mann,
Phys.\ Rev.\ \textbf{125}, 1067 (1962).

\end{thebibliography}
\end{document}